\documentclass[a4paper,11pt]{article}
\pdfoutput=1 

\usepackage{jheppub} 
                     
\usepackage{placeins}

\usepackage{graphicx}
\usepackage{multirow}
\usepackage{amsmath,amssymb,amsfonts}
\usepackage{amsthm}
\usepackage{mathrsfs}
\usepackage[title]{appendix}
\usepackage{xcolor}
\usepackage{textcomp}
\usepackage{subfig}
\usepackage{manyfoot}
\usepackage{booktabs}
\usepackage{algorithm}
\usepackage{algorithmicx}
\usepackage{algpseudocode}
\usepackage{listings}
\usepackage{siunitx}
\usepackage{float}
\usepackage{comment}
\usepackage[capitalise]{cleveref}
\usepackage{enumitem}
\usepackage{tabularx}

\usepackage{xspace}
\usepackage{main-defs}
\usepackage{rotating}
\usepackage{soul}
\usepackage{adjustbox}
\usepackage{pdflscape}
\usepackage{longtable}
\usepackage[utf8]{inputenc}
\usepackage{newunicodechar}
\usepackage{subcaption}
\newunicodechar{−}{\ensuremath{-}}

\newcommand{\cza}{\ensuremath{c_{{\scriptscriptstyle Z}\gamma}}}
\newcommand{\caa}{\ensuremath{c_{\gamma\gamma}}}
\newcommand{\czz}{\ensuremath{c_{{\scriptscriptstyle Z}{\scriptscriptstyle Z}}}}
\newcommand{\cww}{\ensuremath{c_{{\scriptscriptstyle W}{\scriptscriptstyle W}}}}
\newcommand{\kza}{\ensuremath{\kappa_{{\scriptscriptstyle Z}\gamma}}}
\newcommand{\kaa}{\ensuremath{\kappa_{\gamma}}}
\newcommand{\gza}{\ensuremath{g_{{\scriptscriptstyle Z}\gamma}}}
\newcommand{\gaa}{\ensuremath{g_{\gamma\gamma}}}
\newcommand{\gzz}{\ensuremath{g_{{\scriptscriptstyle Z}{\scriptscriptstyle Z}}}}
\newcommand{\gww}{\ensuremath{g_{{\scriptscriptstyle W}{\scriptscriptstyle W}}}}

\title{\boldmath Probing the Effective HZ$\gamma$ Coupling via H$\gamma$ Production at FCC-ee }

\author[a]{Sara Aumiller,}
\author[b]{Lena Herrmann,}
\author[c]{Ken Mimasu,}     
\author[d]{Louis Portalès,}
\author[b]{and Michele Selvaggi}

\affiliation[a]{University of California, Irvine, USA}
\affiliation[b]{European Organization for Nuclear Research (CERN), Geneva, Switzerland}
\affiliation[c]{University of Southampton, UK}
\affiliation[d]{IRFU, CEA, Université Paris-Saclay, Gif-sur-Yvette, France}

\emailAdd{sara.aumiller@cern.ch}
\emailAdd{lena.maria.herrmann@cern.ch}
\emailAdd{ken.mimasu@soton.ac.uk}
\emailAdd{louis.portales@cern.ch}
\emailAdd{michele.selvaggi@cern.ch}

\abstract{

This work presents the first dedicated study of the \eeha\ production mode at the FCC-ee. The analysis exploits all foreseen FCC-ee running scenarios at and above the WW threshold. This study demonstrates that the \eeha\ process provides a novel and direct handle on the effective \vhaa\ and \vhza\ couplings through production, complementary to measurements based on Higgs decays. By combining measurements of multiple Higgs decay channels and center-of-mass energies, we show that an overall precision of 15\% is achievable on the effective \vhza\ coupling. In addition, unlike decay observables, this process is also sensitive to the relative sign of the coupling and allows a full resolution of the corresponding sign ambiguity. We compare the expected sensitivity to the projected bounds from direct measurements of the $\gamma\gamma$ and $Z\gamma$ branching ratios at the HL-LHC and FCC-ee, as well as projections from global fits in the SMEFT framework. 
}

\begin{document} 
\maketitle
\flushbottom

\section{Introduction}

A primary goal of the Higgs physics programme is the precise characterisation of Higgs boson properties. The large integrated luminosity of the High-Luminosity LHC (HL-LHC)~\cite{ATLAS:2025eii} will substantially improve upon LHC Higgs precision measurements~\cite{ATLAS:2022vkf, CMS:2022dwd}, enabling percent-level measurements of the dominant Higgs production and decay rates, enhanced sensitivity to rare decay modes such as $H\to Z\gamma$ and $H\to \mu^+\mu^-$, and direct access to the Higgs self-interaction via di-Higgs production.

The Future Circular Collider in the electron-positron mode (FCC-ee)~\cite{FCC:2025uan, FCC:2018byv, FCC:2025lpp, Blondel:2025kbl} is currently being proposed as the next collider facility at CERN. It consists of a high-luminosity Higgs factory with an extensive physics programme spanning center-of-mass energies between $\sqrt{s}=\SI{90}{\giga\electronvolt}$ and $\sqrt{s}=\SI{365}{\giga\electronvolt} $. A central objective of the FCC-ee programme is to achieve a significant improvement in the precision of Higgs coupling measurements relative to the HL-LHC~\cite{deBlas:2025gyz}.

At the FCC-ee, Higgs bosons are predominantly produced via Higgsstrahlung ($ZH$) and vector-boson-fusion processes ($\nu\nu H$, $\ell\ell H$). By exploiting precise measurements of Higgs decay modes in a wide range of final states~\cite{Selvaggi:2025kmd}, these production channels will enable sensitivity to Higgs couplings ranging from the per-mille to the percent level~\cite{FCC:2025lpp, Blondel:2025kbl, deBlas:2025gyz}. Other rare production modes, such as s-channel Higgs production at $\sqrt{s}=\SI{125}{\giga\electronvolt}$, have also been proposed to probe the first-generation electron Yukawa coupling~\cite{dEnterria:2021xij,Fatehi:2026gnt}.

The effective $H\gamma\gamma$ and $HZ\gamma$ couplings 
($\kaa$ and $\kappa_{Z\gamma}$, see~\cref{sec:theory} for a brief review)
are commonly extracted through the $H\to\gamma\gamma$ and $H\to Z\gamma$ decay branching fractions. While the projected sensitivity on $\kaa$ is expected to be dominated by the HL-LHC, with a projected precision of $1.6\%$~\cite{ATLAS:2025eii}, the sensitivity to $\kza$ will be driven by a combination of HL-LHC and FCC-ee measurements, reaching $4.3\%$ at FCC-ee~\cite{FCC:2025uan, deBlas:2025gyz}. 
In this paper we investigate for the first time in the context of FCC-ee, a complementary Higgs production mode, $e^+e^-\to H\gamma$. This production process is loop-induced in the Standard Model (SM) and can provide direct and complementary sensitivity to both the magnitude and the sign of effective $H\gamma\gamma$ and $HZ\gamma$ interactions. Feasibility studies have previously been performed in the context of the International Linear Collider (ILC) programme~\cite{Aoki:2021khh}, investigating the potential of this channel as a probe of anomalous Higgs couplings.

A further advantage of studying these effective couplings in production is the possibility of exploiting multiple center-of-mass energy scenarios. In particular, the $WW$ threshold run at $\sqrt{s}=\SI{160}{\giga\electronvolt}$, foreseen to deliver an integrated luminosity of $\SI{19.2}{\per\atto\barn}$, already provides a non-negligible $H\gamma$ cross section. This can be combined with measurements at $\sqrt{s}=\SI{240}{\giga\electronvolt}$ ($\SI{10.8}{\per\atto\barn}$) and $\sqrt{s}=\SI{365}{\giga\electronvolt}$ ($\SI{3.12}{\per\atto\barn}$), enhancing the overall sensitivity and allowing one to exploit the characteristic energy dependence of the loop-induced amplitudes and possible EFT contributions.
Finally, since Higgs branching ratios will be precisely constrained by the global FCC-ee Higgs programme, the $H\gamma$ production process can be studied in several Higgs decay modes. This enables a multi-channel approach that maximises sensitivity while providing robust internal cross-checks. 

In this work we present a comprehensive sensitivity study of the $e^+e^-\to H\gamma$ 
production mode in the most relevant final states. On the hadronic modes, we focus on 
the $H\to b\bar{b}$ decay, as other hadronic modes ($H\to gg$, $H\to\tau^+\tau^-$) 
are found to provide negligible sensitivity. For the $H\to WW^*$ decay, we consider 
the semi-leptonic final states $H\to W(\ell\nu)\,W^*(qq)$ and $H\to W(qq)\,W^*(\ell\nu)$ 
with $\ell=e,\mu$, which offer the best compromise between signal yield and 
background rejection.

The structure of the paper is as follows. \Cref{sec:theory} introduces the 
theoretical framework and the effective interactions relevant for $H\gamma$ 
production. \Cref{sec:mc} describes the Monte Carlo simulation setup, the 
background processes considered, and the event reconstruction. \Cref{sec:analysis} 
presents the analysis strategy for the $H\to b\bar{b}$ and semi-leptonic 
$H\to WW^*$ signal channels, and reports the expected signal yields and 
combined sensitivity. In~\cref{sec:interpretation} we discuss the implications 
of the results in the effective field theory interpretation, comparing the projected sensitivity to that from Higgs boson branching ratio measurements at the HL-LHC and FCC-ee 
and to projections from global fits. We conclude in~\cref{sec:conc}.

\section{Theoretical Framework}
\label{sec:theory}
In the SM, the production of a Higgs boson in association with a photon is a rare process. The production of this final state requires charged particles in the initial state, and the tree-level amplitude is proportional to the initial state Yukawa coupling. Since collider initial states typically involve light particles, this coupling is generally very small. 
At hadron colliders, the leading-order contributions from the light quark initial states ($u\bar{u},d\bar{d},s\bar{s}$) are negligible, and the loop-induced $gg\to H\gamma$ amplitude is forbidden by Furry's theorem and $C$-parity of the strong and electromagnetic interactions.  The main tree-level contribution thus comes from the $c\bar{c}$- and $b\bar{b}$-initiated amplitudes, despite their relative PDF suppressions, with the former accounting for about 40\% of the $\mathcal{O}(1$ \si{\femto\barn}$)$ rate at $\sqrt{s}=\SI{13.6}{\tera\electronvolt}$. Electroweak loops contribute at the level of about 10\% at this center-of-mass energy~\cite{Abbasabadi:1997zr}. The $pp\to H\gamma$ process has therefore been proposed as an alternative way to probe the light quark Yukawa couplings~\cite{Aguilar-Saavedra:2020rgo}, as well as modified Higgs boson interactions in Effective Field Theory (EFT) frameworks~\cite{Khanpour:2017inb,Cao:2021trr,Biswas:2022fsr}.

\begin{figure}[t!]
  \centering
  \subfloat[Top triangle loop\label{fig:toploop}]{
\includegraphics[height=0.2\textwidth]{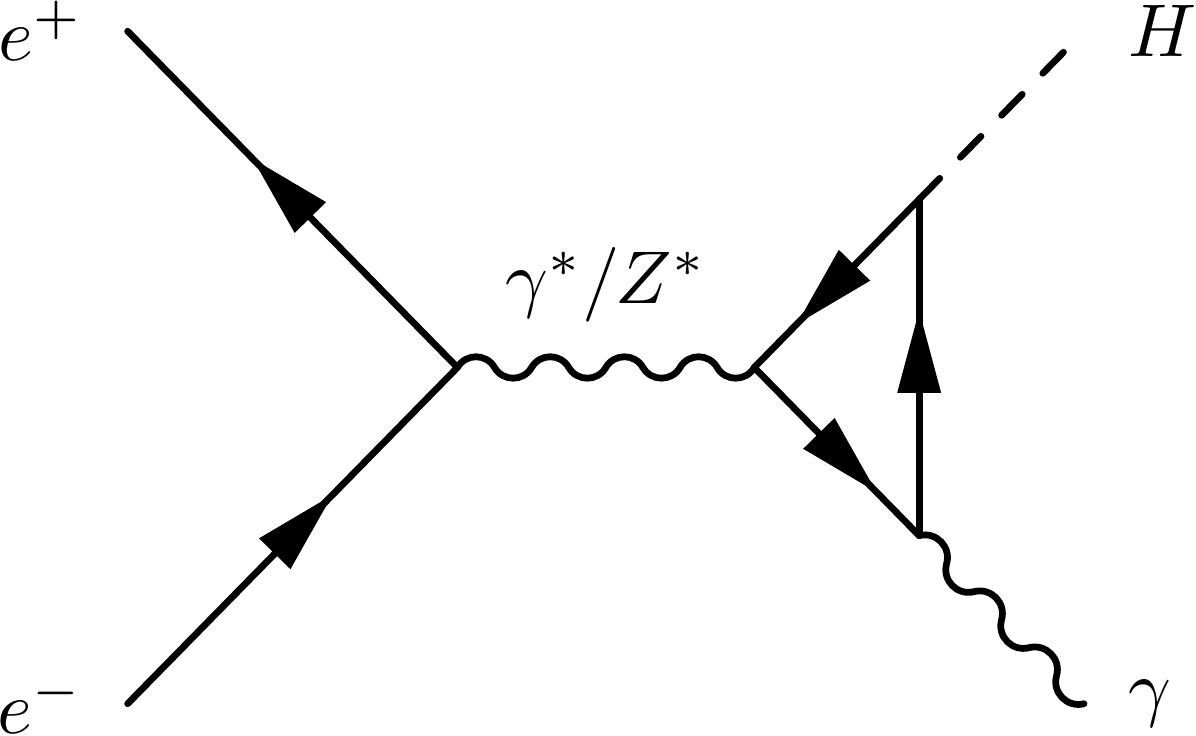}
  }\hspace{1cm}
  \subfloat[$W$ boson triangle loop\label{fig:vloop}]{
\includegraphics[height=0.2\textwidth]{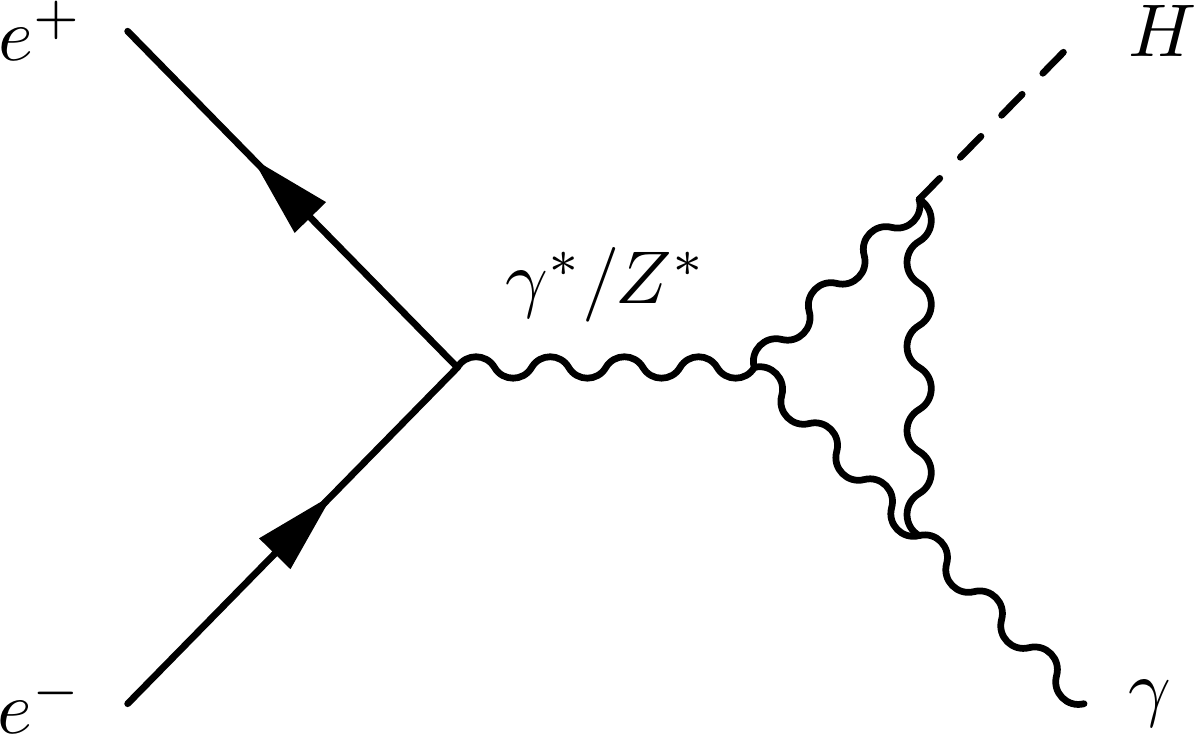}
  }\\
  \subfloat[Box diagram\label{fig:boxdiagram}]{
\includegraphics[height=0.2\textwidth]{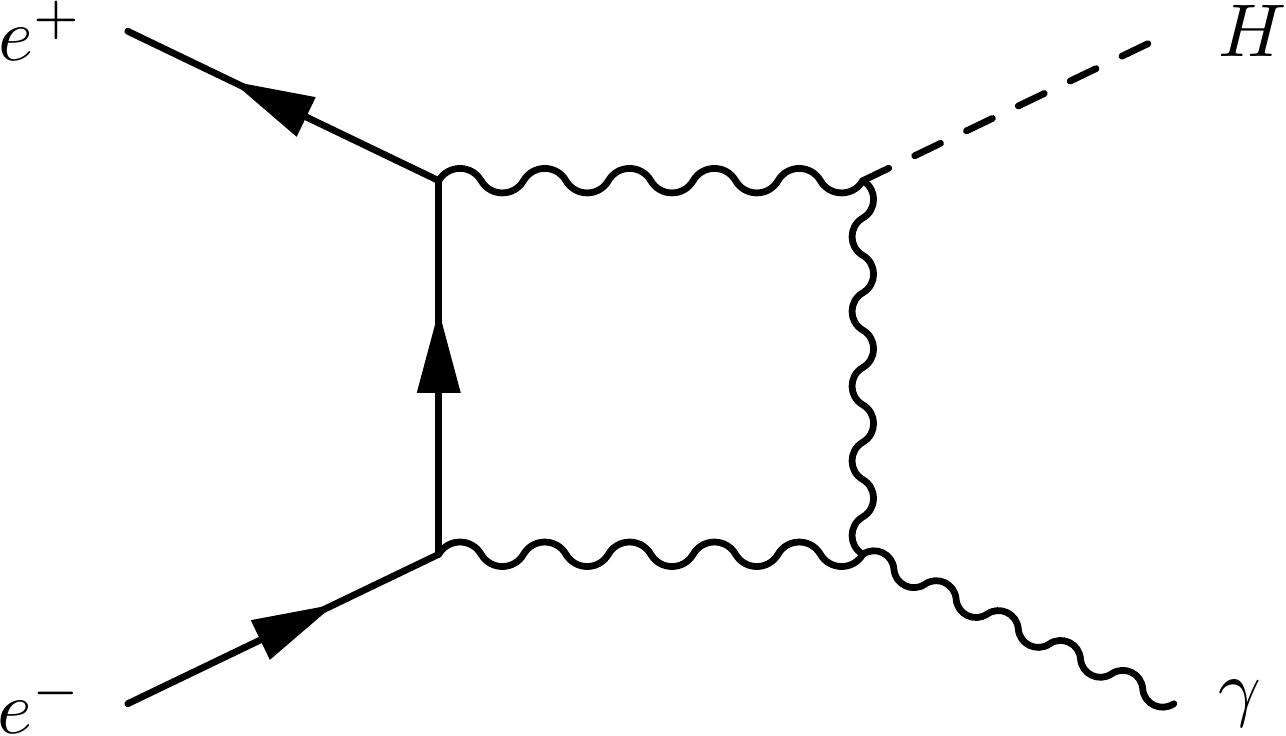}
  }\hspace{0.6cm}
  \subfloat[Anomalous $\gamma\gamma/\gamma Z$ couplings\label{fig:anomalousdiagram}]{
\includegraphics[height=0.2\textwidth]{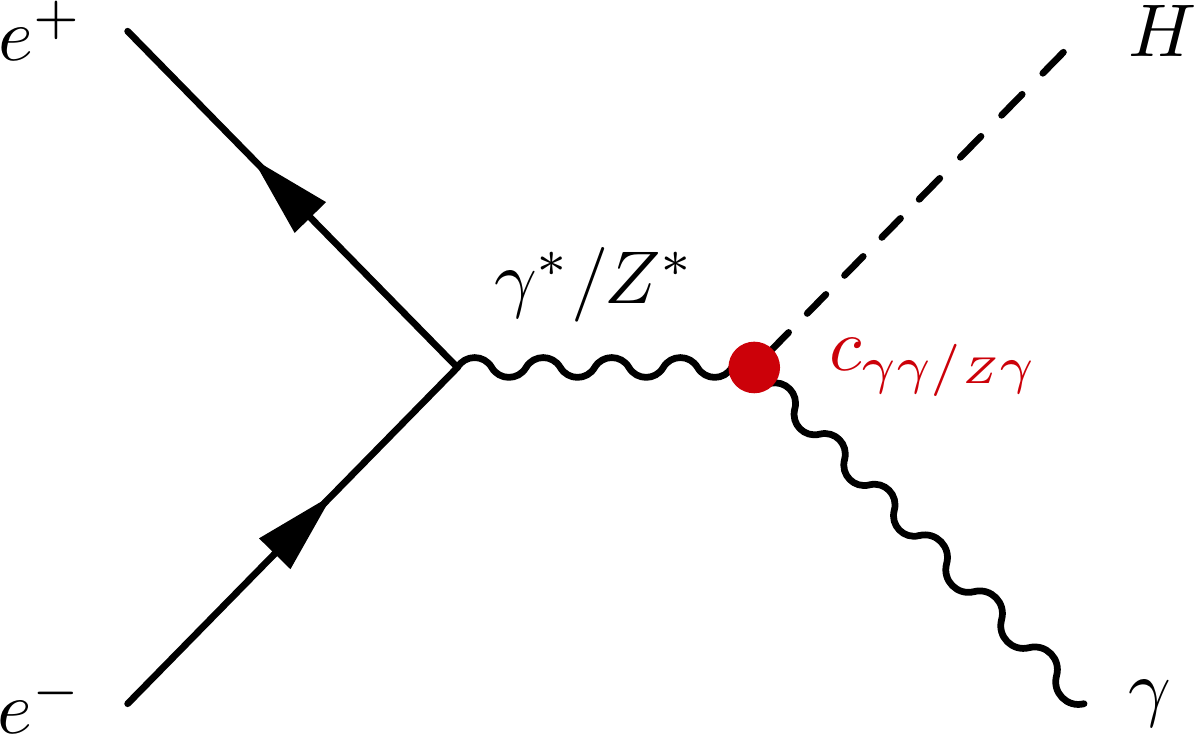}
  }
  \caption{Feynman diagrams for the $e^+e^-\rightarrow H\gamma$ process. Diagrams (a), (b) and (c) show representative one-loop topologies in the SM with the potential insertion of a Higgs coupling modifier to top quarks and $W$-bosons (adapted from Ref.~\cite{Abbasabadi:1995rc}). Diagram (d) shows the tree-level contribution of an anomalous $\gamma\gamma$ or $Z\gamma$ coupling of the Higgs boson as defined in~\cref{eq:L_eff}.}
  \label{fig:feynmanprod}
\end{figure}

\begin{figure}[h!]
    \centering
    \includegraphics[width=12cm]{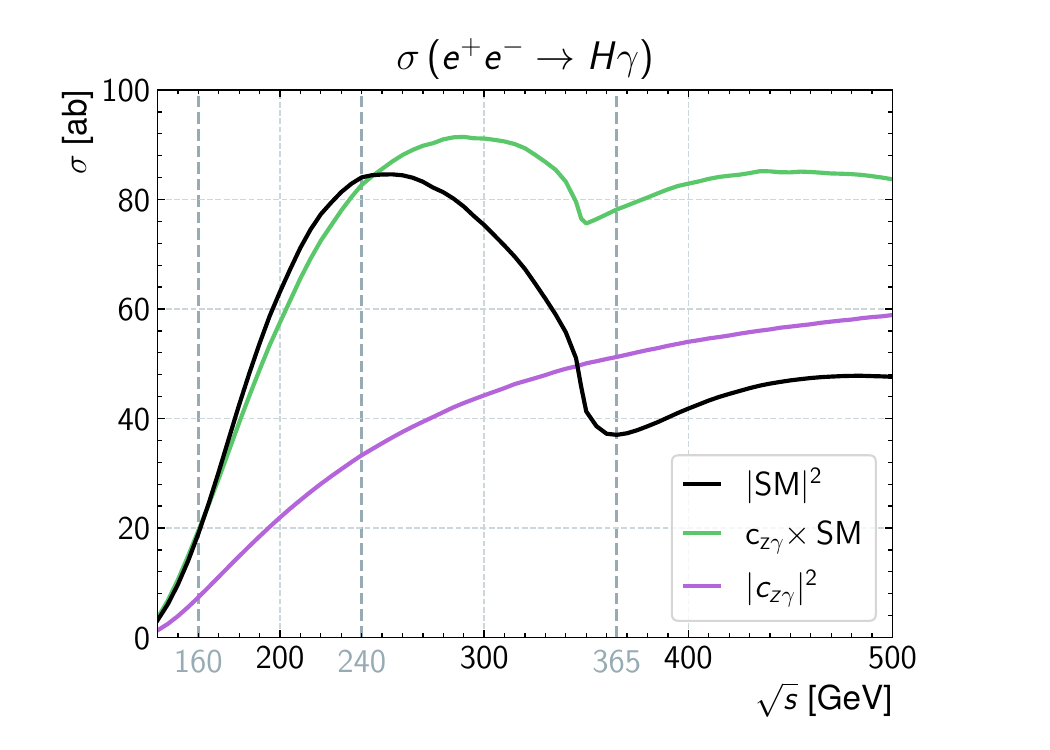}
    \caption{Cross section ($\sigma$) of \eeha\ as a function of center-of-mass energy $\sqrt{s}$ for the SM in black and the contributions from the effective $Z\gamma$ coupling of the Higgs, $\cza$, defined in~\cref{eq:L_eff}. The contribution from the interference between the SM and $\cza$ (proportional to $\cza$) and the pure $\cza$ contribution  (proportional to $\cza^2$) are shown in green and purple, respectively for $\cza=1$. The total (SM+BSM) prediction would correspond to the sum of the three contributions. 
    The dashed gray lines indicate the FCC-ee center-of-mass energies.}
    \label{fig:sigmahgamma}
\end{figure}

At $e^+e^-$ colliders, the small electron Yukawa coupling leads to truly negligible tree-level diagrams, and the dominant contributions to \eeha\ stem from the EW loop diagrams shown in~\cref{fig:feynmanprod}.
The triangle diagrams (\cref{fig:toploop,fig:vloop}) interfere destructively with box diagrams (\cref{fig:boxdiagram}), resulting in a cross-section that is highly sensitive to the center-of-mass energy, as shown in~\cref{fig:sigmahgamma} (black line). One can see a significant reduction of the cross section around the top-pair threshold, where the diagrams involving top quark loops are enhanced through developing an imaginary part. 
From the diagrams in~\cref{fig:feynmanprod}, one can see that the \eeha\ process is sensitive in principle to modified Higgs boson interactions with both fermions and gauge bosons. 

Studies that characterise the Higgs boson couplings often make use of the so-called kappa framework~\cite{LHCHiggsCrossSectionWorkingGroup:2012nn} to model deviations of Higgs boson couplings from SM expectations, in which a scale-factor, $\kappa_i$, is assigned to each  Higgs boson coupling, defined by the ratio of the corresponding Higgs boson production cross-section or partial width to the SM expectation,
\begin{align}
\label{eq:kappa_def}
\frac{\sigma(ii\to h)}{\sigma_{\mathrm{SM}}(ii\to h)}\equiv \kappa_i^2\quad \mathrm{or}
\quad
\frac{\Gamma(h\to jj)}{\Gamma_{\mathrm{SM}}(h\to jj)}\equiv \kappa_j^2\,.
\end{align}
Since loop-induced couplings to massless gauge bosons ($gg,\gamma\gamma,Z\gamma$) are especially relevant for the Higgs boson, associated kappa factors ($\kappa_g$, $\kaa$ and $\kza$, respectively) are also defined for these effective couplings, and can be determined in this framework through signal strengths involving gluon fusion or branching ratios to $\gamma\gamma/Z\gamma$. In a theory where only tree-level SM Higgs couplings are modified, these are entirely determined by the $\kappa$ factors that enter in a given particle's loop contribution to the associated effective coupling (\emph{e.g.} quark couplings for $\kappa_{g}$ ). In general, however, one can consider these factors as independent parameters, representing, \emph{e.g.}, the case where some BSM particles may run in the loops. Such a scenario would correspond to having additional tree-level diagrams that directly mediate $h\gamma\gamma$ and $hZ\gamma$ interactions. 

Since Higgs couplings to massless gauge bosons are not present in the SM Lagrangian, they must be expressed via higher dimensional operators in an effective coupling framework. Neglecting the gluon coupling, these interactions correspond to the dimension-5 operators
\begin{align}
\label{eq:L_eff}
    \mathcal{L}^\mathrm{eff.}\supset \gaa\,\caa \frac{h}{v} F^{\mu\nu}F_{\mu\nu} + \gza\,\cza \frac{h}{v} F^{\mu\nu}Z_{\mu\nu},
\end{align}
where $F_{\mu\nu}$ and $Z_{\mu\nu}$ correspond to the Abelian gauge field strength tensors of the photon and $Z$-boson, respectively. The diagram representing their contributions to the \eeha\ process is shown in~\cref{fig:anomalousdiagram}, where $\caa$ and $\cza$ enter in the diagram with the intermediate photon and $Z$-boson, respectively. The normalisations of the coefficients are set by 
\begin{align}
\label{eq:g_norm}
\gaa = \frac{47\alpha}{18\pi},\quad
\gza = \sqrt{\frac{\alpha G_F M_Z^2}{8\sqrt{2}\pi}}\frac{94\cos\theta_W-13}{9\pi},
\end{align}
where $\alpha$, $G_F$, $M_Z$, and $\theta_W$ denote the electromagnetic fine structure constant, Fermi constant, $Z$-boson mass and Weinberg angle, respectively.
This choice corresponds to the values of the leading-order effective couplings generated by the top quark and $W$-boson loops in the SM, when their masses are taken to infinity~\cite{kniehl1995lowenergytheoremshiggsphysics}. In this way, $c=1$ will lead to an $\mathcal{O}(1)$ modification to the corresponding Higgs boson decay amplitude. Note that we are assuming here that the effective operators are added on top of the SM such that, in the infinite top and $W$-boson mass limit, we would be able to relate the $c$ coefficients to the corresponding $\kappa$
\begin{align}
\kappa_{\gamma/{\scriptscriptstyle Z}\gamma}\simeq
1+c_{\gamma\gamma/{\scriptscriptstyle Z}\gamma} ,
\end{align}
and the SM hypothesis of $\kappa_{\gamma/{\scriptscriptstyle Z}\gamma}=1$ implies $c_{\gamma\gamma/{\scriptscriptstyle Z}\gamma}=0$. As we will see below, the infinite top and $W$-boson mass limit  is not a good approximation when computing Higgs branching fractions to $\gamma\gamma$ and $Z\gamma$ and correction factors must be applied when mapping the $c$ coefficients to the $\kappa$ parameters.

It is useful to frame the measurement of Higgs properties in terms of the $\kappa$'s, but it is also important to remember that such quantities are only
strictly defined for the production and decay of an on-shell Higgs boson. Moreover, the effective operators introduced to capture the couplings to massless gauge bosons are normalised in terms of their relative effect on an on-shell Higgs decay, where the scale is set by $M_H$. Being effective operators, $\cza$ and $\caa$ lead to amplitudes that grow with energy, such that their relative impact changes as a function of it. The coefficient values that lead to $O(1)$ effects in Higgs boson partial widths may give enhanced effects when considering their contribution to higher energy processes such as \eeha\ production at center-of-mass energies above $M_H$. The study of the \eeha\ process at lepton colliders in terms of the Higgs couplings to photons or $Z\gamma$ therefore requires one to go beyond the $\kappa$ framework in general, as described by interaction terms given in Eq.~\eqref{eq:L_eff}, which can be mapped to EFT frameworks such as the Standard Model Effective Field Theory (SMEFT), as discussed in~\cref{sec:EFT}, or Higgs Effective Field Theory (HEFT).

In order to make predictions for the processes of interest, including the modified interactions of Eq.~\eqref{eq:L_eff}, we create a custom \texttt{UFO}~\cite{Alloul:2013bka,Degrande:2011ua,Darme:2023jdn} implementation of $\cza$ and $\caa$, by merging the \texttt{loop\_qcd\_qed\_sm\_Gmu} model~\cite{Hirschi:2015iia} for generating SM loop-induced processes in \texttt{MadGraph5\_aMC@NLO}~\cite{madgraph1} (\texttt{MG5}) with the implementation of said effective vertices from the \texttt{HiggsCharacterisation} model~\cite{Artoisenet:2013puc}. This simple merging of existing implementation is sufficient, as both the Higgs decays to $\gamma\gamma/Z\gamma$ and the $e^+e^-\to H\gamma$ (neglecting $m_e$) are pure loop-induced processes and the new physics couplings only enter at tree-level. Since the new couplings modify the amplitudes linearly, the dependence of partial widths and cross sections takes the form of a quadratic polynomial in $(\cza,\caa)$, where the linear terms come from the interference between the SM and BSM amplitudes, while the quadratic terms come from the square of the BSM amplitude. Using a combination of direct event generation and reweighting, we extract the polynomial dependence of the relevant Higgs boson partial widths and \eeha\ cross sections at $\sqrt{s}=160,240$ and \SI{365}{\giga\electronvolt} as follows:
\begin{align}
\label{eq:Gam_haa}
\Gamma_{h\to\gamma\gamma}&= 9.73\,\left(1 + 1.61\caa + 0.65\caa^2\right)\,\si{\kilo\electronvolt}\,,\\
\label{eq:Gam_hza}
\Gamma_{h\to {\scriptscriptstyle Z}\gamma}&=6.45 \,\left(1 + 1.38\cza + 0.47\cza^2\right)\,\si{\kilo\electronvolt}\,,\\
\label{eq:eeha_160}
\sigma^{160}_{\eeha}&= 18.8\,\left(1 + 0.58\caa + 1.03\cza + 0.20\caa^2 + 0.06\caa\cza + 0.39\cza^2\right)\,\si{\atto\barn}\,,\\
\label{eq:eeha_240}
\sigma^{240}_{\eeha}&= 83.6\,\left(1 + 0.73\caa + 0.98\cza + 0.33\caa^2 + 0.08\caa\cza + 0.40\cza^2\right)\,\si{\atto\barn}\,,\\
\label{eq:eeha_365}
\sigma^{365}_{\eeha}&= 37.0\,\left(1 + 0.50\caa + 2.11\cza + 1.39\caa^2 + 0.30\caa\cza + 1.38\cza^2\right)\,\si{\atto\barn}.
\end{align}
The rates are computed using the following input parameter choices:
\begin{equation}
\begin{gathered}
    M_W =80.419\,\mathrm{GeV},\,\, M_Z = 91.188\,\mathrm{GeV},\,\, M_t = 173.3\,\mathrm{GeV},\,\,\\ M_H = 125\,\mathrm{GeV},\,\,G_F=1.16639\times10^{-5}\,\mathrm{GeV^{-2}},
\end{gathered}
\end{equation}
and using the \texttt{isronlyll} lepton PDF implementation for \texttt{MG5}~\cite{Frixione:2021zdp} with a renormalisation and factorisation scale set to $M_H$ and $\sqrt{s}$ for the partial widths and \eeha\ cross sections, respectively. 

The relative impacts on the partial widths shown in Eqs.~\eqref{eq:Gam_haa} and \eqref{eq:Gam_hza} can be compared with the infinite mass limit expectations discussed above, given the normalisations for the $c$ coefficients. Assuming that a value of one leads to a relative impact $\delta$ on the underlying amplitude leads to the following effect on the partial width:
\begin{align}
\label{eq:scaling}
\mathcal{A}=\mathcal{A}_\mathrm{SM}\left(1+\delta\,c\right)\rightarrow \sigma \simeq \Gamma_\mathrm{SM}\left(1+2\delta c + \delta^2c^2\right),
\end{align}
where the approximate equality accounts for possible differences in the phase space dependence of the matrix elements between the SM and the $c$'s.
For an $\mathcal{O}(1)$ effect on the amplitude, we have $\delta\sim1$, and hence an $\mathcal{O}(2)$ effect from the linear term and $\mathcal{O}(1)$  effect from the quadratic term. In the limit of small $c$ we recover the usual factor of one half relation between the relative precision of a cross section or branching ratio measurement and the constraint on $c$, \emph{i.e.}, the relative constraining power on the corresponding $\kappa$
.
The numerical prefactors of the linear and quadratic terms in the partial widths indeed follow this pattern but indicate a slight suppression of the BSM contribution with respect to the naive expectation. This is likely due to finite mass effects and the associated kinematics of phase space integration that lead to a different SM amplitude compared to the infinite mass limit coefficients of Eq.~\eqref{eq:g_norm}. One can then match the predictions for the partial widths to the amplitude 
scaling of Eq.~\eqref{eq:scaling} to obtain a precise mapping between 
the $c$ coefficients and the $\kappa$ parameters:

\begin{align}
\label{eq:kappa_relation}
\kaa = 1+0.81\caa + \mathcal{O}(\caa^2),\quad \kza = 1+ 0.69 \cza + \mathcal{O}(\cza^2).
\end{align}

Turning to the relative impacts on the \eeha\ process in Eqs~\eqref{eq:eeha_160}--\eqref{eq:eeha_365}, there is evidence of more suppression compared to the partial widths. This is somewhat expected due to the additional box diagrams, which are not present in the branching fractions and do not depend on $\cza,\caa$. The linear prefactors of $\caa$ vary between 0.5 and 0.73, peaking at $\sqrt{s}=\SI{240}{\giga\electronvolt}$ , while the linear coefficients of $\cza$ have a stronger relative impact, with prefactors around one for 160 and \SI{240}{\giga\electronvolt} and about two for \SI{365}{\giga\electronvolt}. The latter is evidence for the energy growth of the BSM amplitude. Looking at the quadratic pieces, given the presence of two coefficients here, the cross term appears but is always rather small compared to the individual squared terms which have prefactors between 0.2 and 0.4 for 160 and \SI{240}{\giga\electronvolt} and around 1.4 for \SI{365}{\giga\electronvolt}, showing, again, the energy growing effect of the dimension-5 coupling. 

We can further analyse the energy behaviour by looking at the green and purple curves in~\cref{fig:sigmahgamma} that depict the linear and quadratic contributions of $\cza$ as a function of $\sqrt{s}$, compared to the SM curve in black. At low energies, up to about 240 GeV, we see that the interference term essentially tracks the SM rate which grows rather quickly with energy. The SM cross section then turns over as the destructive interference between the triangle and box diagrams becomes more severe, reaching its minimum around the top pair threshold. The interference term also features the same turnover but is relatively less suppressed thanks to the energy growth of the underlying BSM amplitude. The quadratic term, instead, has a monotonic growth with energy, since there are only two relevant, tree-level diagrams with identical momentum dependence in the high energy limit.

Overall, the \eeha\ process displays an enhanced sensitivity to $\cza$ over $\caa$ and, given the tight constraints expected on the latter from HL-LHC, we do not expect it to play a significant role in our analysis. The actual sensitivity of this process is slightly lower but remains at a similar level to that of the $H\to Z\gamma$ partial width at $\sqrt{s}=160,\SI{240}{\giga\electronvolt}$, while at \SI{365}{\giga\electronvolt} it has an enhanced relative impact. It is therefore worthwhile to investigate whether one can obtain additional information on this coupling by measuring the \eeha\ cross section at the various possible energy stages of the FCC-ee machine. With the largest SM cross section at $\sqrt{s}=\SI{240}{\giga\electronvolt}$, the nominal, expected integrated luminosity of \SI{10.8}{\per\femto\barn} of this run would yield 900 signal events. In a background-free scenario, this suggests a potential statistical sensitivity of 3-4\% on the total rate, which is better in relative terms than the $\sim$14\% projected on the $Z\gamma$ branching fraction of the Higgs from the HL-LHC~\cite{ATLAS:2025eii}. However, being loop-induced, the rarity of this process means that it has to contend with several background processes, as we discuss in the next sections, which ultimately drive the sensitivity to this process.
As we will see, there is an interplay between the relative impact of $\cza$, the absolute scale of the SM cross section, which peaks strongly around $\sqrt{s}$=\SI{240}{\giga\electronvolt}, and the energy dependence of the various background contributions in each channel that is crucial to take into account in order to robustly project our sensitivity to this cross section.

\section{Detector Simulation and Monte Carlo Event Generation}
\label{sec:mc}

For our sensitivity projection, we simulate signal and background events at the three different center-of-mass energies at FCC-ee, \ecm~=~\SI{160}{\giga \electronvolt}, \SI{240}{\giga \electronvolt} and \SI{365}{\giga \electronvolt}. The samples are scaled to the integrated luminosities of the baseline FCC-ee scenario with four interaction points, and data collected only at the nominal beam energy values, assuming no $WW^*$ or $t\bar{t}$ threshold scans around \ecm~=~\SI{160}{\giga \electronvolt} and \SI{365}{\giga \electronvolt}. The Monte Carlo events for signal are generated with \texttt{MadGraph5\_aMC@NLO} \texttt{v3.5.7}~\cite{madgraph1, madgraph2} (\texttt{MG5}), and \texttt{Whizard} \texttt{v3.0.3}~\cite{whizard}  is used for background events. Although \texttt{Whizard} is the standard event generator for FCC-ee, \texttt{MG5} is better suited to generate loop-induced processes, which justifies its choice for \eeha\ events. The decay of on-shell resonances and the hadronization is handled with \texttt{Pythia}~\texttt{v8.311}~\cite{pythia8} for signal processes, and with \texttt{Pythia}~\texttt{v6.427}~\cite{pythia6} for backgrounds.
Production-related configurations are taken from the so-called \texttt{winter2023} campaign~\cite{central_production}, used for a large majority of the prospective analyses documented in the 2025 FCC feasibility report~\cite{FCC:2025lpp}, that includes realistic estimates of the beam parameters (energy spread and size of the luminous region). All the samples used were produced via the central production system~\cite{fcc-sw-note}. 

The study is performed assuming the IDEA~\cite{IDEA_det} detector concept performance. Given the distinctive presence of a mono-chromatic photon for this channel (see discussion later, \cref{sec:kinematics}), the electromagnetic energy resolution is a key driver of the sensitivity. In its current incarnation, a relative energy resolution of $3 \% /\sqrt{E} \oplus 1 \%$ is assumed for the IDEA detector. The IDEA detector concept is well integrated and tested in the fast simulation framework  \texttt{Delphes}~\cite{k4simdelphes, delphes} using the IDEA detector card~\cite{delphes_card_idea}. \texttt{Delphes} provides a list of particle flow candidates in the form of charged hadrons, neutral hadrons, photons, electrons, and muons. These particle flow objects are subsequently clustered into jets using the exclusive Durham $k_T$ algorithm~\cite{Catani:1991hj,Fastjet}, with the number of jets \(N\) chosen according to the targeted final state.
The resulting jets are assigned flavour probabilities using a transformer based tagger~\cite{Bedeschi:2022rnj,jettagger}.
For each jet, the tagger predicts probabilities for the seven flavour hypotheses \(b\), \(c\), \(s\), \(u\), \(d\), \(\tau\), and gluon, which are used directly as inputs to the analysis.

\subsection{The Signal Process}

The $H\gamma$ signal is generated at one loop in the SM using \texttt{MG5} 
with loop-induced matrix elements, employing the \texttt{loop\_qcd\_qed\_sm\_Gmu} model, 
without any generator-level cuts, and with initial-state radiation (ISR) enabled. 
For the WW threshold and $t\bar{t}$ runs, we assume that the full integrated luminosities 
of \SI{19.2}{\per\atto\barn} and \SI{3.12}{\per\atto\barn} are collected at fixed 
center-of-mass energies of \SI{160}{\giga\electronvolt} and \SI{365}{\giga\electronvolt}, 
respectively. In practice, both runs involve a threshold scan spanning a range of energies, 
but the impact of this simplification on the present analysis is negligible.
The expected signal yields at the three center-of-mass energies considered are summarized 
in~\cref{tab:num_signal_events}.
%
%
\begin{table}[h!]
     \centering
     \begin{tabular}{c c c c}
        \ecm\ [GeV] & $\sigma(H \gamma)$ & int. luminosity &  \# events \\
         \hline
        160 &   \SI{18.9}{\atto\barn} & \SI{19.2}{\per\atto\barn} & 363\\
        240 &  \SI{83.9}{\atto\barn} &\SI{10.8}{\per\atto\barn} & 906 \\
        365 & \SI{36.6}{\atto\barn} & \SI{3.12}{\per\atto\barn} & 114\\
        \hline
     \end{tabular}
    \caption{Signal cross-section, integrated luminosity and expected number of signal events for different energies at FCC-ee}
    \label{tab:num_signal_events}
\end{table}

Dedicated samples are generated for each Higgs decay channel considered in the 
analysis: $H \rightarrow b\bar{b},\ c\bar{c},\ WW^*,\ gg,\ \tau^+\tau^-$ using SM branching ratios.

\subsection{The Background Processes}
\label{sec:bkg_proc}

The \eeha\ signal must be extracted against several background processes whose 
cross sections exceed that of the signal by several orders of magnitude. 
The backgrounds are grouped into the following categories:

\begin{itemize}
  \item $\gamma\,\ell^+\ell^-$: radiative dilepton production 
    $e^+e^-\to\gamma\,\ell^+\ell^-$ with $\ell=e,\mu,\tau$, 
    generated as three separate samples. The $\gamma e^+e^-$ process, 
    dominated by radiative Bhabha scattering, constitutes the largest 
    background contribution prior to any event selection. The 
    $\gamma\mu^+\mu^-$ and $\gamma\tau^+\tau^-$ rates follow the 
    corresponding $Z$ decay branching fractions.
  \item $\gamma\,q\bar{q}$: radiative hadronic production 
    $e^+e^-\to\gamma\,f\bar{f}$. The $\gamma\,b\bar{b}$ and 
    $\gamma\,c\bar{c}$ contributions are generated as separate samples 
    due to their relevance to the \eeha\ signal extraction, while 
    $f\in\{u,d,s\}$ are combined into a single sample.
\item $\gamma\,W+X$: radiative production in association with a $W$ boson, 
    generating $\gamma\,q\bar{q}W$ and $\gamma\,\ell\nu W$ final states 
    with $\ell=e,\mu,\tau$. Only one $W$ boson is  required to be on shell 
    in the generation, so that off-shell contributions are 
    included. This is particularly important at \ecm~=~\SI{160}{\giga\electronvolt}, 
    where the center-of-mass energy is only marginally above the $WW$ 
    threshold and a significant fraction of $W$ bosons are produced off shell.
    For $\ell=e$, 
    large $t$-channel contributions from single $W$ production 
    ($e^+e^-\to We\nu\gamma$) are included, which do not involve a second 
    (off-shell) $W$ boson. These processes constitute the dominant irreducible backgrounds to the 
    $\gamma WW^*$ signal.
  \item $\gamma\,Z+X$: radiative production in association with a $Z$ boson. 
    Again, only one $Z$ boson is required to be on shell. Off-shell contributions are included for consistency 
    with the $\gamma\,W+X$ treatment and to ensure adequate modeling at 
    \ecm~=~\SI{160}{\giga\electronvolt}. The $\gamma\,q\bar{q}Z$ final 
    state can mimic the $\gamma\,WW^*$ signal in the semi-leptonic channel when one lepton from $Z\to\ell^+\ell^-$ falls outside the detector acceptance, producing 
    apparent missing momentum. 
\item $\gamma\,H+X$: non-signal single Higgs production modes in which 
    an ISR photon mimics the signal photon. The dominant contribution is 
    $\nu\bar{\nu}H$ production via $WW$ fusion and $ZH$ with $Z\to\nu\bar{\nu}$, 
    generated inclusively as $e^+e^-\to\nu\bar{\nu}H$ without an explicit 
    photon in the final state. The signal-like topology arises when a hard 
    ISR photon is radiated, producing a final state with a photon, missing 
    momentum from the neutrinos, and the Higgs decay products. This background 
    is considered at \ecm~=~\SI{240}{\giga\electronvolt} and 
    \SI{365}{\giga\electronvolt}, and is found to be negligible after the 
    full analysis selection is applied.
   \item $t\bar{t}$: top quark pair production, relevant only at 
    \ecm~=~\SI{365}{\giga\electronvolt}.
\end{itemize}

Generator-level kinematic cuts are applied to all background processes  involving a final-state photon, to help matrix element integration convergence and to optimally populate the analysis phase space. The invariant mass of each $f\bar{f}$ pair is required to exceed \SI{5}{\giga\electronvolt}. 
The photon energy\footnote{ISR excluded}, when explicitly requested, is required to exceed 15, 50, and \SI{140}{\giga\electronvolt} at \ecm~=~160, 240, and \SI{365}{\giga\electronvolt}, respectively.
These thresholds are motivated by the signal kinematics: as discussed in \cref{sec:analysis}, the two-body nature of the \eeha\ process produces a monochromatic photon with momentum well above these values at each 
energy stage, such that the generator-level cuts introduce no bias on the signal-like phase space while substantially reducing the size of the generated background samples. A minimum separation of 
$\Delta R(\gamma,f)>0.01$ between the photon and any final-state fermion is imposed, and the polar angle of all final-state particles is required to satisfy $|\cos\theta|<0.998$.

A summary of all background processes and the 
inclusive cross sections after generator-level cuts is given 
in~\cref{tab:background_processes}.

\begin{table}[htb]
  \centering
  \renewcommand{\arraystretch}{1.15}
  \footnotesize
  \begin{tabular}{llccc}
    \toprule
    Category & Process &
    $\sigma$(160\,GeV) [pb] & $\sigma$(240\,GeV) [pb] & $\sigma$(365\,GeV) [pb] \\
    \midrule
    \multirow{3}{*}{$\gamma\,\ell^+\ell^-$}
      & $\gamma e^+e^-$      & $921$  & $196$   & $131$   \\
      & $\gamma\mu^+\mu^-$   & $2.82$ & $0.803$ & $0.382$ \\
      & $\gamma\tau^+\tau^-$ & $2.60$ & $0.770$ & $0.362$ \\
    \midrule
    \multirow{3}{*}{$\gamma\,q\bar{q}$}
      & $\gamma b\bar{b}$              & $8.06$ & $2.35$ & $0.985$ \\
      & $\gamma c\bar{c}$              & $7.45$ & $2.16$ & $0.954$ \\
      & $\gamma q\bar{q}$ ($q=u,d,s$) & $24.2$ & $6.98$ & $2.98$  \\
    \midrule
    $\gamma\,W{+}X$ & $\gamma q\bar{q}^{\prime}W$, $\gamma\ell\nu W$
      & $0.037$ & $0.230$ & $0.020$ \\
    $\gamma\,Z{+}X$ & $\gamma q\bar{q}Z$, $\gamma\ell^+\ell^-Z$
      & $1.11$  & $0.361$ & $0.089$ \\
    $\gamma\,H{+}X$ & $\nu\bar{\nu}H$ (ZH$+$VBF)
      & ---  & $0.046$ & $0.054$ \\
    $t\bar{t}$ & $t\bar{t}$
      & --- & --- & $0.461$ \\
    \bottomrule
  \end{tabular}
  \caption{Background processes considered in the analysis, grouped by  category, with the inclusive cross section at each center-of-mass energy.}
  \label{tab:background_processes}
\end{table}

\section{Experimental Analysis}
\label{sec:analysis}

The analysis strategy proceeds in several steps, common to all decay channels 
considered. First, a decay-mode-agnostic preselection is defined, exploiting 
the two-body kinematics of the \eeha\ process and the characteristic monochromatic photon recoiling against the Higgs boson. This common preselection provides a starting point with high signal efficiency and rejects a large amount of reducible backgrounds independently of the Higgs decay mode. Since \eeha\ is a production process, all Higgs decay modes are in principle accessible, and a multi-channel approach can be pursued to maximize the overall sensitivity. In this work, we focus on $H\to b\bar{b}$ and the semi-leptonic $H\to WW^*$ modes, which offer the best sensitivity. Other hadronic channels, namely $H\to gg$ and $H\to\tau^+\tau^-$, were investigated but found to provide negligible sensitivity and are excluded from the combination, as discussed in~\cref{sec:Hjj}. The study of additional decay modes, such as $H\to ZZ^*$ or $H\to\gamma\gamma$, is left for future work. After the common preselection, a channel dependent strategy is applied in each case. High level discriminating observables are identified that exploit the kinematic and topological differences between signal and background, and are used to train a multivariate BDT classifier. The output of the BDT provides the final discriminating observable, which is used as input to a binned profile likelihood fit for signal extraction.

\subsection{Event kinematics}
\label{sec:kinematics}

The \eeha\ process is characterized by two-body kinematics at Born level, which provides powerful discrimination against backgrounds. Energy-momentum conservation leads to a monochromatic photon with momentum:
\begin{equation}
  p_\gamma = \frac{s - m_H^2}{2\sqrt{s}} \, ,
  \label{eq:p_mono}
\end{equation}
taking the values $p_\gamma = 31.2$, $87.4$, and \SI{161.1}{\giga\electronvolt} 
at $\sqrt{s}=160$, $240$, and \SI{365}{\giga\electronvolt}, respectively. 
The recoil mass, defined as
\begin{equation}
  m_{\rm recoil}^2 =
  \bigl( (P_{e^+}+P_{e^-}) - P_\gamma \bigr)^2 \, ,
  \label{eq:m_recoil}
\end{equation}
peaks at $m_H$ for signal events, independently of the Higgs decay mode. In the $\gamma\,W+X$ and $\gamma\,Z+X$ backgrounds, the photon features a broad momentum spectrum, while in $\gamma\,q\bar{q}$ production the photon is monochromatic but at a value determined by $\sqrt{s}$ and $m_Z$. 

\subsection{Common Event Preselection}
\label{sec:ana_strat}

The goal of the common preselection is to exploit the features of the \eeha\ topology, namely the presence of a hard, isolated 
photon recoiling against hadronic or leptonic activity from the Higgs decay, without targeting any specific Higgs decay mode. This loose selection is designed to strongly suppress the dominant radiative backgrounds while retaining high signal efficiency, leaving the residual discrimination to the selections described in the following subsections.

\begin{figure}[t]
  \centering
  \begin{minipage}[t]{0.49\textwidth}
    \centering
    \includegraphics[width=\linewidth]{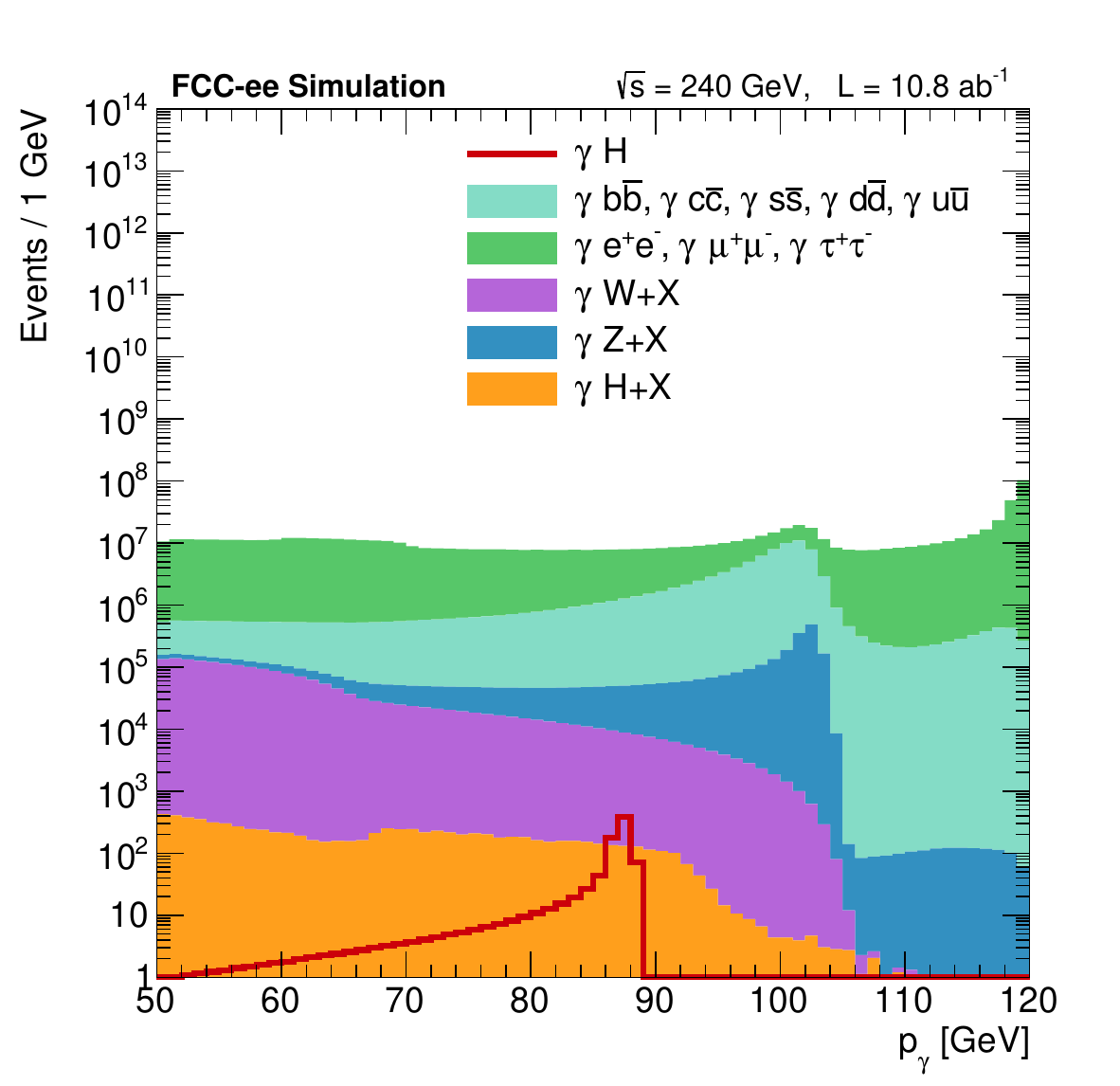}
  \end{minipage}
  \begin{minipage}[t]{0.49\textwidth}
    \centering
    \includegraphics[width=\linewidth]{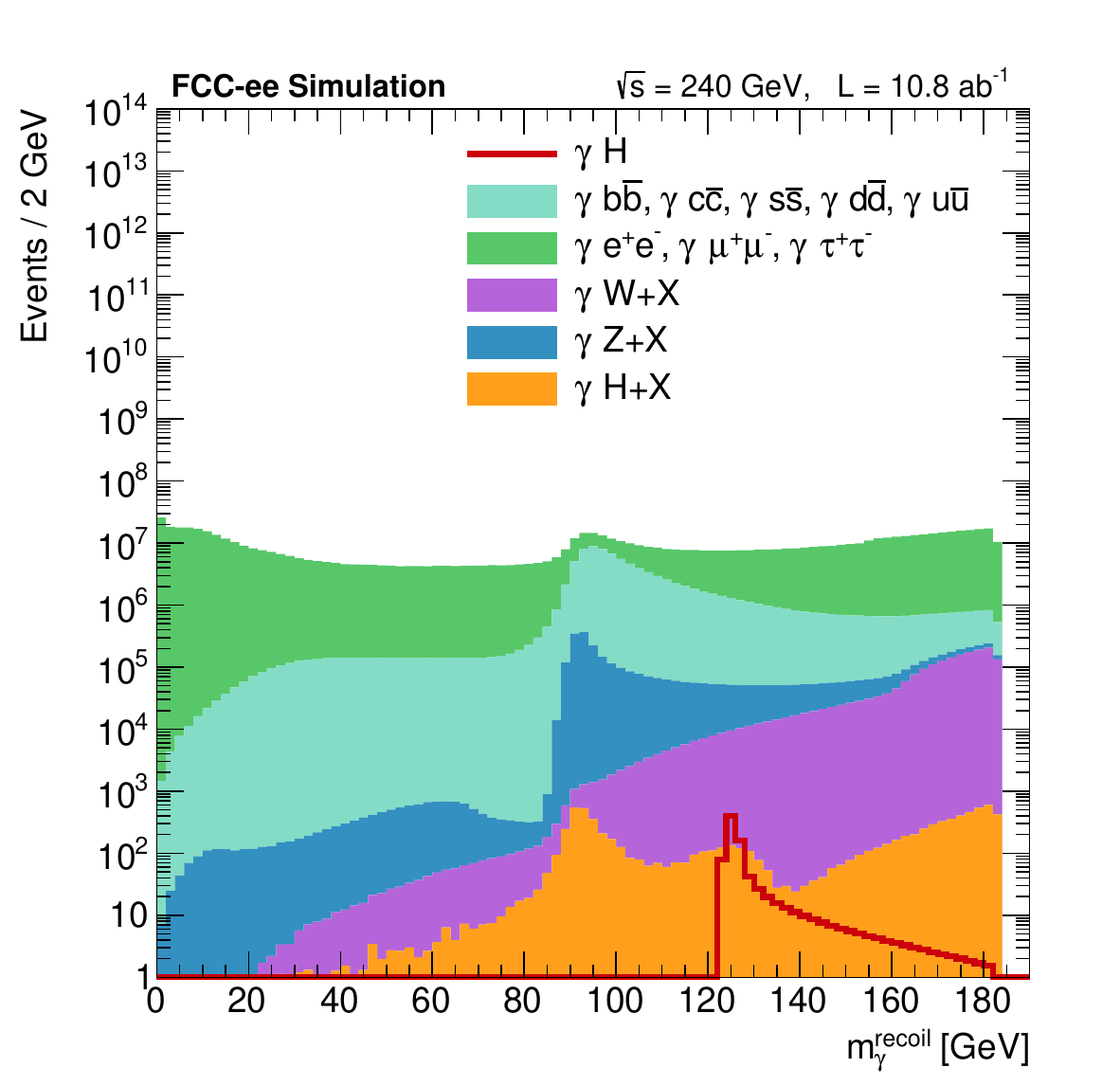}
  \end{minipage}\hfill

\caption{Photon momentum $p_\gamma$ (left) and recoil mass $m_\mathrm{recoil}$ (right), as defined in \cref{eq:p_mono,eq:m_recoil}, after requiring at least 
one isolated photon with $p_\gamma > \SI{50}{\giga\electronvolt}$ at $\sqrt{s}=\SI{240}{\giga\electronvolt}$. The signal (solid line) is shown overlaid on the stacked background contributions.}
  \label{fig:mrec_pa_nosel_240}
\end{figure}

Photon isolation is computed using charged hadrons within a cone of 
$\Delta R = 0.2$ around the photon direction. The relative isolation variable is defined as
\[
\mathrm{iso}(\gamma)=\frac{\sum_{\text{ch}} p_i}{p_\gamma},
\]
where the sum runs over the momenta of charged particles inside the cone. Events are required to contain at least one isolated photon with $\mathrm{iso}(\gamma)<0.2$. The highest energy isolated photon is taken as the signal candidate. A minimum photon momentum of $p_\gamma > 15$, $50$, and 
\SI{140}{\giga\electronvolt} is required at $\sqrt{s}=160$, $240$, and \SI{365}{\giga\electronvolt}, respectively, exploiting the hard, monochromatic photon expected from the two-body signal kinematics to suppress radiative backgrounds with soft or collinear photons. The photon momentum and recoil mass distributions are shown for signal and backgrounds in~\cref{fig:mrec_pa_nosel_240} for $\sqrt{s}=\SI{240}{\giga\electronvolt}$, and in~\cref{app:recoil_obs} for the other energy stages.

In addition, at least six reconstructed tracks ($N_\mathrm{trk} > 5$) are required among the remaining particles, excluding the isolated photon. This requirement reflects the expected hadronic or leptonic activity from the Higgs decay and efficiently rejects radiative dilepton backgrounds ($\gamma e^+e^-$ and $\gamma\mu^+\mu^-$), which are characterized by very low track multiplicities, and suppresses a large fraction of the $\gamma\tau^+\tau^-$ contribution.

The effect of the common preselection is summarized in~\cref{tab:common_preselection} for $\sqrt{s}=\SI{240}{\giga\electronvolt}$, with the corresponding yields for the other center-of-mass energies shown in~\cref{app:preselection}. After the common preselection, 87.2\% of the signal is selected at $\sqrt{s}=\SI{240}{\giga\electronvolt}$, corresponding to 786 events, while the total background is reduced to $\mathcal{O}(10^8)$ events, yielding a signal-to-background ratio of $\mathcal{O}(10^{-5})$. The $\gamma\,\ell^+\ell^-$ background is suppressed by three orders of magnitude by the track multiplicity requirement. Similar signal efficiencies and background rejections are achieved at the other center-of-mass energies.

\begin{table}[t]
  \centering
  \renewcommand{\arraystretch}{1.15}
  \footnotesize
  \resizebox{\textwidth}{!}{
  \begin{tabular}{lccccc|c}
    \toprule
                   selection &              $\gamma\,q\bar q$ &         $\gamma\,\ell^+\ell^-$ &                  $\gamma\,W+X$ &                  $\gamma\,Z+X$ &                  $\gamma\,H+X$ &                     $\gamma H$ \\
\midrule
           $\rm all\,events$ & $1.24 \times 10^{8}$ (100.0)\% & $2.13 \times 10^{9}$ (100.0)\% & $2.49 \times 10^{6}$ (100.0)\% & $3.90 \times 10^{6}$ (100.0)\% & $2.20 \times 10^{6}$ (100.0)\% & $9.01 \times 10^{2}$ (100.0)\% \\
$\rm p_{\gamma}\,>\,50\,GeV$ &  $9.77 \times 10^{7}$ (78.8)\% &  $8.47 \times 10^{8}$ (39.7)\% &  $1.95 \times 10^{6}$ (78.3)\% &  $3.12 \times 10^{6}$ (79.8)\% &   $1.36 \times 10^{4}$ (0.6)\% &  $8.63 \times 10^{2}$ (95.8)\% \\
         $\rm N_{trk}\,>\,5$ &  $9.75 \times 10^{7}$ (78.6)\% &   $1.41 \times 10^{5}$ (0.0)\% &  $1.63 \times 10^{6}$ (65.5)\% &  $2.41 \times 10^{6}$ (61.7)\% &   $1.01 \times 10^{4}$ (0.5)\% &  $7.86 \times 10^{2}$ (87.2)\% \\

    \bottomrule
  \end{tabular}
  }
  \caption{Event yields and relative efficiencies after the common preselection at $\sqrt{s}~=~\SI{240}{\giga\electronvolt}$. The preselection consists of an isolated photon requirement
  with $p_\gamma>\SI{50}{\giga\electronvolt}$ and a minimum of six reconstructed tracks, excluding the isolated
  photon.}
  \label{tab:common_preselection}
\end{table}

\subsection{The $H\to WW^*$ Channel}
\label{sec:HWW}

\begin{figure}[htb]
  \centering
  \begin{minipage}[t]{0.49\textwidth}
    \centering
    \includegraphics[width=\linewidth]{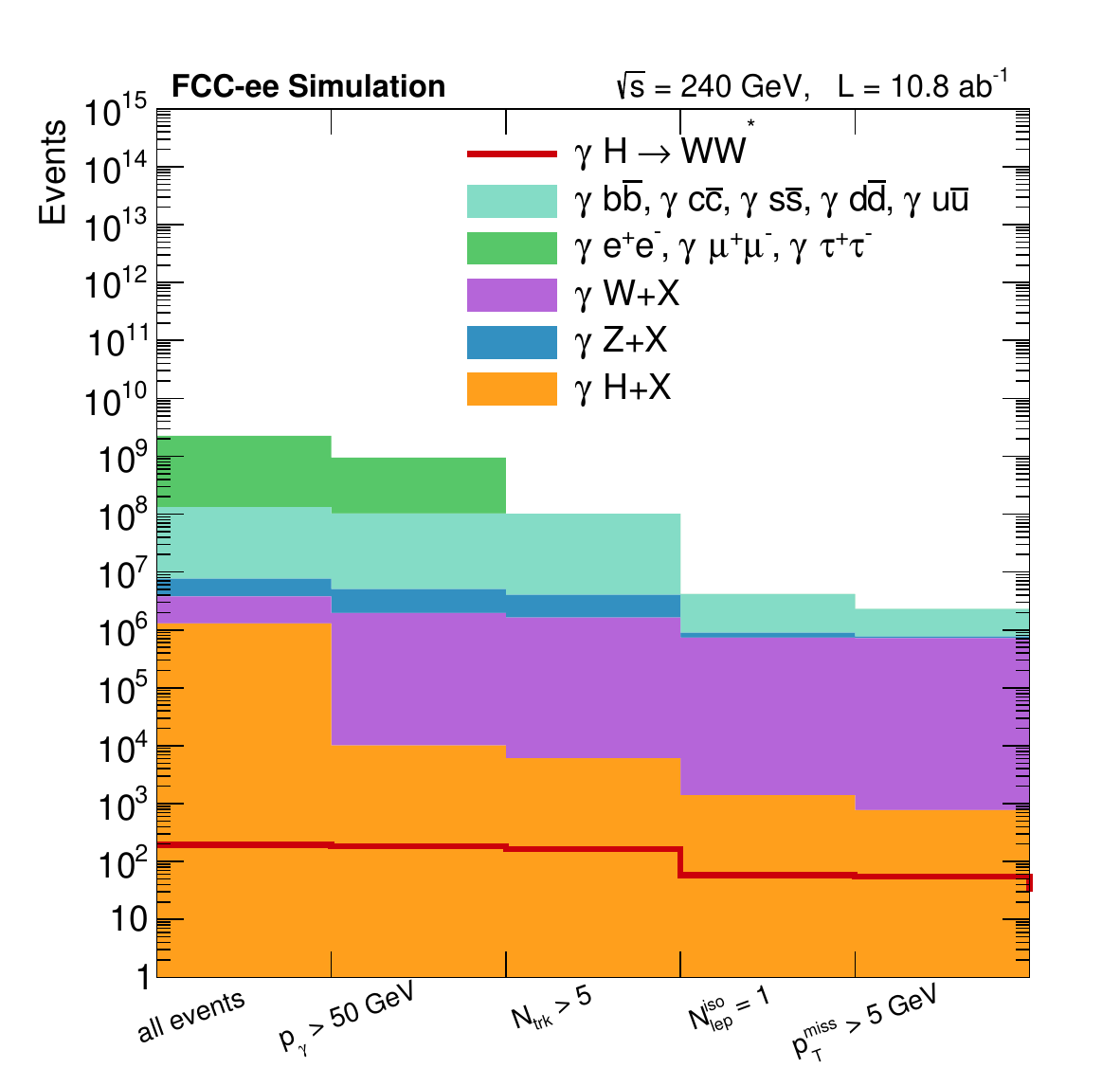}
  \end{minipage}\hfill
  \begin{minipage}[t]{0.49\textwidth}
    \centering
    \includegraphics[width=\linewidth]{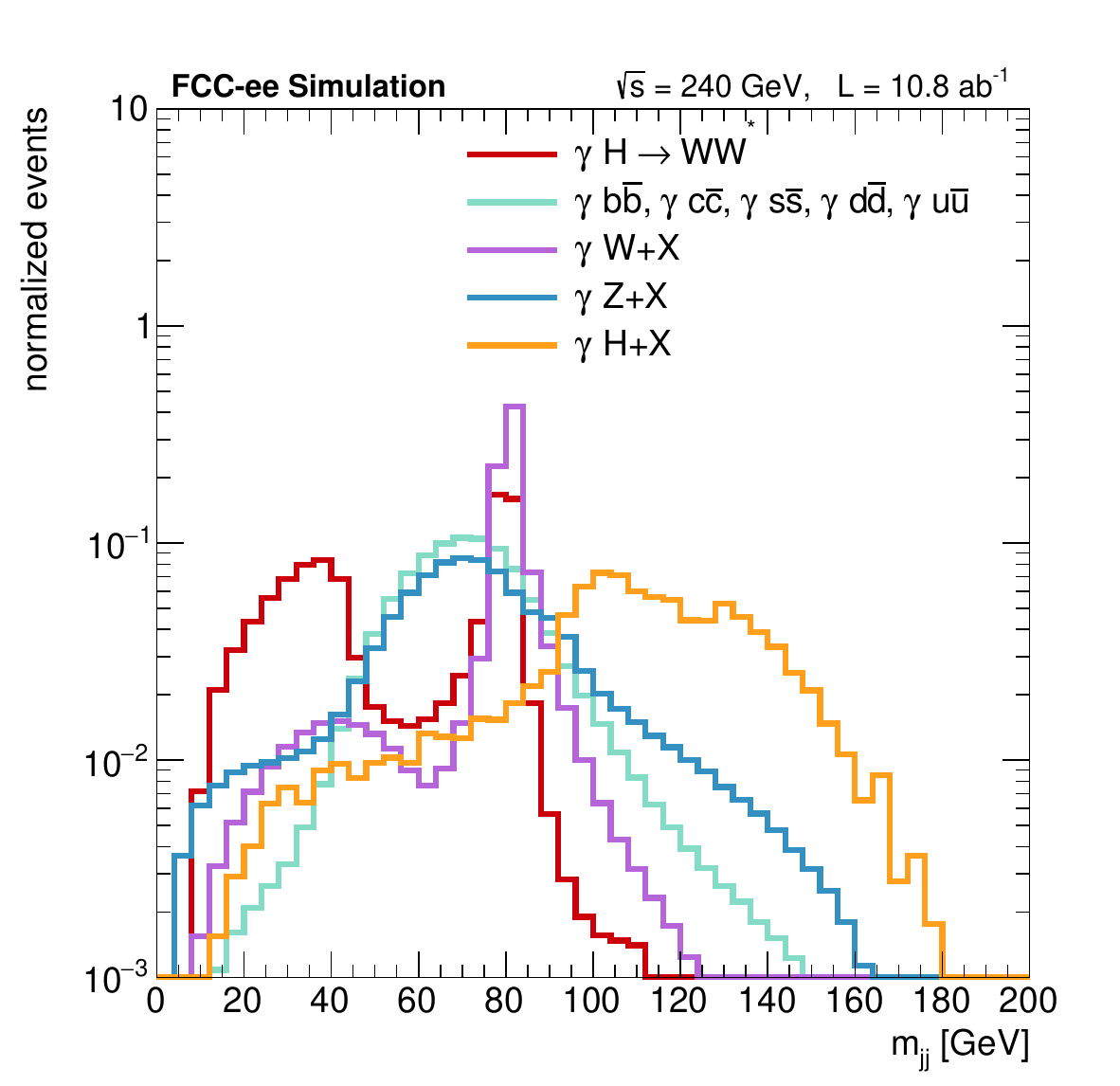}
  \end{minipage}

    \caption{
    Left: Cumulative yields for the signal and dominant background processes in the $WW^*$ final state. Right: dijet invariant-mass distribution after channel-specific selection, illustrating the
    separation between the off-shell ($5<m_{jj}<60$~GeV) and on-shell ($m_{jj}>60$~GeV) modes used to define orthogonal analysis categories.
    }
  \label{fig:hww_presel_240}
\end{figure}

The Higgs decay into a pair of $W$ bosons constitutes the second largest 
branching fraction in the SM, with $\mathrm{BR}(H\to WW^*)=21.4\%$. 
This analysis focuses on the semi-leptonic final states
\(
H\to W(\ell\nu)\,W^*(qq)
\)
and
\(
H\to W(qq)\,W^*(\ell\nu)
\)
with $\ell=e,\mu$, which together account for approximately 29\% of all 
$WW^*$ decays. Leptonic $\tau$ decays, $\tau\to\ell\nu\bar{\nu}$, are also 
included in the analysis, although the event selection is not explicitly 
optimized for this contribution. These semi-leptonic final states are particularly attractive as they feature a distinctive topology with the 
presence of an isolated lepton and missing transverse momentum from the neutrino, which provides effective handles to reject the overwhelming purely hadronic $\gamma\,q\bar{q}$ backgrounds. The two reconstructed jets allow the hadronic $W$ decay to be reconstructed. Moreover, the knowledge of the initial state four-momentum in $e^+e^-$ collisions allows the missing momentum to be fully determined, enabling a complete reconstruction of the leptonic $W$ decay and, in turn, of the full $H\to WW^*$ candidate. 

\subsubsection{$H \rightarrow WW^*$ channel preselection}

The event selection builds upon the common preselection described in~\cref{sec:ana_strat}. In addition, events are required to contain exactly one isolated lepton and a minimum missing transverse momentum,
\[
N^{\mathrm{iso}}_{\ell} = 1,
\qquad
p_{\mathrm{T}}^{\mathrm{miss}} > 5~\mathrm{GeV},
\]
which efficiently suppresses purely hadronic and fully leptonic backgrounds. 
The use of the \(p_{\mathrm{T}}^{\mathrm{miss}} \), rather than the total missing momentum \(p^{\mathrm{miss}} \), helps to mitigate the contribution of $Z+X$ background with radiative return, which can feature large \(p_{\mathrm{z}}^{\mathrm{miss}} \) while having vanishing \(p_{\mathrm{T}}^{\mathrm{miss}} \). The lepton (electron or muon) is required to be isolated (the isolation criteria being identical to that of the photon, defined in~\cref{sec:ana_strat}) and to have a minimum momentum of 5 GeV. 
After removing the photon and the isolated lepton, the remaining reconstructed particles are clustered into 2 jets using the exclusive Durham $k_{\text{T}}$ algorithm, as discussed in~\cref{sec:mc}. The invariant mass of the resulting dijet system \(m_{jj}\) is used as the key observable to characterize the $WW^*$ topology and to further discriminate against non-resonant backgrounds. 

\Cref{fig:hww_presel_240} (left) shows the effect of the $WW^*$ channel 
preselection on signal and background yields at $\sqrt{s}=\SI{240}{\giga\electronvolt}$, 
with the corresponding yields for $\sqrt{s}=\SI{160}{\giga\electronvolt}$ and 
\SI{365}{\giga\electronvolt} shown in~\cref{app:hwwpreselection}. The remaining 
dominant backgrounds at this stage, ordered by expected yield, are 
$\gamma\,q\bar{q}$ (dominated by $\gamma\,b\bar{b}$ and $\gamma\,c\bar{c}$), 
$\gamma\,W+X$, $\gamma\,Z+X$, and, to a smaller extent at 
$\sqrt{s}\geq\SI{240}{\giga\electronvolt}$, $\gamma\,H+X$. The right panel 
shows the reconstructed dijet invariant-mass distribution, illustrating the 
two kinematic regimes corresponding to the on-shell and off-shell $W$ decays.

The dijet invariant mass provides a handle to separate on-shell and off-shell semi-leptonic $WW^*$ contributions and to ensure orthogonality between them. This separation is motivated by the different background composition in the two regions, as illustrated in~\cref{fig:hww_presel_240} (right). Treating the on-shell and off-shell topologies independently allows to optimise background suppression in each category. Events with \(\SI{5}{\giga\electronvolt} < m_{jj} < \SI{60}{\giga\electronvolt}\) are assigned to the $W(\ell\nu)\,W^*(qq)$ category, targeting the off-shell hadronic $W^*$ contribution, while events with \(m_{jj} > \SI{60}{\giga\electronvolt}\) are classified as $W(qq)\,W^*(\ell\nu)$ candidates. Within this approach, 53\% (47\%) of events are assigned to the on-shell (off-shell) category.

\subsubsection{$H \rightarrow W(\ell \nu)\,W^*(qq)$ Channel}
\label{sec:HWW_lvqq}

This analysis targets the semi-leptonic Higgs decay mode
\(H \rightarrow W(\ell\nu)\,W^*(qq)\), where the on-shell $W$ boson decays leptonically and the off-shell $W^*$ decays
hadronically. The final state is characterised by two reconstructed jets compatible with an off-shell W* decay. The expected signal yields, applying the channel dependent selection described above and the \(\SI{5}{\giga\electronvolt}~<~m_{jj}~<~\SI{60}{\giga\electronvolt}\) requirement are 13, 31, and 3 events at \(\sqrt{s}=160\), 240, and \SI{365}{\giga\electronvolt}, respectively.

\begin{figure}[t]
  \centering

  \begin{minipage}[t]{0.49\textwidth}
    \centering
    \includegraphics[width=\linewidth]{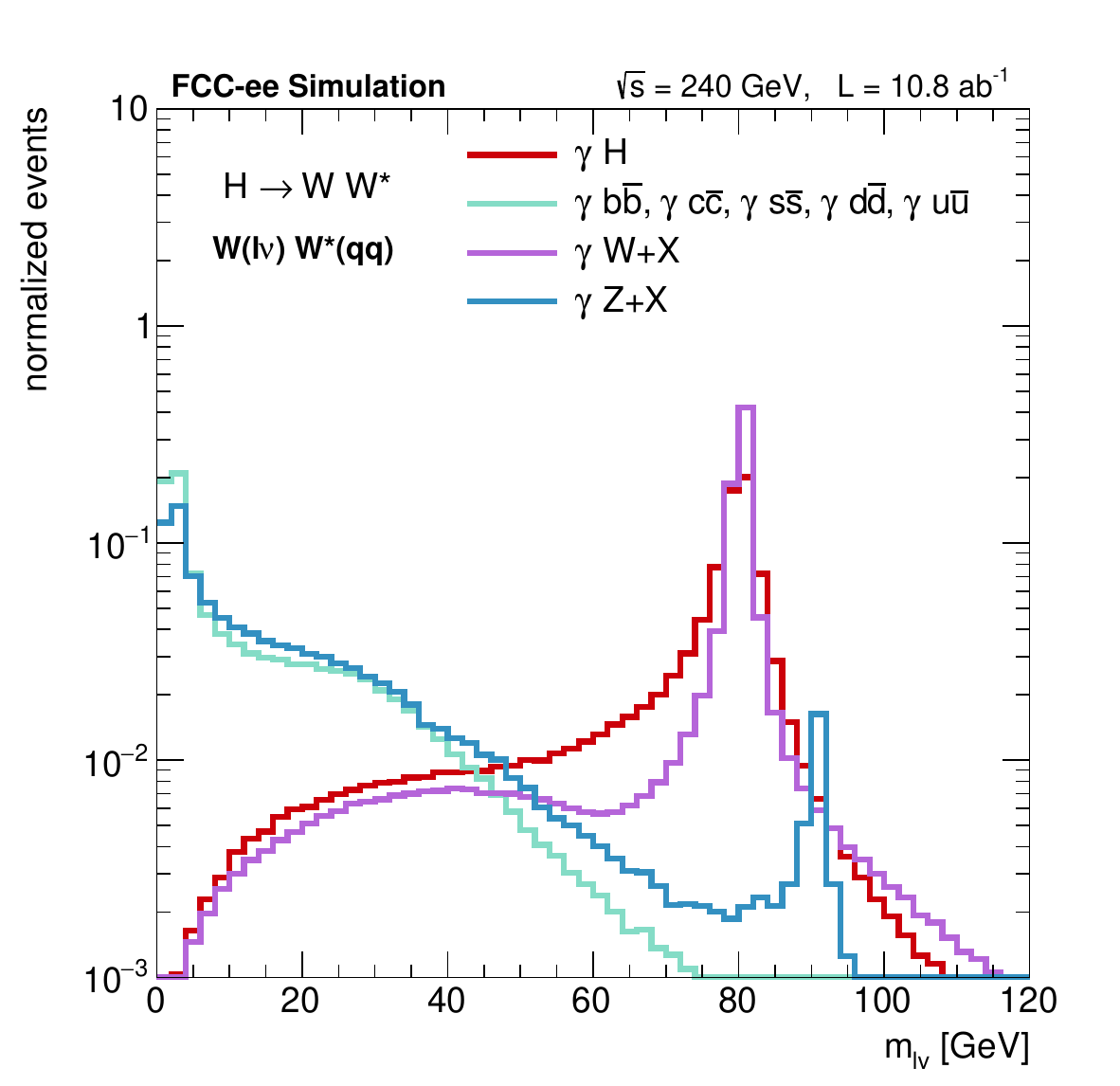}\\
  \end{minipage}\hfill
  \begin{minipage}[t]{0.49\textwidth}
    \centering
    \includegraphics[width=\linewidth]{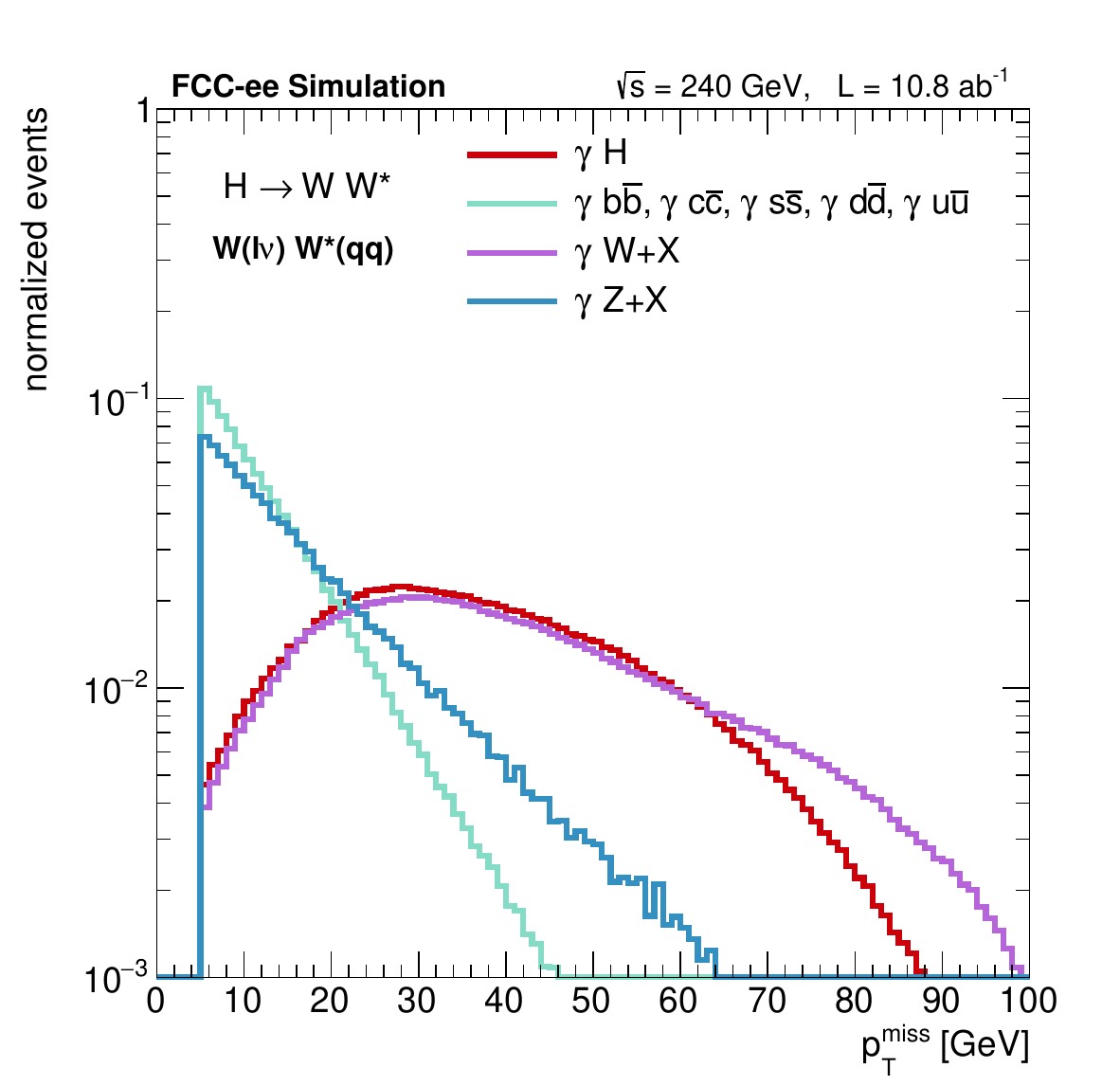}\\
  \end{minipage}

  \vspace{0.25cm}

  \begin{minipage}[t]{0.49\textwidth}
    \centering
    \includegraphics[width=\linewidth]{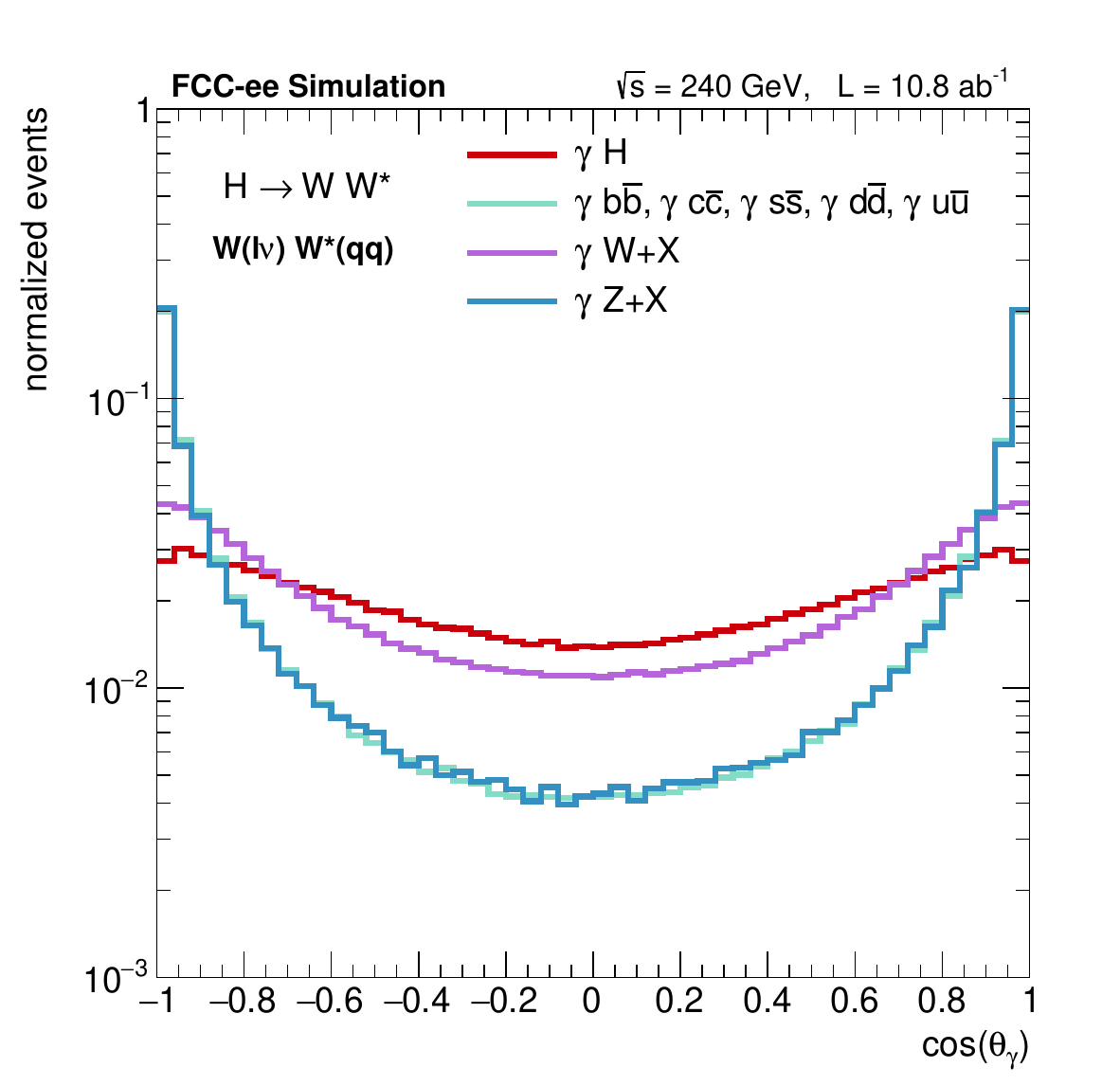}\\
  \end{minipage}\hfill
  \begin{minipage}[t]{0.49\textwidth}
    \centering
    \includegraphics[width=\linewidth]{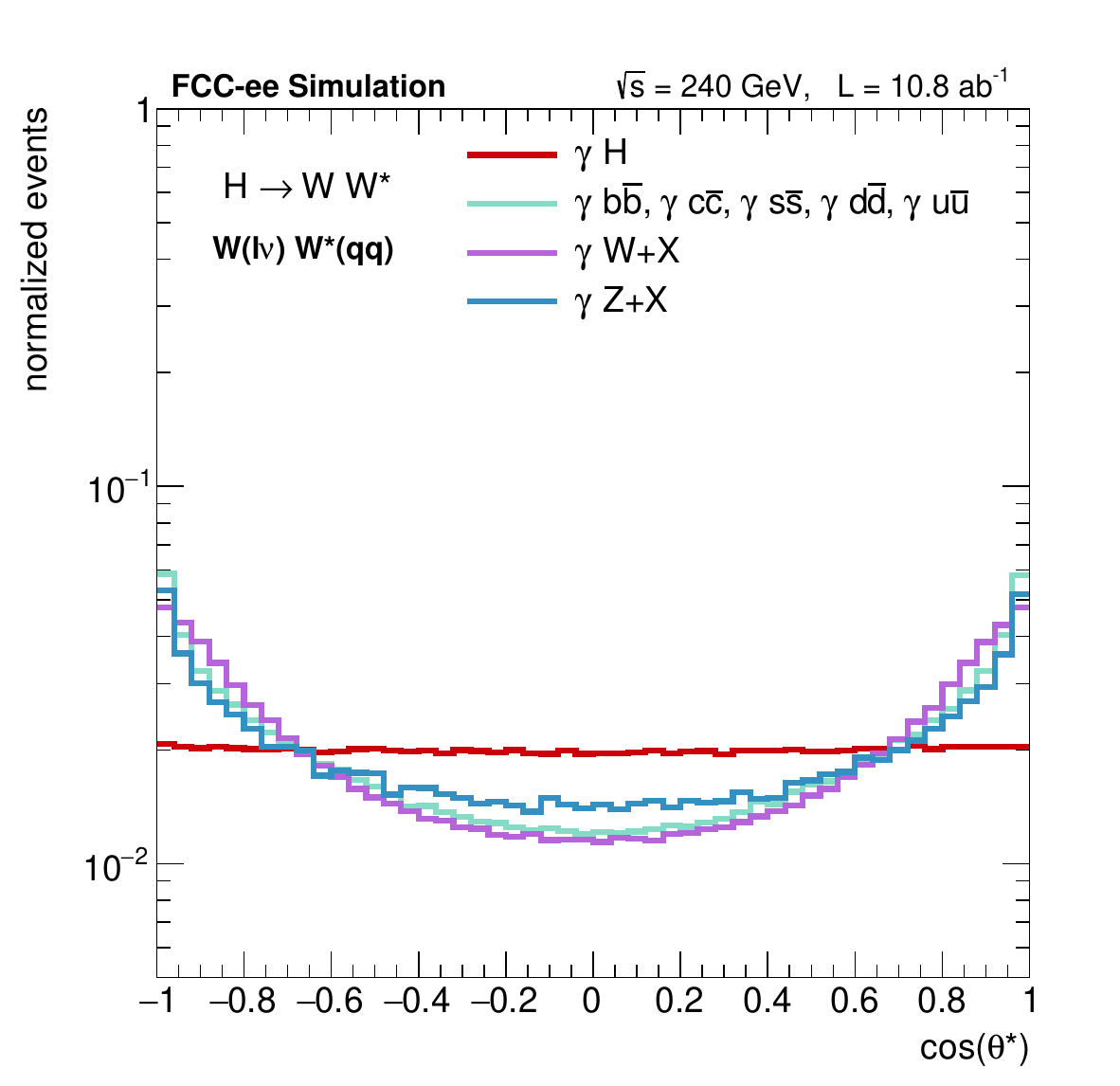}\\
  \end{minipage}

    \caption{
    Distributions of the most discriminating input variables for the
    $H\rightarrow~W(\ell\nu)\,W^*(qq)$ analysis at \(\sqrt{s}=\SI{240}{\giga\electronvolt}\), shown after
    the $H \rightarrow WW^*$ channel selection.
    Top left: reconstructed invariant mass of the leptonic $W$ candidate.
    Top right: missing transverse momentum.
    Bottom left: the photon polar angle \( \mathrm{cos(\theta_\gamma)}\).
    Bottom right: the polar angle of the
$W$ boson in the Higgs rest frame  \( \mathrm{cos(\theta^*})\).
    }
    
  \label{fig:hww_lvqq_vardisc_240}
\end{figure}

A multiclass Boosted Decision Tree (BDT) based on the XGBoost algorithm~\cite{DBLP:journals/corr/ChenG16} is trained to discriminate the signal against all major background classes simultaneously. 
Rather than treating the $\gamma\,W+X$ and $\gamma\,Z+X$ categories as single classes, they are split into separate leptonic and hadronic training classes \(\gamma\,\ell\nu W\) and \(\gamma\,q\bar{q}W\), and \(\gamma\,\ell^+\ell^- Z\) and \(\gamma\,q\bar{q}Z\), since this more closely matches the kinematics of the on-shell and off-shell $WW^*$ contributions in the signal. This allows the BDT to independently exploit the different final-state topologies, increasing the separation power between signal and background.

In total 5 (6) background classes are used at \(\sqrt{s}=\SI{160}{\giga\electronvolt}\) (\(\sqrt{s}\geq\SI{240}{\giga\electronvolt}\)), with the following nomenclature: \textit{QQ} denotes the \(\gamma(b\bar{b}+c\bar{c})\) background, \(\gamma\,\ell\nu W\) and \(\gamma\,q\bar{q}W\) are denoted \textit{\(\ell\nu W\)} and \textit{\(q\bar{q}W\)}, and \(\gamma\,\ell^+\ell^- Z\) and \(\gamma\,q\bar{q}Z\) are denoted \textit{\(\ell\ell Z\)} and \textit{\(q\bar{q}Z\)}. At $\sqrt{s}\geq\SI{240}{\giga\electronvolt}$, the \(\gamma\,H(\nu\bar{\nu})\) process plays a non-negligible role and is included as an additional background class, labeled \(\mathrm{\nu\nu H}\).

The input variables include the four-momenta of the photon, the isolated lepton, the two reconstructed jets, and the missing momentum system, together with the four-momentum of the photon recoil system and the reconstructed $W$ candidates. 
These kinematic variables exploit the characteristic double resonant nature of the signal. Angular variables defined in the Higgs rest frame, specifically the polar angles of the hadronic and leptonic $W$ decay products, exploit the scalar nature of the Higgs boson. The isotropic $W$ decay angular distribution in the Higgs rest frame provides strong separation against the $\gamma\,W+X$ continuum, where the $W$ is produced with a non-trivial spin configuration. The Durham jet clustering scales 
$y_{23}$ and $y_{34}$, which characterize the jet activity in the event, and the number of reconstructed particles excluding the photon are provided as additional inputs.

The lepton flavor is also included as a discriminating variable. Since $t$-channel single $W$ production contributes only to the $\gamma\,e\nu W$ final state and not to $\gamma\,\mu\nu W$, the background composition differs between the electron and muon channels, and the BDT can exploit this asymmetry.

For each jet, the probabilities associated with the seven flavour hypotheses predicted by the jet tagger described in~\cref{sec:mc} are provided as input. This information is particularly powerful because $W$ bosons decay preferentially to $ud$ and $cs$ quark pairs, while the dominant $\gamma\,q\bar{q}$ backgrounds are flavour diagonal and enriched in $b$ and $c$ quarks from 
$Z$ decays. The BDT can therefore exploit the flavour correlation between the two jets to discriminate signal from background. 

A complete list of all input variables is provided in~\cref{app:BDTinput}. The most discriminating observables include the invariant mass of the (on-shell) leptonic $W$ candidate, \(p_{\mathrm{T}}^{\mathrm{miss}}\), the photon polar angle
\(\cos\theta_\gamma\), and the polar angle of the $W$ boson in the Higgs rest frame \(\cos\theta^*\). The normalized distributions of these variables at $\sqrt{s}=\SI{240}{\giga\electronvolt}$ are shown in~\cref{fig:hww_lvqq_vardisc_240}.

\begin{figure}[t]
  \centering

  \begin{minipage}[t]{0.49\textwidth}
    \centering
    \includegraphics[width=\linewidth]{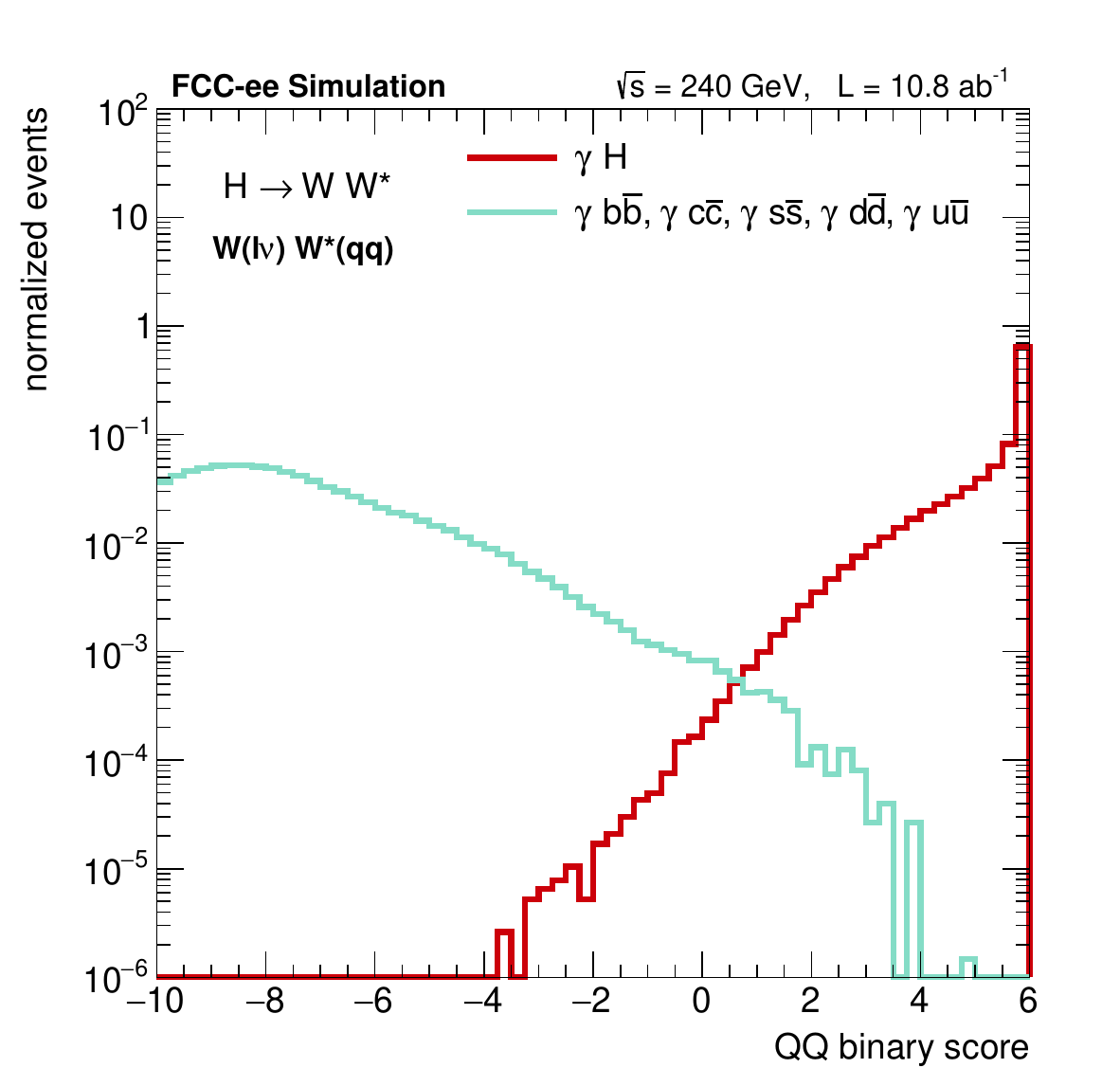}\\
  \end{minipage}\hfill
  \begin{minipage}[t]{0.49\textwidth}
    \centering
    \includegraphics[width=\linewidth]
    {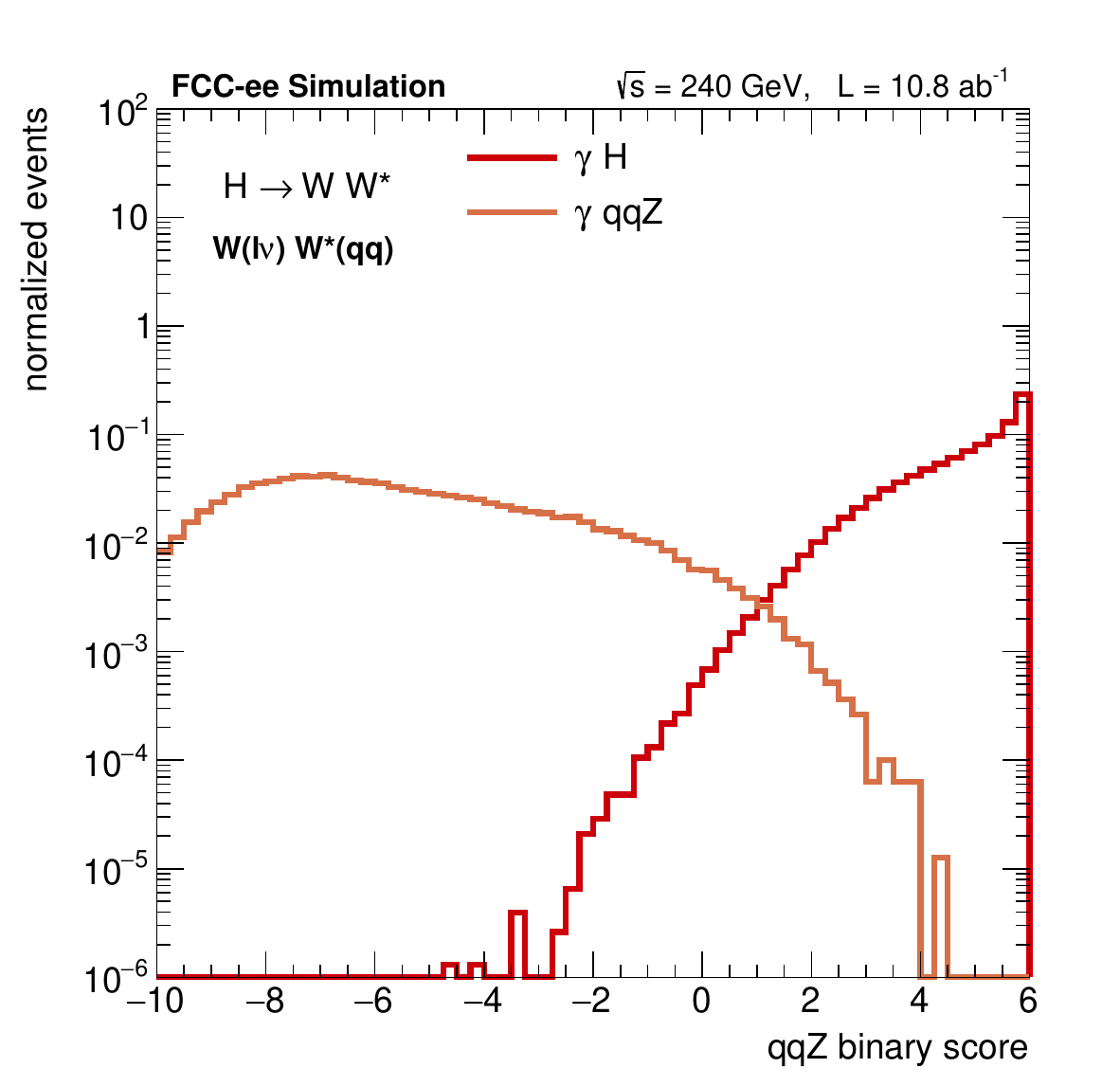}\\
  \end{minipage}

  \vspace{0.25cm}

  \caption{
  Normalised BDT discriminant scores for the $H\rightarrow~W(\ell\nu)\,W^*(qq)$ selection at $\sqrt{s}=\SI{240}{\giga\electronvolt}$, trained against the indicated background in each panel.}
  \label{fig:hww_lvqq_binarydisc_240}
\end{figure}

The multiclass BDT outputs are used in a two step strategy. Reducible backgrounds such as \(\gamma\,b\bar{b}\) and \(\gamma\,c\bar{c}\), \(\gamma \ell \nu W\), \(\gamma q\bar q Z\), and \(\gamma \ell^+\ell^-\ Z\), are suppressed by applying dedicated cuts on their corresponding BDT discriminants. 
For example, \cref{fig:hww_lvqq_binarydisc_240} shows the normalized BDT discriminant scores for QQ and the signal (left) and qqZ and the signal (right). A cut on the QQ score $> 4.25$ and the qqZ score $> 4.0$ effectively rejects the background while keeping most of the signals. 
The remaining irreducible background, dominated by the $\gamma\,q\bar{q}W$ continuum, is not further reduced by selection cuts. Instead, its dedicated BDT discriminant output is used directly as the observable in the binned profile likelihood fit for signal extraction. After the full selection at $\sqrt{s}=\SI{240}{\giga\electronvolt}$, 23 signal events remain against a total background of 2039, dominated by the $\gamma\,q\bar{q}W$ continuum, 
At $\sqrt{s}=\SI{160}{\giga\electronvolt}$ and \SI{365}{\giga\electronvolt}, the signal yields are reduced to 10 and 2 events against backgrounds of 3782 
and 56.

\begin{figure}[t]
  \centering
  \begin{minipage}[t]{0.49\textwidth}
    \centering
    \includegraphics[width=\linewidth]{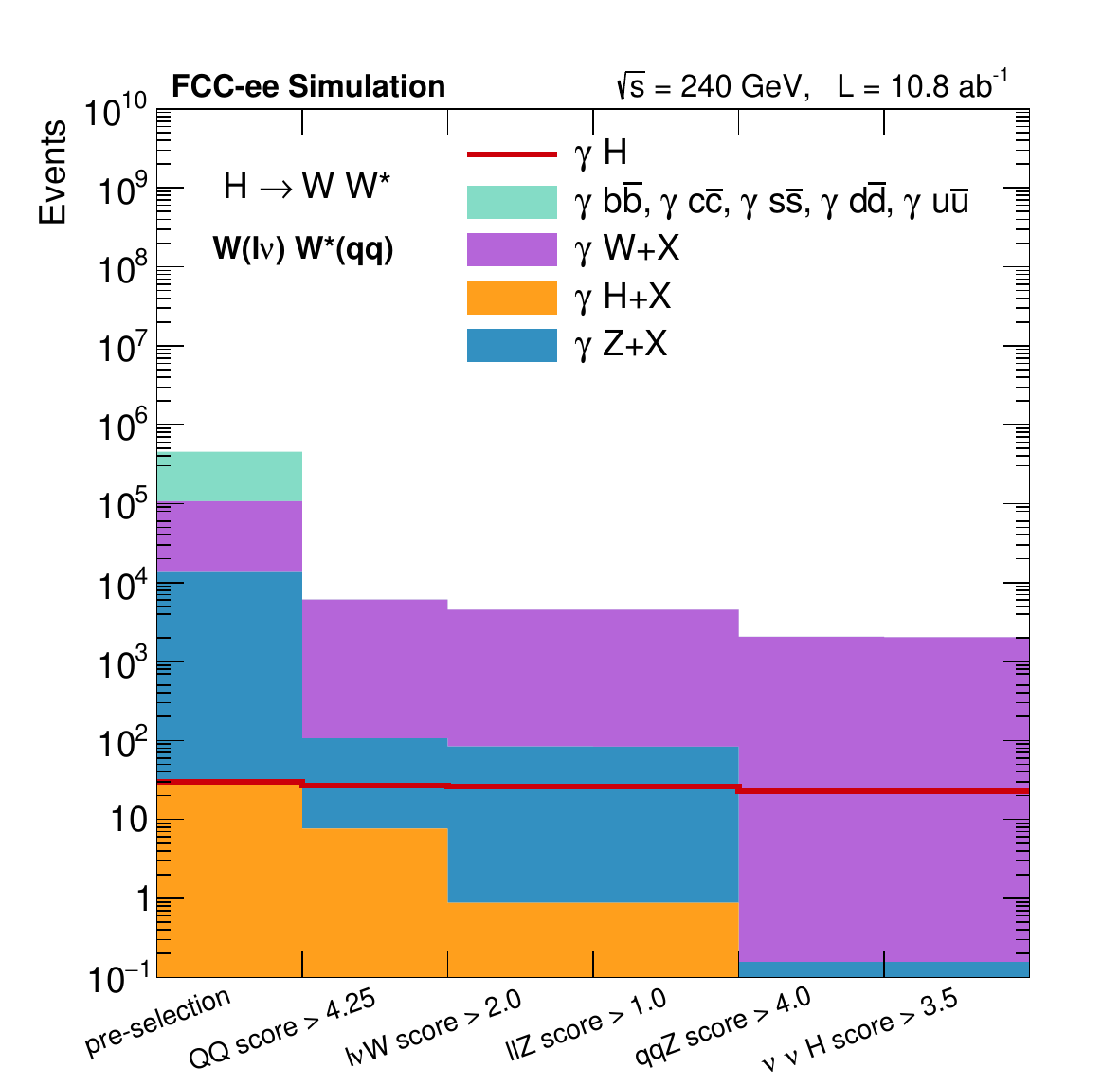}
  \end{minipage}\hfill
  \begin{minipage}[t]{0.49\textwidth}
    \centering
    \includegraphics[width=\linewidth]{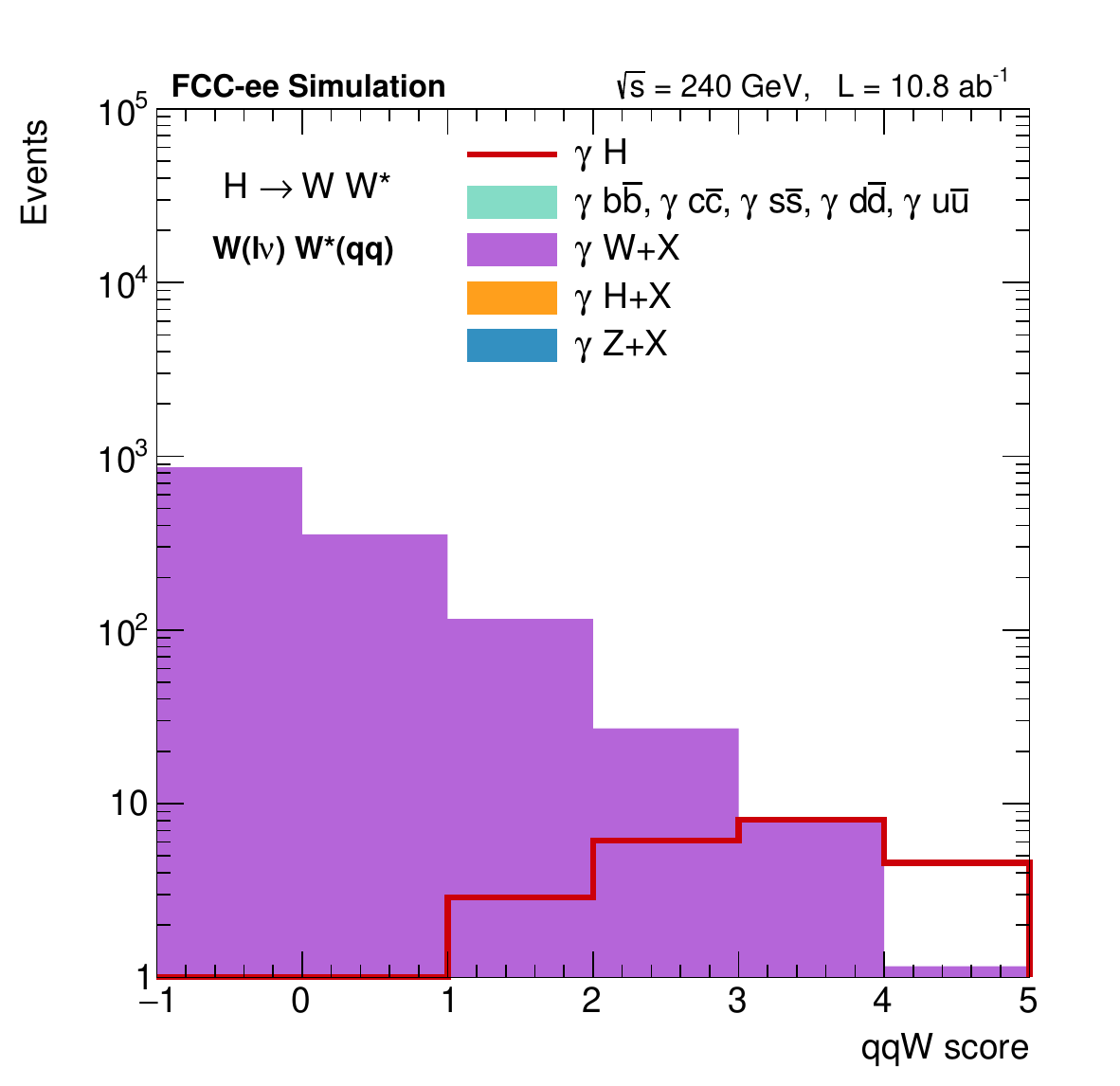}
  \end{minipage}

\caption{
Final selection for the $H\rightarrow W(\ell\nu)\,W^*(qq)$ analysis at
\(\sqrt{s}=240\)~GeV.
Left: cumulative cut flow for the signal and dominant background classes in the multiclass BDT based selections.
Right: BDT discriminant against the remaining \(\gamma q\bar q W\) background, used as input to the binned likelihood fit.
}

\label{fig:hww_lvqq_final_240}
\end{figure}

The complete cut flow up to the final selection is presented in the left panel of~\cref{fig:hww_lvqq_final_240}, while the right panel shows the final \(q\bar q W\) discriminant used as input to the likelihood fit, exhibiting a signal enhanced region at large discriminant values. The BDT is retrained independently for each center-of-mass energy to account for the different signal and background kinematics. The corresponding figures for $\sqrt{s}=\SI{160}{\giga\electronvolt}$ and $\sqrt{s}=\SI{365}{\giga\electronvolt}$ are shown in~\cref{app:hwwpreselection}.

\subsubsection{$H \rightarrow W(qq)\,W^*(\ell\nu)$ Channel}
\label{sec:HWW_qqlv}

The analysis of the complementary semi-leptonic channel
\(H \rightarrow W(qq)\,W^*(\ell\nu)\)
closely follows the strategy described for the
$H \rightarrow W(\ell\nu)\,W^*(qq)$ final state.
The same channel dependent selection is used. Events are assigned to this category by requiring an on-shell hadronic $W$ decay,
\[
m_{jj} > \SI{60}{\giga\electronvolt},
\]
which ensures orthogonality with the off-shell hadronic topology.

\begin{figure}[t]
  \centering

  \begin{minipage}[t]{0.49\textwidth}
    \centering
    \includegraphics[width=\linewidth]{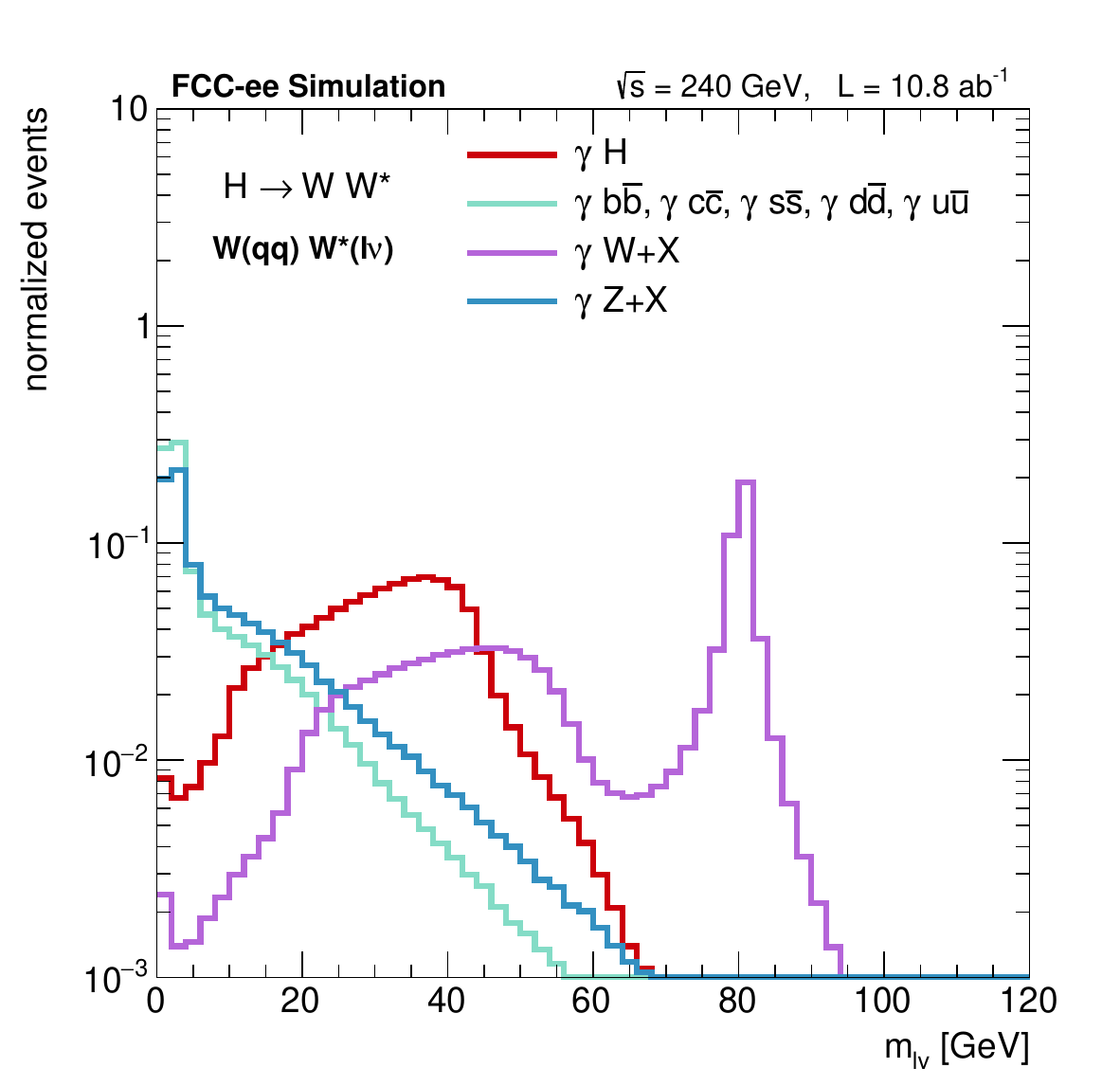}
  \end{minipage}\hfill
  \begin{minipage}[t]{0.49\textwidth}
    \centering
    \includegraphics[width=\linewidth]{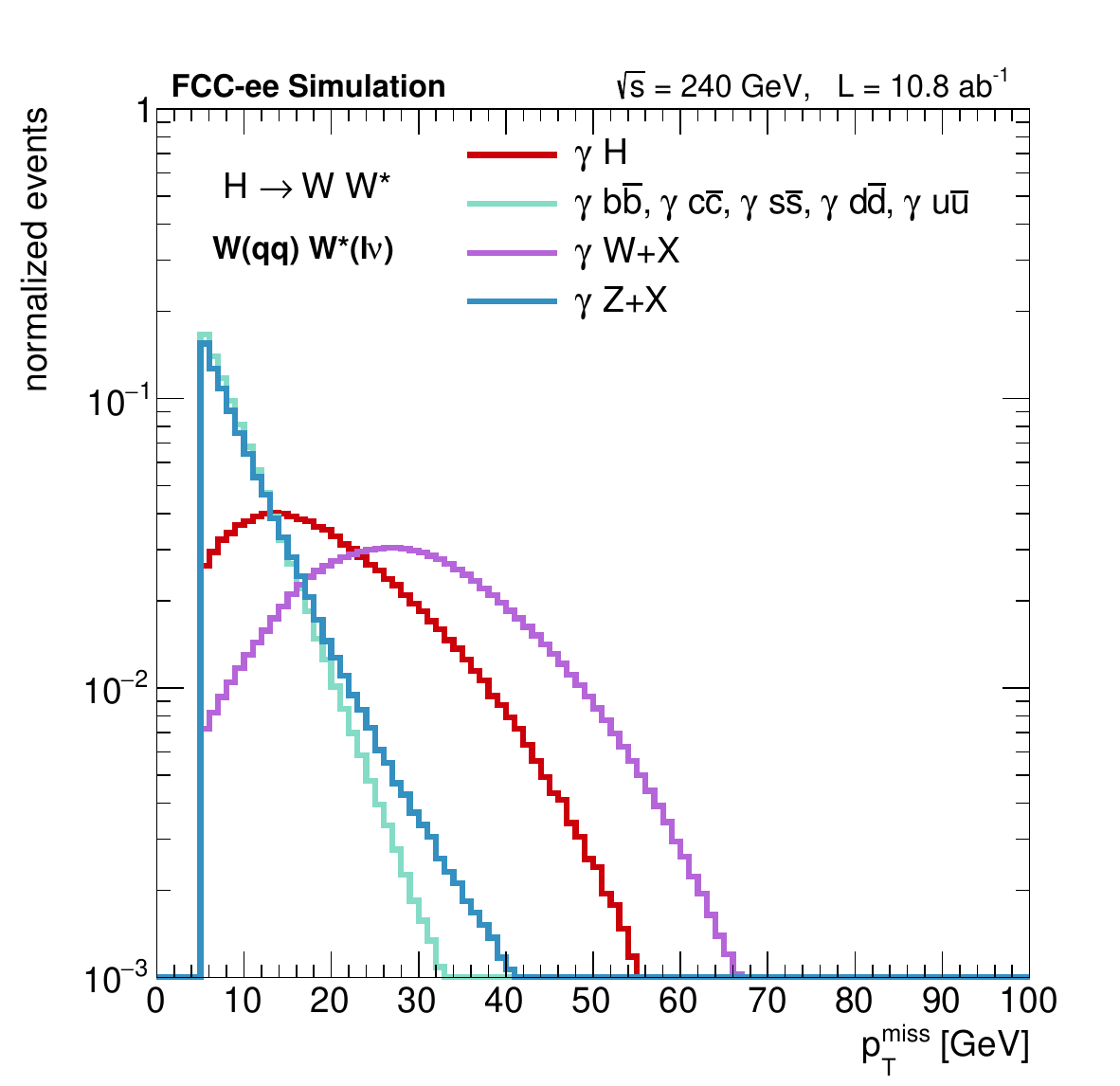}
  \end{minipage}

  \caption{
  Distributions of discriminating variables for the
  $H\rightarrow W(qq)\,W^*(\ell\nu)$ analysis at \(\sqrt{s}=240\)~GeV after preselection.
  Left: reconstructed invariant mass of the leptonic $W^*$ candidate.
  Right: missing transverse momentum.
  }
  \label{fig:hww_qqlv_vardisc_240}
\end{figure}

The leptonic $W^*$ decay produces softer leptons and \(p_{\mathrm{T}}^{\mathrm{miss}} \). As a consequence, radiative hadronic backgrounds, in particular \(\gamma\,b\bar{b}\) and \(\gamma\,c\bar{c}\), that can produce semi-leptonic decays can easily mimic the signal. This is illustrated in~\cref{fig:hww_qqlv_vardisc_240}, which shows the distributions of the reconstructed leptonic $W^*$ mass and of \(p_{\mathrm{T}}^{\mathrm{miss}}\) after preselection. Therefore, compared to the on-shell analysis, this channel is more difficult to isolate.

As in the on-shell analysis, reducible backgrounds are suppressed through cuts on their corresponding binary BDT outputs. After the full selection at $\sqrt{s}=\SI{240}{\giga\electronvolt}$, 14 signal events remain against a total background of 1616, dominated by the $\gamma\,\ell\nu W$ continuum. At $\sqrt{s}=\SI{160}{\giga\electronvolt}$ and $\SI{365}{\giga\electronvolt}$, 
the signal yields are reduced to 8 and 1 events against backgrounds of 2702 and 50, respectively.

In contrast to the $W(\ell\nu)\,W^*(qq)$ channel, the dominant irreducible background after these selections is associated with the
\(\gamma\,\ell\nu W\) topology. Consequently, the final statistical discrimination is performed using the \(\ell\nu W\) BDT output, which is employed as the input observable to the binned
likelihood fit.

\begin{figure}[t]
  \centering
  \begin{minipage}[t]{0.49\textwidth}
    \centering
    \includegraphics[width=\linewidth]{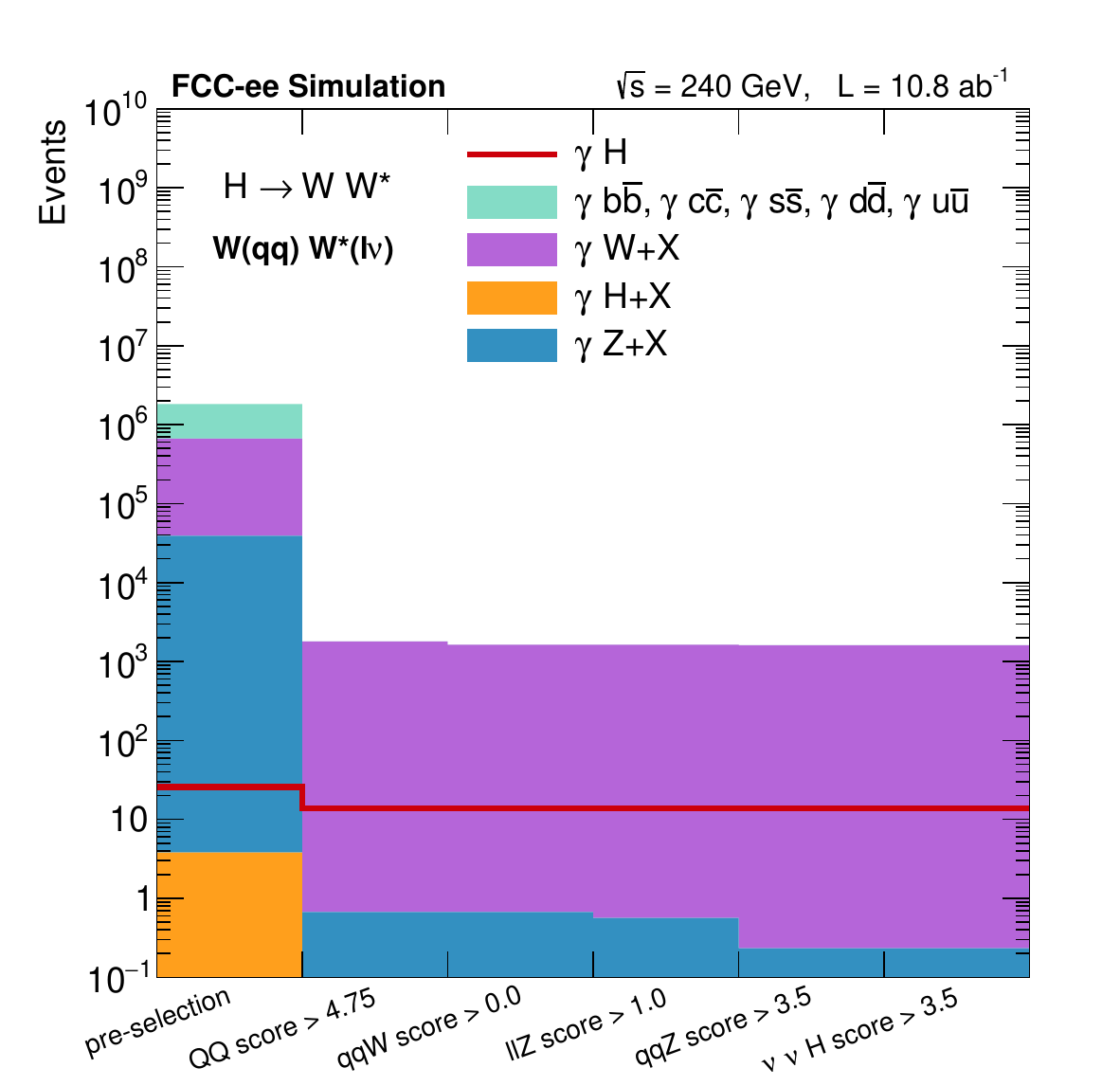}
  \end{minipage}\hfill
  \begin{minipage}[t]{0.49\textwidth}
    \centering
    \includegraphics[width=\linewidth]{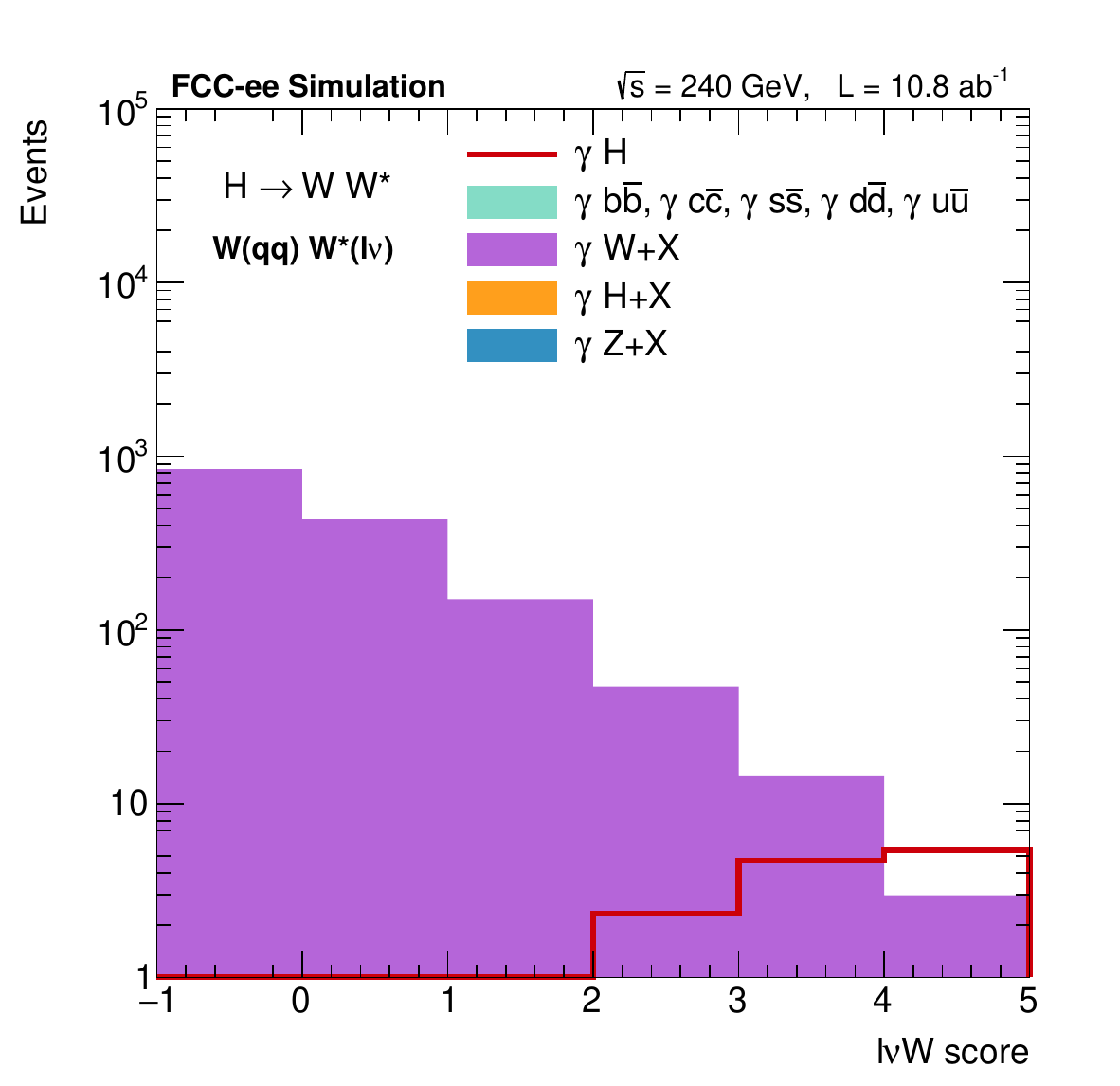}
  \end{minipage}

  \caption{
  Final selection for the $H\rightarrow W(qq)\,W^*(\ell\nu)$ analysis at
  \(\sqrt{s}=\SI{240}{\giga\electronvolt}\). Left: cumulative cut flow for the signal and dominant background classes. Right: BDT discriminant against the remaining \(\gamma\,\ell\nu W\) background, used as input to the binned likelihood fit.
  }
  \label{fig:hww_qqlv_final_240}
\end{figure}

The complete cut flow and the final \(\ell\nu W\) discriminant are shown in~\cref{fig:hww_qqlv_final_240}. Due to the softer kinematics and the increased overlap with hadronic backgrounds,
the achievable signal-to-background ratio is smaller than in the on-shell analysis. The corresponding figures for $\sqrt{s}=\SI{160}{\giga\electronvolt}$ and $\sqrt{s}=\SI{365}{\giga\electronvolt}$ are shown in~\cref{app:hwwpreselection}.

\subsection{The $H \to b\bar{b}$ channel}
\label{sec:Hjj}

Higgs decays to two hadronic jets account for a combined branching fraction of approximately 74\%, including $H\to b\bar{b}$ (58\%), $H\to gg$ (8\%), $H\to c\bar{c}$ (3\%), and $H\to\tau^+\tau^-$ (6\%). In this study we focus exclusively on the $H\to b\bar{b}$ channel, which offers the best combination of signal yield and background rejection owing to the large branching fraction 
and the availability of efficient $b$-tagging. The $H\to gg$ channel was investigated but found to be non-competitive. Indeed, although the $\gamma\,q\bar{q}$ background does not contain a genuine $Z\to gg$ contribution, the limited separation between gluon and quark initiated jets provided by current tagging algorithms makes the dominant background difficult to suppress, resulting in an expected precision of $\mathcal{O}(500\%)$ at $\sqrt{s}=\SI{240}{\giga\electronvolt}$. The $H\to\tau^+\tau^-$ channel in the fully hadronic $\tau$ decay mode was found to be even less sensitive. Both channels are therefore excluded from the current combination.

\subsubsection{$H \rightarrow b\bar{b}$ channel preselection}

 \begin{figure}[t]
  \centering
  \begin{minipage}[t]{0.49\textwidth}
    \centering
    \includegraphics[width=\linewidth]{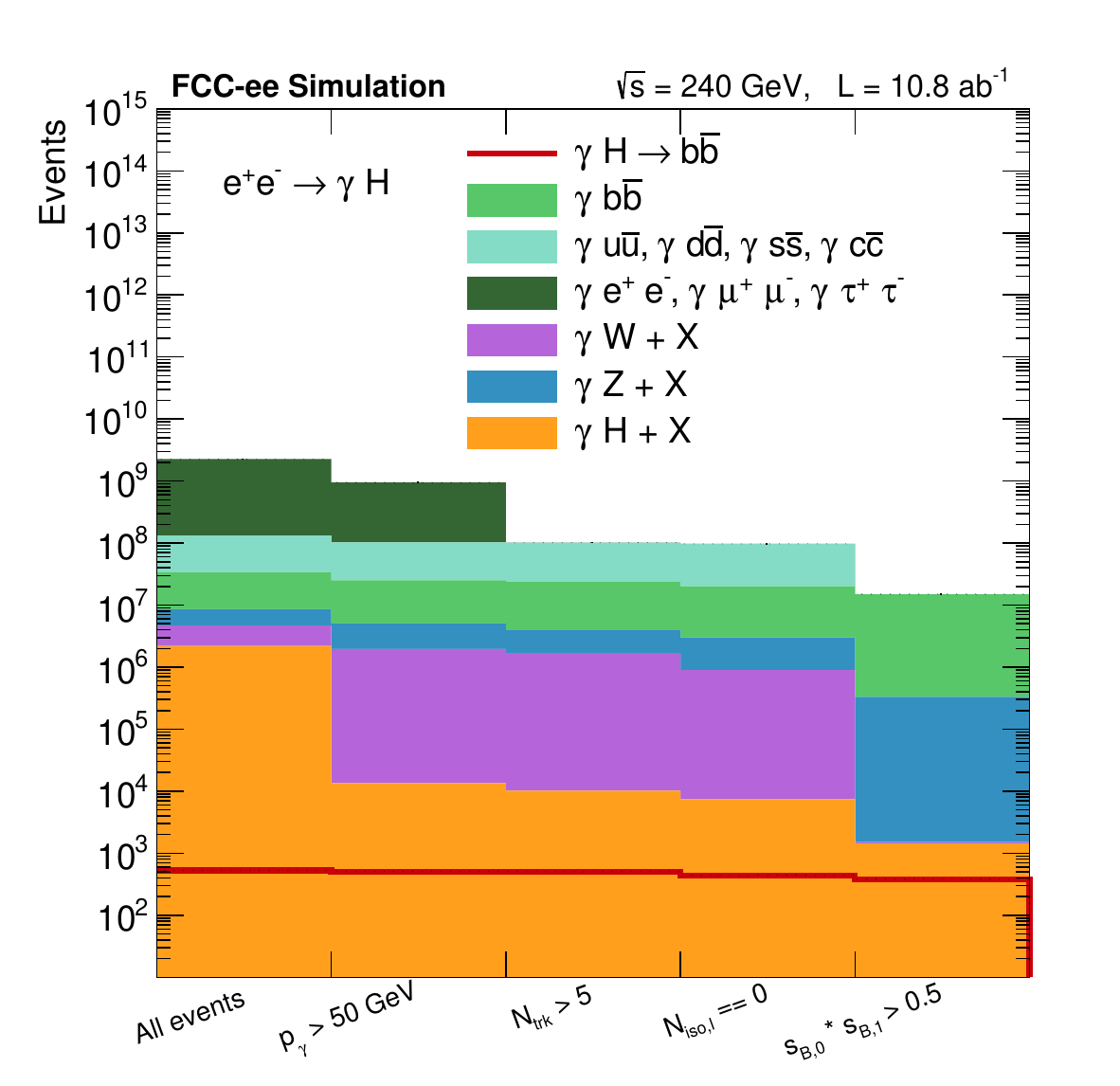}
  \end{minipage}\hfill
  \begin{minipage}[t]{0.49\textwidth}
    \centering
    \includegraphics[width=\linewidth]{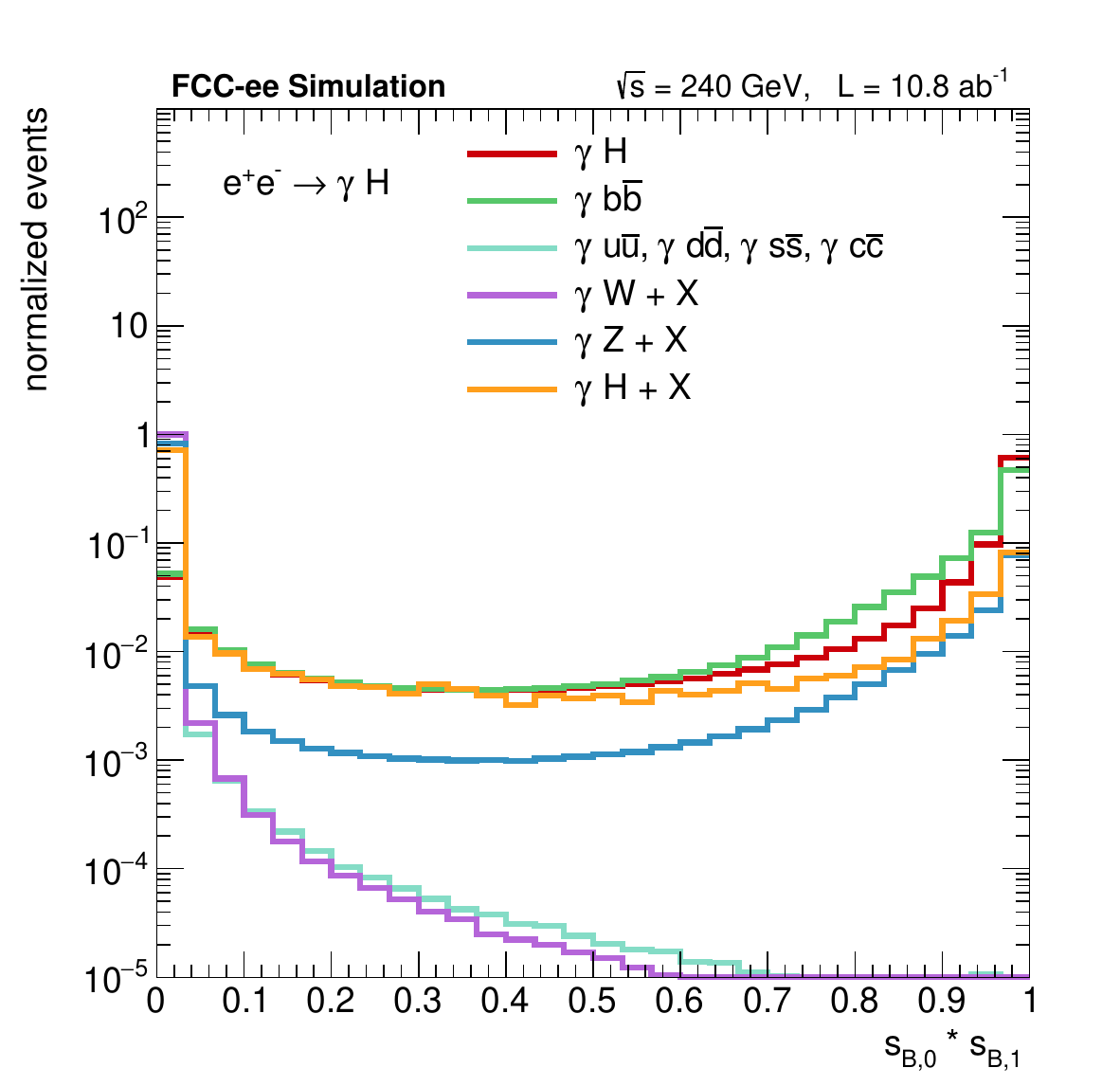}
  \end{minipage}
\caption{Cumulative yields for the signal and background processes in the $b\bar{b}$ final state (left). Product of b-jet-like scores after preselection and lepton veto, illustrating the central channel selection cut (right).}
  \label{fig:Hbb_specific_cuts}
\end{figure}

To ensure orthogonality with the semi-leptonic $H\to WW^*$ analysis 
described in~\cref{sec:HWW}, events containing at least one isolated lepton are vetoed. The preselection criteria, discussed in~\cref{sec:ana_strat}, are designed to isolate the \eeha\ production mode and are applied in combination with the isolated lepton veto. 

After removing the isolated photon, the remaining reconstructed particles are clustered into two jets. The jet flavour tagging scores defined in~\cref{sec:mc} are used to identify the partonic flavor of each jet. After the common preselection and lepton veto, the dominant backgrounds are $\gamma\,q\bar{q}$ (overwhelmingly $\gamma\,b\bar{b}$ and $\gamma\,c\bar{c}$), followed by 
$\gamma\,Z+X$ and $\gamma\,H+X$. The $\gamma\,W+X$ background, which involves non diagonal flavour combinations from $W$ decays, is efficiently suppressed by the $b$-tagging requirement. The final channel specific selection observable is defined as the product of the $b$-tagging probabilities of the two reconstructed jets, shown in the right panel of~\cref{fig:Hbb_specific_cuts}. Requiring this product to be greater than 0.5 retains the bulk of the $\gamma\,b\bar{b}$  
background while strongly suppressing all other contributions, leaving $\gamma\,b\bar{b}$ as the dominant background by almost two orders of magnitude for all investigated center-of-mass scenarios. At $\SI{240}{\giga\electronvolt}$ a total of 380 signal and $15 \times 10^{6}$ background events remain. For $\SI{160}{\giga\electronvolt}$ and $\SI{365}{\giga\electronvolt}$ the signal yields 152 and 37 events after the channel preselection, the background $96 \times 10^{6}$ and $1.3 \times 10^{6}$ , respectively. The cut flow of the $\SI{240}{\giga\electronvolt}$ scenario, showing the corresponding event yields at each selection stage are shown in the left panel of~\cref{fig:Hbb_specific_cuts}. The corresponding event selection for $\sqrt{s}=\SI{160}{\giga\electronvolt}$ and $\sqrt{s}=\SI{365}{\giga\electronvolt}$ are shown in~\cref{fig:fitobs_160,fig:fitobs_365}.

\subsubsection{Multi-variate analysis}

Since the $\gamma b\bar{b}$ final-state topology closely resembles that of the signal, further discrimination based solely on a simple kinematic event selection would lead to a significant loss in signal efficiency. Therefore, a BDT binary classifier is employed to enhance the separation between signal and most dominant background ($\gamma b\bar{b}$). Rather than imposing sequential one dimensional kinematic selections, the BDT exploits the full multivariate information of the event, including correlations among discriminating observables. The XGBoost framework~\cite{DBLP:journals/corr/ChenG16} BDT implementation is used. The classifier is trained on a balanced dataset of $O(10^6)$
 signal ($\gamma$H($b\bar{b}$)) and background ($\gamma b\bar{b}$) events. 
The choice of input observables follows the strategy outlined in Ref.~\cite{darkphoton_study} and are chosen to exploit the kinematic and topological differences between signal and background. They can be grouped into three physically motivated categories: (a) Global event kinematics, (b) angular and topological variables, and (c) Higgs mass reconstruction and recoil observables.

\paragraph{Global event kinematics}

Momentum conservation implies a balanced event topology when all final-state particles are reconstructed. Apparent imbalances, quantified by the missing momentum $p^\mathrm{miss}$ and missing transverse momentum $p_{\mathrm{T}}^\mathrm{miss}$, 
arise from undetected particles or limited detector acceptance and therefore provide discriminating power. The energy ratio of the two reconstructed jets $E_{j_1}/E_{j_0}$, where $j_0$ and $j_1$ are the leading and subleading jets, respectively, provides complementary information on the energy sharing between them. In signal events, the jets originate from a Higgs decay, while in the dominant $\gamma\,b\bar{b}$ background they arise from a $Z$ boson, leading 
to different boost, energy configurations and spin.

\paragraph{Angular and topological observables}

The production mechanisms of signal and background processes lead to different angular configurations of the final-state jets. This information is encoded in the polar and azimuthal angles of the two jets, cos($\theta(j_{0,1})$) and cos($\phi(j_{0,1})$), as well as in their three-dimensional opening angle
  \begin{equation}
         \cos(\alpha(jj))=\frac{\Vec{p_{j_0}}\cdot \Vec{p_{j_1}}}{|\Vec{p_{j_0}}||\Vec{p_{j_1}}|}.
     \end{equation} 
The opening angle is particularly sensitive to the boost of the parent particle and therefore provides additional discrimination between Higgs decays and $\gamma b\bar{b}$ production.

\paragraph{Higgs reconstruction and recoil observables}

The signal process features a resonant di-jet system and a monochromatic photon recoiling against the Higgs boson. This is captured by the di-jet invariant mass $m_{jj}$, the recoil mass $m_\mathrm{recoil}$, and the photon momentum $p_\gamma$, which are the most discriminating variables in this category. To combine the information from the reconstructed Higgs mass and the expected photon energy into a single variable, a signal compatibility variable is also constructed as
\begin{equation}
  \chi^2_H = \frac{(m_{jj}-m_H)^2}{\sigma^2(m_{jj})} + 
  \frac{\left(E_{\gamma} - \frac{s - m_H^2}{2\sqrt{s}}\right)^2}{\sigma^2(E_\gamma)},
\end{equation}
which quantifies the compatibility of an event with the signal hypothesis, accounting for detector resolution. The Higgs mass resolution $\sigma(m_{jj})$ is estimated from the full width at half maximum (FWHM) of the reconstructed dijet mass distribution, while the photon energy resolution is parametrized as $\sigma(E_\gamma) = 0.03\sqrt{E_\gamma}$, consistent with the assumed IDEA electromagnetic calorimeter performance. In practice, $\chi^2_H$ provides limited additional discrimination, as it combines observables already supplied individually to the BDT. \\

A list of all input variables is provided in \cref{app:BDTinput}.
The discriminating power of the variables depends on the Higgs boost and therefore on the center-of-mass energy. Consequently, the BDT is retrained for each $\sqrt{s}$ configuration. For the $\sqrt{s}=\SI{240}{\giga\electronvolt}$ selection, the most discriminating variables are the di-jet mass, missing transverse momentum, photon momentum, jet opening angle, and missing momentum. The strong separation from $m_{jj}$ arises from the resonant peaks at the Higgs and Z masses.

 \begin{figure}[]
   \centering

  \begin{minipage}[t]{0.5\textwidth}
    \centering
    \includegraphics[width=0.9\linewidth]{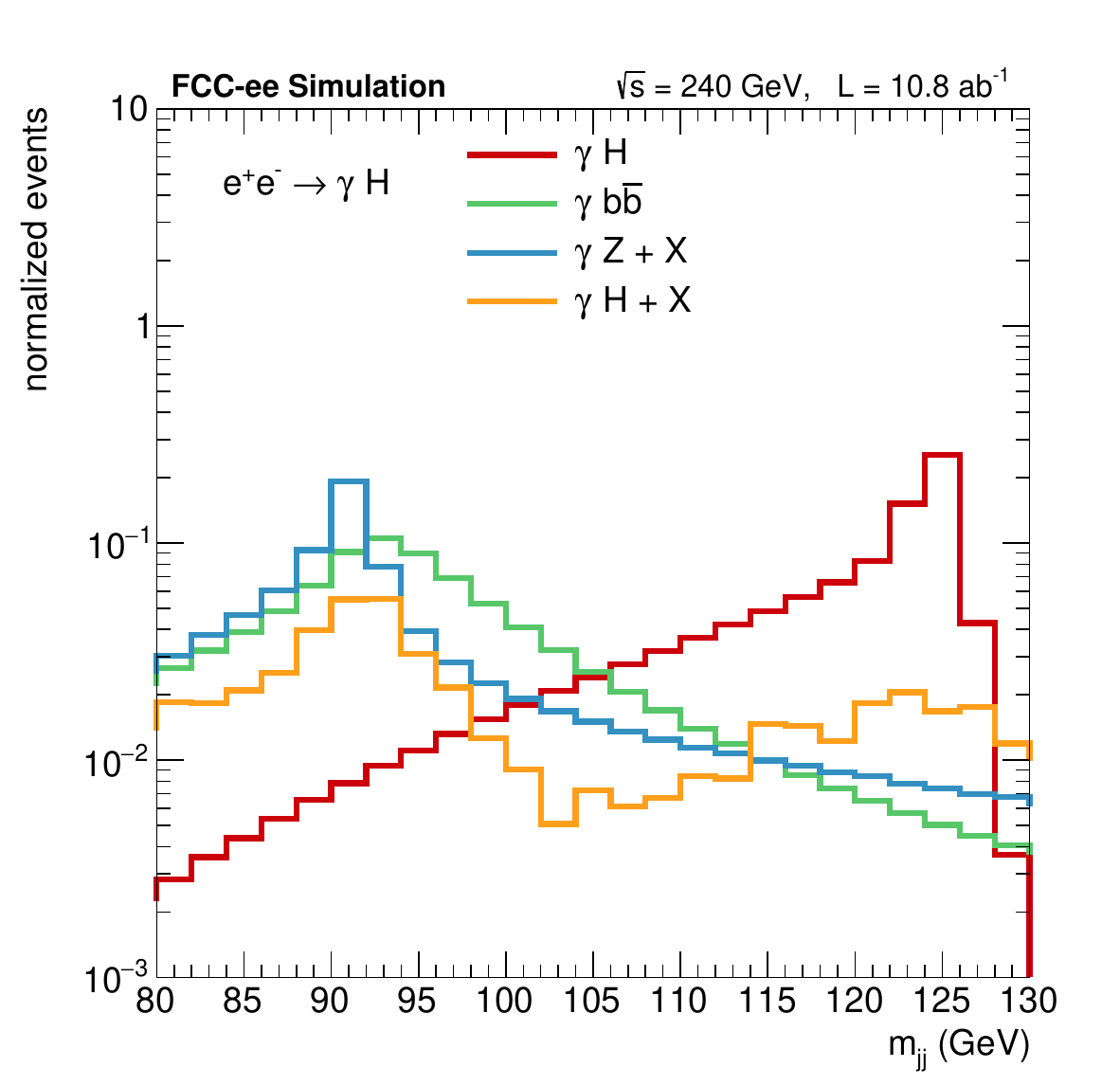}\\
  \end{minipage}\hfill
  \begin{minipage}[t]{0.5\textwidth}
    \centering
    \includegraphics[width=0.9\linewidth]{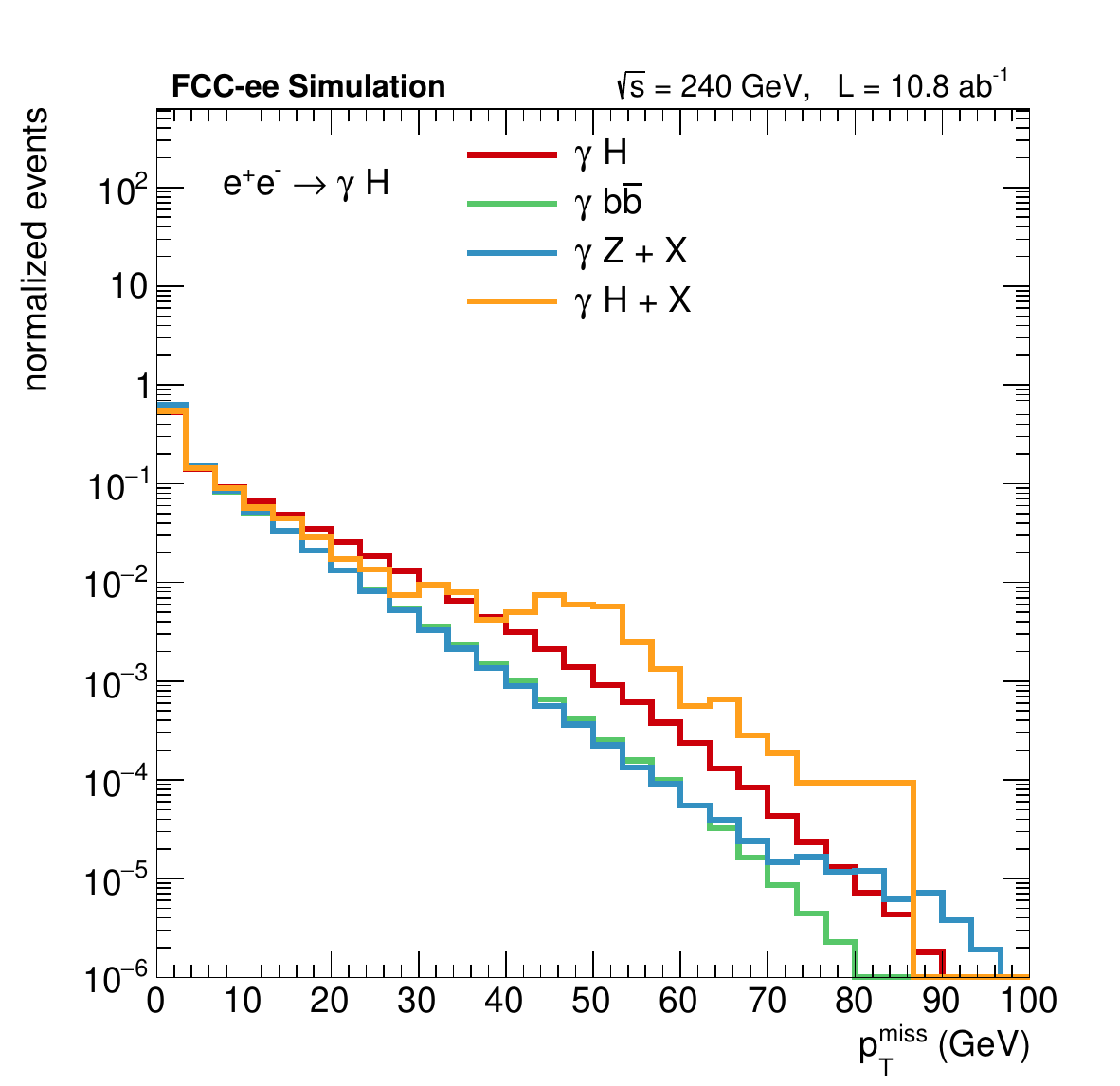}\\
  \end{minipage}

  \vspace{0.25cm}

  \begin{minipage}[t]{0.5\textwidth}
    \centering
    \includegraphics[width=0.9\linewidth]{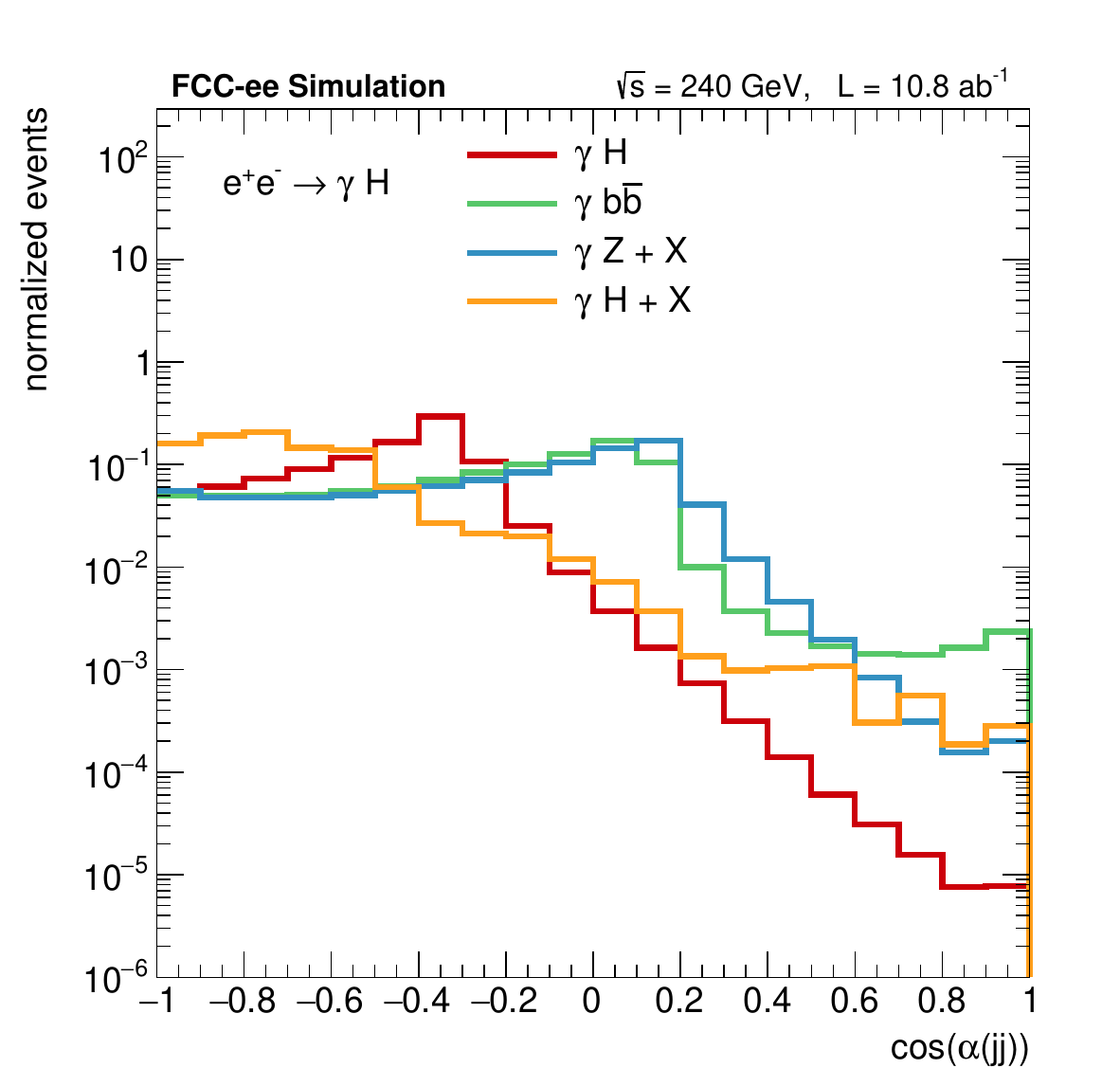}\\
  \end{minipage}\hfill
  \begin{minipage}[t]{0.5\textwidth}
    \centering
    \includegraphics[width=0.9\linewidth]{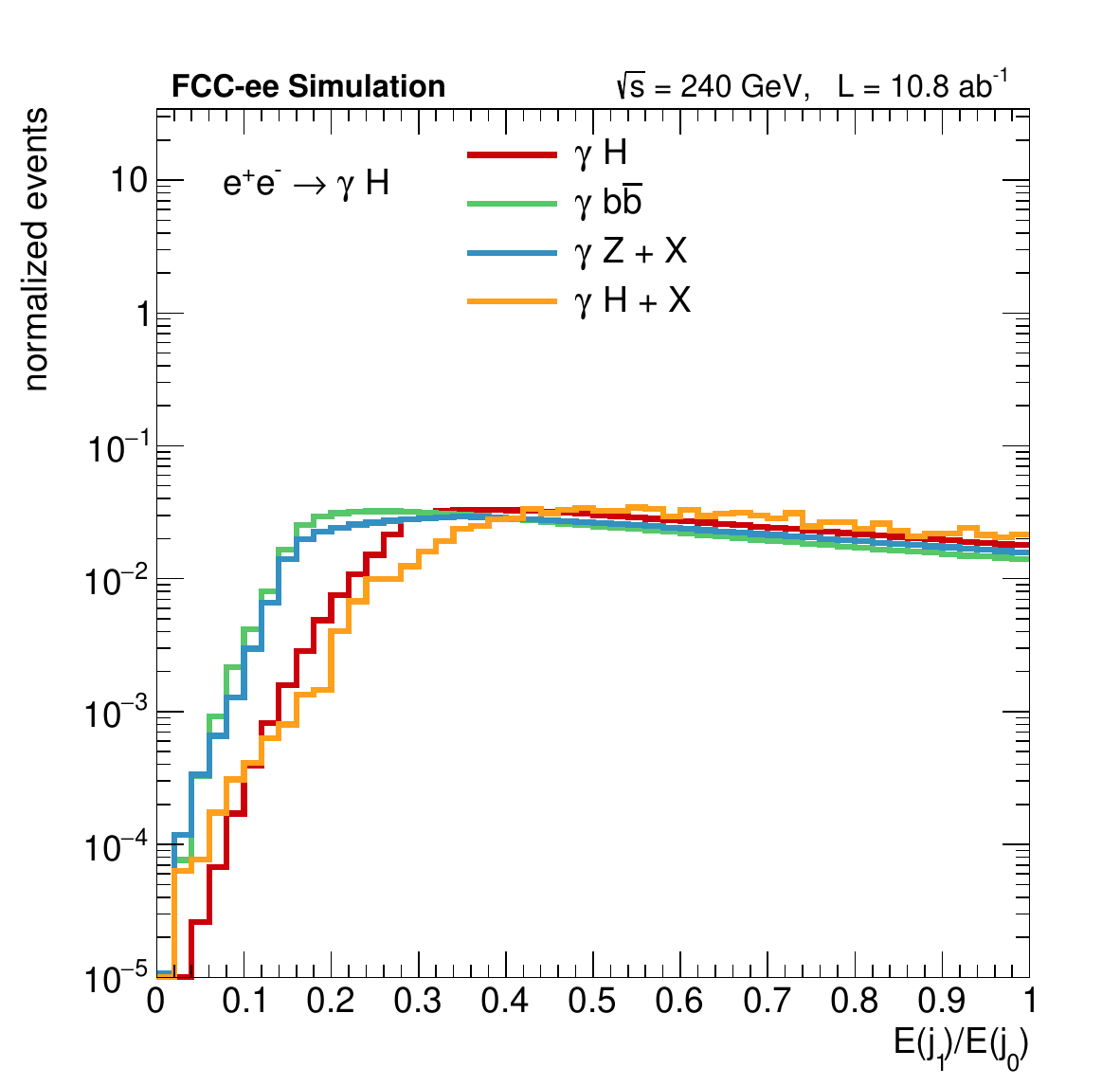}\\
  \end{minipage}
\caption{Distributions of representative BDT input variables for the $H\to b\bar{b}$ analysis at $\sqrt{s}=\SI{240}{\giga\electronvolt}$, shown after the channel-specific selection. Top left: dijet mass $m_{jj}$. Top right: missing transverse momentum $p_{\mathrm{T}}^\mathrm{miss}$. 
Bottom left: three-dimensional opening angle of the two jets $\cos(\alpha_{jj})$. Bottom right: jet energy ratio $E_{j_1}/E_{j_0}$.}
  \label{fig:BDT_inputs}
\end{figure}

\begin{figure}[htb]
  \centering
  \includegraphics*[width=0.5\textwidth]{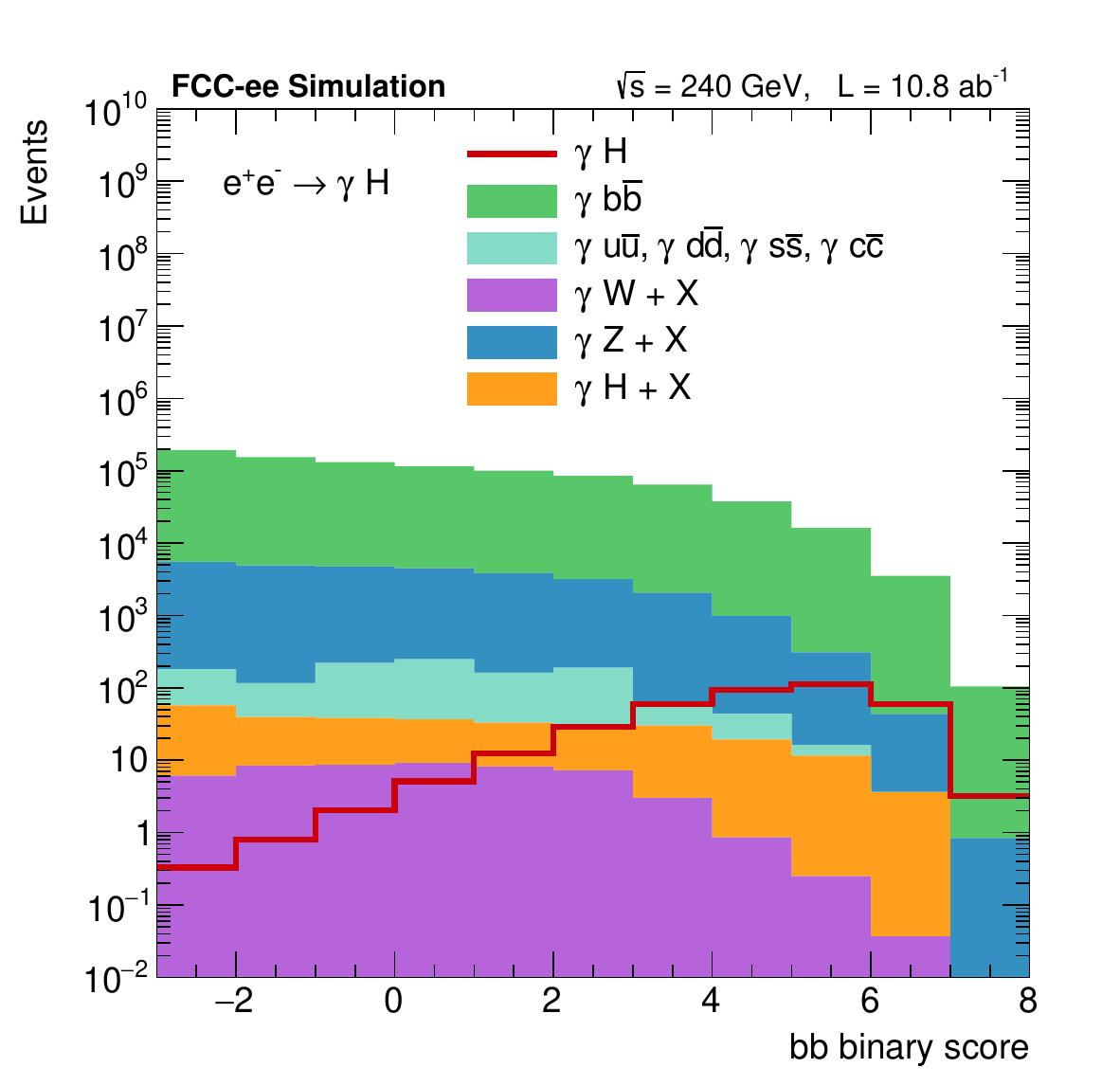}
  \caption{BDT discriminant for the $H\to b\bar{b}$ analysis 
  at $\sqrt{s}=\SI{240}{\giga\electronvolt}$, used as input to the binned likelihood fit.}
  \label{fig:BDT_score_bb}
\end{figure}

The distributions of a subset of input variables are shown in~\cref{fig:BDT_inputs} to illustrate the kinematic and topological differences between signal and 
background. The BDT is retrained independently for each center-of-mass energy to account for the different signal and background kinematics. The trained classifier output is used as the final observable in the statistical analysis. To improve fit stability, the logit transformation of the BDT score is applied, 
which avoids compression of events near the score boundaries and allows for improved binning while preserving the discriminating power, as shown in~\cref{fig:BDT_score_bb}. The corresponding discriminants for $\sqrt{s}=\SI{160}{\giga\electronvolt}$ and $\sqrt{s}=\SI{365}{\giga\electronvolt}$ are shown in~\cref{fig:fitobs_160,fig:fitobs_365}.

\subsection{Results and Combination}
\label{sec:expresults}

For all the analysis channels described above, the signal extraction is performed using a binned profile likelihood fit to a single observable. The parameter of interest (POI) is the signal strength \(\mu_{\gamma H}\), defined as the ratio of the measured \(\sigma(e^+e^-\to\gamma H)\times\mathrm{BR}\) to its SM prediction. 

The fit is performed on Asimov data sets, corresponding to pseudo-data generated under the SM expectation.
For each analysis category, binned templates are prepared for the signal and all background processes and used as input to the likelihood. In each category, the likelihood function is constructed as the product of Poisson probabilities over all bins, comparing the observed yield to the expected sum of signal and background contributions. 

As nuisance parameters, one normalisation factor per background process is included in the fit. Since background normalisations will be predicted to high accuracy at the FCC-ee, and the luminosity will be known to better than 1\%, a conservative prior uncertainty of 1\% is assigned to the background yields. Shape uncertainties are neglected in the present study. Statistical uncertainties arising from the finite size of simulated samples are
incorporated using the Beeston--Barlow method~\cite{Barlow:1993dm}. Given the limited number of dominant background processes and the large Monte Carlo
samples foreseen at the FCC-ee, such statistical uncertainties are expected to be negligible in a full experimental analysis.

Additional systematic effects related to background modelling, event selection, and flavour tagging are not included. At the FCC-ee, large control samples, in particular using radiative Z events, will allow these uncertainties to be constrained well below the level of the quoted sensitivities.

Standalone fits are performed for each considered channel: $H \rightarrow b \bar{b}$, $H \rightarrow W(\ell \nu)W^*(qq)$, and $H \rightarrow W(qq)W^*(\ell \nu)$. Since all analyzed final states are mutually orthogonal by construction, the individual measurements can be combined to obtain the overall precision on the \(\gamma H\) production process at the different FCC-ee centre-of-mass energies \(\sqrt{s}=160\), 240 and \SI{365}{\giga\electronvolt}. In addition, while the results are expressed in terms of $\sigma\times\mathrm{BR}$, the Higgs branching fractions will be known with high precision from the global FCC-ee Higgs programme, so the signal strengths can be directly interpreted as measurements of the $\gamma H$ production cross section.

The resulting relative precisions on \(\sigma(e^+e^-\to\gamma H)\times\mathrm{BR}(H\to XX)\) are summarised in~\cref{table:results}. At \SI{240}{\giga\electronvolt}, where both the signal cross section and the integrated
luminosity are largest, the best sensitivity is achieved, with a combined precision of 26\%. At \SI{160}{\giga\electronvolt} and \SI{365}{\giga\electronvolt}, the expected precision is reduced due to the smaller signal yields, but remains relevant for the re-interpretation in terms of constraints on the effective $\gamma\gamma/Z\gamma$ couplings, as will be discussed in~\cref{sec:interpretation}.

\begin{table}[t]
  \centering
  \renewcommand{\arraystretch}{1.15}
  \footnotesize
  \resizebox{\textwidth}{!}{
  \begin{tabular}{lccc|c}
    \toprule
    $\sqrt{s}$
    & $H\to b\bar b$
    & $H\to W(\ell\nu)\,W^*(qq)$
    & $H\to W(qq)\,W^*(\ell\nu)$
    & \textbf{Combination} \\
    \midrule
    \SI{160}{\giga\electronvolt}
    & 214\%
    & 115\%
    & 132\%
    & \textbf{80\%} \\
    \SI{240}{\giga\electronvolt}
    & 71\%
    & 34\%
    & 46\%
    & \textbf{26\%} \\
    \SI{365}{\giga\electronvolt}
    & 255\%
    & 122\%
    & 203\%
    & \textbf{97\%} \\
    \bottomrule
  \end{tabular}
  }
  \caption{
  Relative precision
  \(\delta(\sigma_{\gamma H}\times\mathrm{BR})/(\sigma_{\gamma H}\times\mathrm{BR})\)
  for the investigated exclusive \(\gamma H\) channels and their combination at
  different FCC-ee centre-of-mass energies.
  }
  \label{table:results}
\end{table}

\section{Interpretation}
\label{sec:interpretation}
We now use the projected sensitivities of \eeha\ cross section measurements at the three FCC-ee center-of-mass energies to determine their potential to constrain anomalous Higgs boson couplings to photons and Z bosons, namely $\caa$ and $\cza$. The sensitivities derived should be put into the context of our expected knowledge of the Higgs couplings from the HL-LHC and from the FCC-ee programme. 

The most recent, combined ATLAS and CMS projections from the HL-LHC are obtained in the $\kappa$-framework and quote relative sensitivities on $\kaa$ and $\kza$ of 1.8\% and 6.8\%, respectively~\cite{ATLAS:2025eii}. We translate these sensitivities into bounds on the corresponding branching ratios as follows. Given the $\mathcal{O}(10^{-3})$ branching fractions into $\gamma\gamma/Z\gamma$, we can safely neglect the impact of corrections to the corresponding partial width on the total width, and identify the relative partial width corrections to relative modifications of their corresponding branching ratio. Taking the usual $\kappa^2$ scaling, we find the value of the uncertainty that would lead to a $\chi^2=1$ for a branching ratio measurement, assuming the SM value is observed, giving a relative precision of 14.1\% and 3.6\%, for $\mathrm{BR}_{Z\gamma}$ and $\mathrm{BR}_{\gamma\gamma}$, respectively. As expected, the error on the branching ratio is approximately twice the error on the $\kappa$ factor.

For the FCC-ee, projected sensitivities on the $ZH$ and VBF production cross sections times branching ratios were determined in a recent study~\cite{Selvaggi:2025kmd}. Neglecting any BSM effects in production, we combine the quoted sensitivities from the two production modes at $\sqrt{s}=\si{240}$ and $\SI{365}{\giga\electronvolt}$, obtaining relative precisions of 9.5\% and 3.4\% on $\mathrm{BR}_{Z\gamma}$ and $\mathrm{BR}_{\gamma\gamma}$, respectively. We summarize the results in~\cref{tab:BR_projections}.
\begin{table}[htb]
  \centering
  \renewcommand{\arraystretch}{1.15}
  \footnotesize
  \begin{tabularx}{\textwidth}{>{\raggedright\arraybackslash}X 
  >{\centering\arraybackslash}X 
  c
  >{\centering\arraybackslash}X 
  >{\centering\arraybackslash}X 
  >{\centering\arraybackslash}X 
  >{\centering\arraybackslash}X}
    \toprule
    \multirow{2}{*}{$\Delta\mathrm{BR}$(\%)} &
    \multirow{2}{*}{HL-LHC} &
    FCC-ee(240)&
    \multicolumn{2}{c}{FCC-ee(365)} &
    FCC-ee  &
    All
    \\
     & & ZH & ZH & WWH & combined & combined\\
    \midrule 
    $\gamma\gamma$& 3.6  & 3.6  & 13 & 15 & 3.4 & 2.5 \\
    $Z\gamma$     & 14.1 & 11.8 & 22 &  23 & 9.5 & 7.9 \\
    \bottomrule
  \end{tabularx}
  \caption{Summary of projected relative precisions in percent on the $\gamma\gamma$ and $Z\gamma$ branching fractions of the Higgs boson from the HL-LHC~\cite{ATLAS:2025eii} and two center-of-mass energies of the FCC-ee~\cite{Selvaggi:2025kmd}.}
  \label{tab:BR_projections}
\end{table}
We will use these projected sensitivities as a baseline, comparing our projected bounds from \eeha\ to those obtained from assuming a model independent measurement of the branching ratios as per the combined number in~\cref{tab:BR_projections}. We emphasize, however that this comparison is not strictly consistent, since the HL-LHC projections are obtained in the $\kappa$-framework and the FCC-ee projections assume SM production rates, while our interpretation is of EFT type, and differs from the $\kappa$-framework as defined in~\cref{eq:kappa_def}. Nevertheless it provides a useful point of reference for the \eeha\ process. 

Our statistical analysis uses as inputs the combined relative precision on the \eeha\ cross sections at each energy stage, assuming that three independent signal strength measurements are performed, observing the SM value with an uncertainty as per~\cref{table:results}. We construct a simple $\chi^2$ function
\begin{align}
    \chi^2_{H\gamma}(\caa,\cza) = \sum_{E} \frac{\left(1-\mu^{H\gamma}_{E}(\caa,\cza)\right)^2}{(\delta_{E}^\mu)^2},
\end{align}
where $E=160,240,\SI{365}{\giga\electronvolt}$ denotes the centre of mass energy of the FCC-ee run, $\mu^{H\gamma}_E$ the predicted \eeha\ signal strength parameterizing the relative impacts of $(\cza,\caa)$ given in~\cref{eq:eeha_160,eq:eeha_365} and $\delta_{E}^\mu$ is the relative precision for that energy run from~\cref{table:results}. The ``1'' in the numerator is the observed value that we assume to lie exactly at the SM point. 

\subsection{2D Bounds from \eeha}
\begin{figure}[t]
    \centering
    \includegraphics[width=0.49\textwidth]{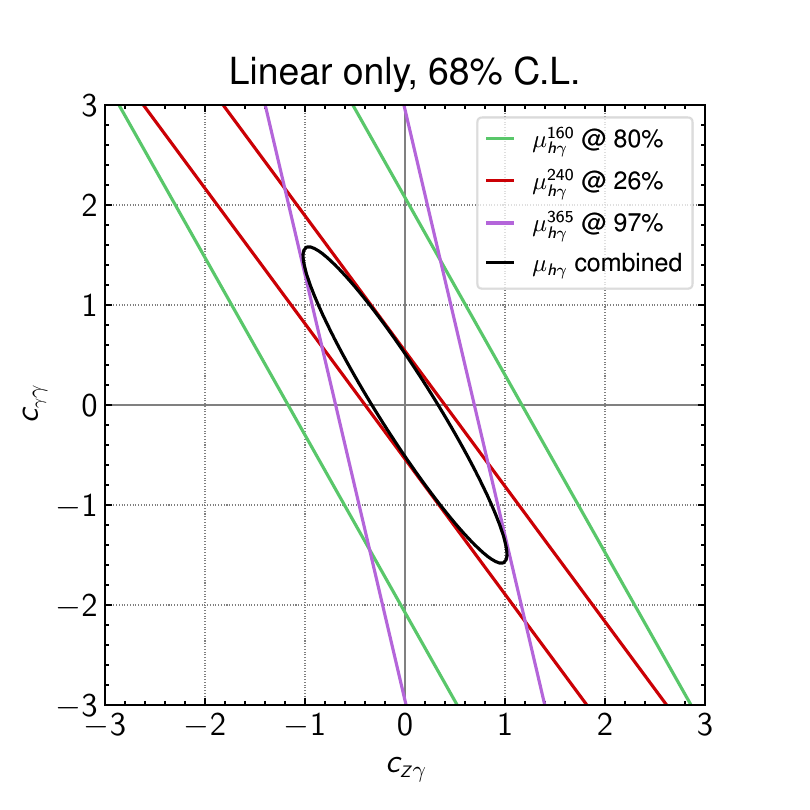}
    \includegraphics[width=0.49\textwidth]{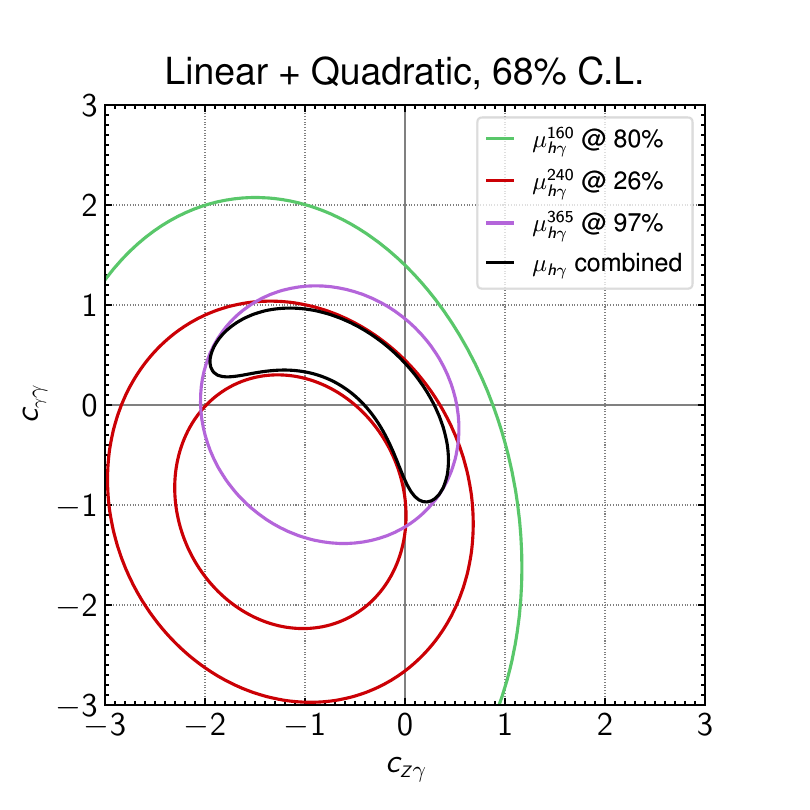}
    \caption{Projected 68\% C.L. allowed regions in the $(\cza,\caa)$ plane from \eeha\ cross section measurements at the three FCC-ee centre-of-mass energy runs, as per the sensitivities given in~\cref{table:results}. Bounds from the $\sqrt{s}=160,240$ and 365 GeV runs are shown in green, red and purple, respectively, with their combination shown in black. The left panel shown the bounds when including only the interference contribution while the right panel shows the bounds when both the interference and squared contributions are included in the relative impact over the SM.}
    \label{fig:hgamma_bounds_2D}
\end{figure}

~\cref{fig:hgamma_bounds_2D} shows the projected 68\% C.L. allowed regions in the $\cza,\caa$ plane, delimited by isocontours of $\chi^2=2.24$. The left and right panels show the cases where only the interference, and both interference and squared contributions are included, respectively. Individual sensitivities for the 160, 240 and \SI{365}{\giga\electronvolt} runs are show in green, red and purple, respectively and the black contour denotes the combination. 
At the linear-only level, each measurement constrains a specific direction in parameter space, and is blind to the orthogonal direction. Thanks to the energy dependence of the cross section the combination of two measurements at different energies leads to a closed fit, and the main sensitivity is driven by the combination of the 240 and 365 GeV runs. The 240 GeV run provides the best relative sensitivity to the \eeha\ cross section, while the enhanced sensitivity of the 365 GeV run, leads to a larger impact than the 160 GeV run, despite having a relative sensitivity to the cross section that is three times worse. Moreover, the direction it constrains is less aligned with that of the 240 GeV run, leading to a bigger impact when combined. When including quadratic effects, the shape of the likelihood is clearly distorted, indicating a significant impact of these corrections, but the overall message that the sensitivity is driven by the combination of the 240 and 365 GeV runs is unchanged. Overall the measurements yield an $\mathcal{O}(1)$ sensitivity to the $\cza,\caa$ parameters.

\subsection{Comparison to $\gamma\gamma$ and $Z\gamma$ Branching Ratio Measurements \label{sec:BRcomparison}}
\begin{figure}[t]
    \centering
    \includegraphics[width=0.49\textwidth]{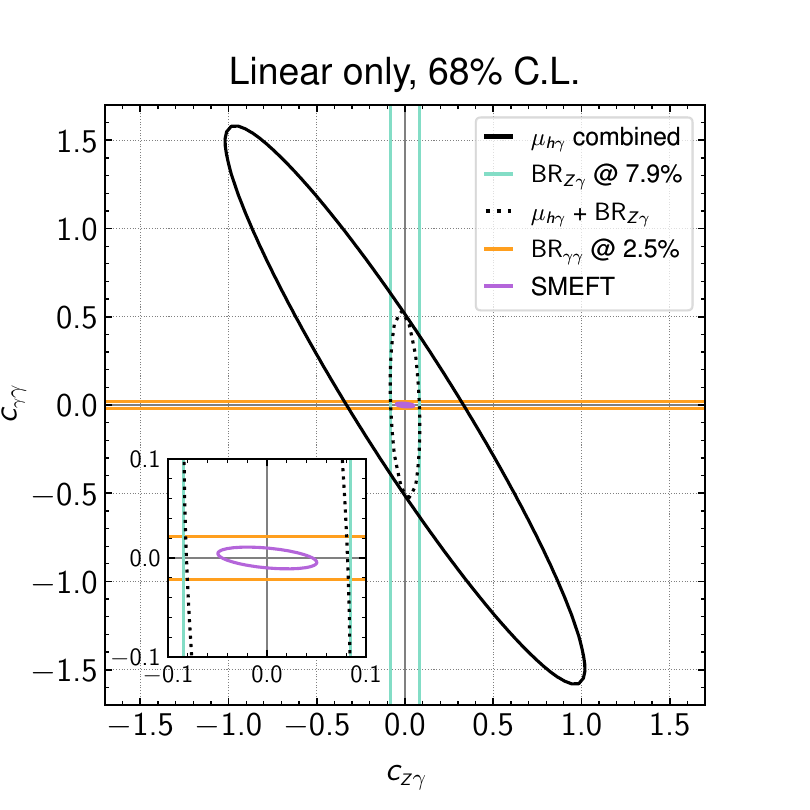}
    \includegraphics[width=0.49\textwidth]{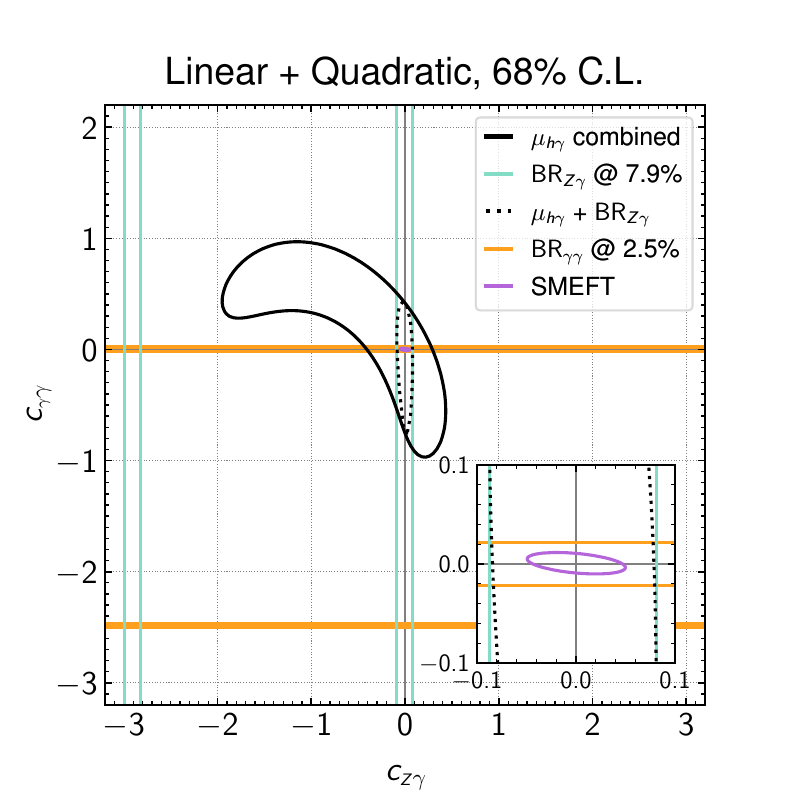}
    \caption{Comparison of the combined 68\% C.L. sensitivity in the $(\cza,\caa)$ plane via FCC-ee \eeha\ cross section measurements from~\cref{fig:hgamma_bounds_2D} (black) with the combined HL-LHC and FCC-ee projected sensitivities to the individual parameters from Higgs branching ratio measurements into $Z\gamma$ (teal) and $\gamma\gamma$ (orange). The black region shows the combination of \eeha\ and $\mathrm{BR}_{Z\gamma}$. The purple region represents our estimate of the projected bounds in the SMEFT interpretation discussed in~\cref{sec:EFT}. As in~\cref{fig:hgamma_bounds_2D}, the left and right panels show bounds including interference only and interference plus squared contributions, respectively.}
    \label{fig:combined_bounds_2D}
\end{figure}

~\cref{fig:combined_bounds_2D} shows the combined projected \eeha\ sensitivity in black alongside the respective bounds on the individual couplings, derived assuming a measurement of the corresponding $Z\gamma$ and $\gamma\gamma$ branching ratios, in teal and orange respectively, assuming combined HL-LHC and FCC-ee sensitivities from~\cref{tab:BR_projections}. Again the left and right panels given the sensitivity assuming linear contributions only and including both linear and quadratic contributions, respectively.
In the vicinity of the SM point at the origin, we see that the expected sensitivity from the \eeha\ process that we have derived has a weaker constraining power than the individual branching ratio measurements, particularly in the $\caa$ direction, owing to the tight constraint from $\mathrm{BR}_{\gamma\gamma}$. It is nevertheless complementary in simultaneously probing both couplings, in contrast with the branching ratio measurements. Moreover, in the right panel, including the quadratic contributions reveals the presence of secondary minima in the branching ratio likelihoods, where the quadratic terms cancel with the linear ones. Taking the branching ratio measurements alone leads to a four-fold degeneracy in regions that are compatible with the SM prediction. This degeneracy is fully lifted by the measurement of \eeha, thanks to the measurement being performed at several different center-of-mass energies. As we will discuss in the next section, this is driven primarily by the combination of measurements performed at $\sqrt{s}=$240 and \SI{365}{\giga \electronvolt}.

\subsection{1D Sensitivity to $\cza$}
\begin{figure}[t]
    \centering
    \includegraphics[width=\textwidth]{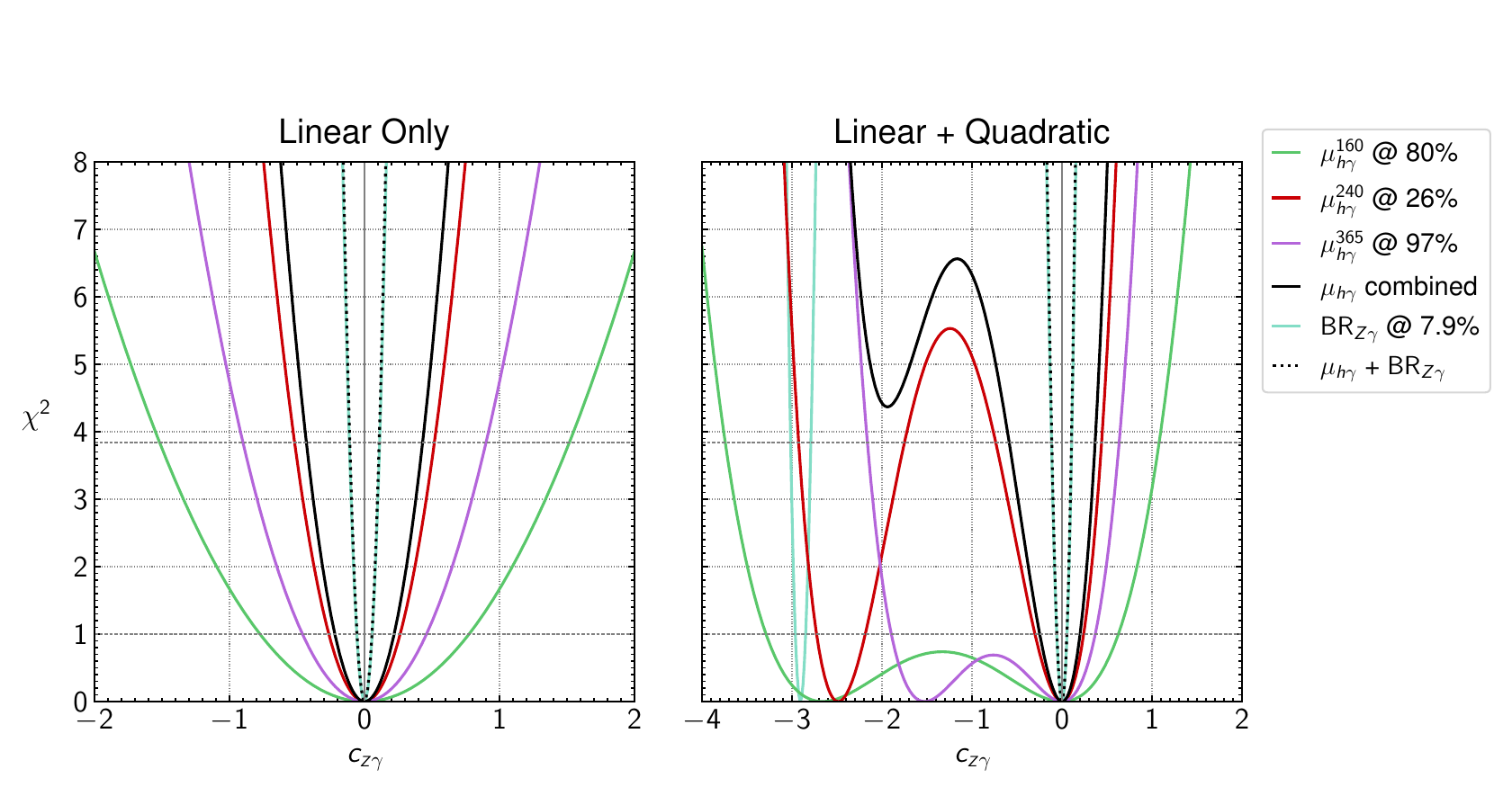}
    \caption{$\chi^2$ values as a function of $\cza$ for the projected sensitivities of \eeha\ measurements, as in~\cref{fig:combined_bounds_2D}. Also included is the corresponding $\chi^2$ of a projected 7.9\% measurement of $\mathrm{BR}_{Z\gamma}$ in teal and its combination with the three \eeha\ measurements in dotted black. }
    \label{fig:combined_bounds_1D}
\end{figure}

Given the tight constraints on $\caa$ from the diphoton branching ratio measurements, we focus now on the one-dimensional case, taking $\cza$ alone, with $\caa=0$. ~\Cref{fig:combined_bounds_1D}, shows the $\chi^2$ function contributions from the various projected \eeha\ signal strength measurements along with their combination in solid black. The projected sensitivity on the $H\to Z\gamma$ sensitivity of 7.9\% is plotted in teal, and the combination of the two in dotted black. 
Both panels show that, in the vicinity of the SM point ($\cza=0$), the sensitivity is dominated by $H\to Z\gamma$ measurement. As mentioned in the previous section, the dependence on the parameters feature a second minimum for negative $\cza$ where the interference and quadratic terms cancel to recover the SM prediction. Each measurement, being performed at a fixed energy scale, possesses such a double minimum structure, but their combination lifts the second minimum beyond the $\Delta\chi^2=3.84$ threshold, effectively excluding it. In particular, the combination of only the \eeha\ measurements is sufficient to exclude the second minimum, and this capacity is driven by the combination of the $\sqrt{s}=240$ and $\SI{365}{\giga\electronvolt}$ measurements. Combining a single measurement at \SI{240}{\giga\electronvolt} with the $H\to Z\gamma$ measurement would also be sufficient to achieve this. \Cref{tab:1D_bounds} summarizes the 68\% C.L. bounds on $\cza$, comparing the combined \eeha\ measurements with the $H\to Z\gamma$ precision, and their combination.
\begin{table}[htb]
  \centering
  \renewcommand{\arraystretch}{1.15}
  \footnotesize
  \resizebox{\textwidth}{!}{
  \begin{tabular}{lccc}
    \toprule
    $\mathbf{\cza}$ \textbf{bound, 68\% C.L.}&$\mu_{H\gamma}$ & $\mathrm{BR}_{Z\gamma}$ & combination \\
    \midrule
    
    Interference only &
    $[-0.22, 0.22]$ &
    $[-0.057,0.057]$& 
    $[-0.055,0.055]$\\
    \midrule
    Interference + square &
    $[-0.25, 0.20]$&
    $[-3.0, -2.8]\cup [-0.059,0.056]$& 
    $[-0.057,0.054]$\\
    \bottomrule
  \end{tabular}
  }
  \caption{Projected 68\% C.L. bounds on $\cza$ from combined \eeha\ signal strength measurements at the three energy runs of FCC-ee, the 7.9 \% measurement of the $H\to Z\gamma$ branching fraction, and their combination. The first row shows the results when considering only the interference contribution of the effective operator, while the second row shows the bounds obtained when including both the interference an squared contributions.}
  \label{tab:1D_bounds}
\end{table}
The \eeha\ measurements can be seen to lead to a $\sim4$\% improvement in the $\cza$ bound in the interference case as well as the bound around the SM point when considering interference and squared contributions. For illustrative purposes, we also translate the $\cza$ bounds on the effective $\kza$ (\cref{tab:1D_bounds_kappa}) that scales the $H\to Z \gamma$ branching fraction according to Eq.~\eqref{eq:kappa_def} using Eq.~\eqref{eq:kappa_relation}, with the relative precision on this quantity given in~\cref{tab:1D_bounds}.
\begin{table}[htb]
  \centering
  \renewcommand{\arraystretch}{1.15}
  \footnotesize
  \resizebox{\textwidth}{!}{
  \begin{tabular}{lccc}
    \toprule
    $\mathbf{\kza}$ \textbf{bound, 68\% C.L.} &$\mu_{H\gamma}$ & $\mathrm{BR}_{Z\gamma}$ & combination \\
    \midrule
Interference only &
    $[0.85, 1.15]$ &
    $[0.960,1.040]$& 
    $[0.962,1.038]$\\
    \midrule
    Interference + square &
    $[0.83, 1.14]$&
    $[-1.045,-0.965]\cup [0.960,1.039]$& 
    $[0.961,1.038]$\\
    \bottomrule
  \end{tabular}
  }
  \caption{Bounds on the effective $\kza$ translated from~\cref{tab:1D_bounds}.}
  \label{tab:1D_bounds_kappa}
\end{table}

Overall, we find that \eeha\ has a lower sensitivity to $\cza$ compared to the direct measurement of the $Z\gamma$ branching fraction when considering the combined capacity of the HL-LHC and FCC-ee programs. This is perhaps not so surprising despite the enhanced relative sensitivity of this process, given its rarity compared to the usual Higgs production processes at FCC-ee. The much larger statistics offered by the latter make up for the slightly weaker sensitivity that they offer on $\cza$. Nevertheless, the measurement of \eeha\ is able to lift a second, degenerate minimum present when constraining $\cza$ with only a single branching ratio measurement. This process can be interpreted as offering a stand-alone sensitivity to $\kza$ of 15\%, which is a reasonable achievement given its rarity, and could serve as a valuable cross-check when probing this coupling, as the measurement is independent of both the $HZZ$ and $HWW$ couplings that enter in $ZH$ and VBF production modes, respectively.

\subsection{EFT Interpretation\label{sec:EFT}}
So far, we have been comparing our derived sensitivity from the \eeha\ process with bounds derived from the projected measurements of Higgs branching fractions into $\gamma\gamma$ and $Z\gamma$ that were derived under certain assumptions. As previously discussed, our interpretation in terms of $\cza$ and $\caa$ is inherently in the EFT framework, 
since it employs higher dimension operators to model the interactions, and leverages the energy growth of the underlying amplitudes. The HL-LHC and FCC-ee projections were rather derived assuming a $\kappa$-formalism that deals in coupling modifiers, and with $M_H$ setting the energy scale at which the $\gamma\gamma$ and $Z\gamma$ couplings are probed. These analyses also allow for the possibility of an invisible decay channel for the Higgs boson that can be constrained by direct searches. The total width, however, cannot be measured directly in a model-independent way. In the $\kappa$-framework, the identical dependence on $\kappa_Z$ of the $ZH$ cross section and the branching ratio to $ZZ^\ast$ means that the total width can be resolved indirectly but this is not true in a general BSM scenario. In the SMEFT, for example, additional inputs from the VBF cross section and $WW^\ast$ branching ratio are required to measure the width indirectly~\cite{Yan:2021tmw}.

A more fitting comparison of our sensitivity would be against the results of global EFT analyses determining the projected sensitivities on the set of relevant operators that can be probed with the HL-LHC and FCC-ee datasets. In the dimension-6  Warsaw basis of the SMEFT~\cite{Buchmuller:1985jz,Grzadkowski:2010es}, for example, there are three independent operators that contribute to $\cza$ and $\caa$
\begin{align}
    \mathcal{L}_\mathrm{SMEFT}^{(6)}\supset
    \frac{C_{WW}}{\Lambda^2}\, \phi^\dagger\phi\, W_I^{\mu\nu}W^I_{\mu\nu} +
    \frac{C_{BB}}{\Lambda^2}\, \phi^\dagger\phi\, B^{\mu\nu}B_{\mu\nu} + 
    \frac{C_{\phi WB}}{\Lambda^2}\, \phi^\dagger\tau_I\phi\, W^I_{\mu\nu}B^{\mu\nu},
\end{align}
where $W^I_{\mu\nu}$ and $B_{\mu\nu}$ are the field strength tensors for the $SU(2)_L$ and $U(1)_Y$ gauge bosons, respectively and $\Lambda$ is an arbitrary cutoff parameter, typically associated with the scale of new physics.
After EW symmetry breaking, the operators contribute to single and double Higgs couplings to all EW gauge bosons, including the $ZZ$ and $WW$ analogues of $\cza$ and $\caa$. The mapping between the Wilson coefficients and the anomalous couplings is
\begin{align}
\gaa\caa & = \frac{v^2}{\Lambda^2}\left(
\sin^2\theta_W C_{WW} + 
\cos^2\theta_W  C_{BB} -
\frac{1}{2}\sin2\theta_WC_{HWB}
\right)\,,\\
\gza\cza & = \frac{v^2}{\Lambda^2}\left(
\sin2\theta_W(C_{WW} + C_{BB}) -
\cos2\theta_WC_{HWB}
\right)\,,\\
\gzz\czz & = \frac{v^2}{\Lambda^2}\left(
\cos^2\theta_W C_{WW} + 
\sin^2\theta_W  C_{BB} +
\frac{1}{2}\sin2\theta_WC_{HWB}
\right)\,,\\
\gww\cww & = 2\frac{v^2}{\Lambda^2} C_{WW}\,,
\end{align}
where $\gaa$ and $\gza$ are defined in~\cref{eq:g_norm}, and $\gzz$ and $\gww$ are  normalisation parameters that we do not use in our analysis. The three operators contribute in a correlated way to several couplings that can be constrained by the suite of HL-LHC and FCC-ee measurements, including processes such as $e^+e^-\to ZH$, which can be measured extremely precisely compared to \eeha. $C_{HWB}$, for example, can additionally be constrained by EW precision measurements. In the more general framework of the HEFT, which is defined in the broken EW phase, the correlations between the gauge-Higgs couplings are not present, and the anomalous coupling parameters can be considered as independent. Note that in our interpretation we only consider the leading, tree-level, contributions to \eeha\ from the associated operators. Modifications of the Higgs boson to the top quark and the $W$ boson (as modeled by $\cww$), among others, can also affect the 1-loop amplitudes but these effects are loop-suppressed compared to those that we consider here and we therefore neglect them in what follows.

EFT constraints are best derived in global fits including many observables and taking into account as many relevant operators as possible. One such analysis was published recently by the SMEFiT collaboration~\cite{terHoeve:2025gey}, in which they perform a global fit over 51 operators, focusing specifically on HL-LHC and FCC-ee projections, and including renormalisation group running effects, assuming a new physics scale of 5 TeV. The analysis takes into account the combination of LEP data and a suite of HL-LHC projections for top, Higgs and Diboson measurements, as well as FCC-ee projections for runs at the $Z$-pole, 161, 240, 350, and \SI{365}{\giga\electronvolt}. Relevant FCC-ee processes include light fermion pair-production above the $Z$-pole, Higgs production and decay via $ZH$ and VBF, Diboson production and top-pair production~\cite{Celada:2024mcf}. The $H\gamma$ process is not included in the analysis. 

We translate the resulting marginalised posterior distributions in the space of $C_{WW}, C_{BB}$ and $C_{HWB}$ into the $\cza,\caa$ plane\footnote{We would like to thank Luca Mantani for providing the SMEFiT samples from this study.} and construct a likelihood in this space using a multivariate normal distribution, taking into account the full covariance matrix in the Wilson coefficients from the fit. We use this likelihood to draw 68\% C.L. allowed regions representing the projected global constraining power of the combined HL-LHC and FCC-ee datasets in the SMEFT framework, projected to the space of $\gamma\gamma$ and $Z\gamma$ couplings of the Higgs.

The resulting contours are shown in~\cref{fig:combined_bounds_2D}. From the zoomed insets, we can compare the global SMEFT interpretation to the  individual branching ratio measurements. The SMEFT analysis, on one hand, allows many Wilson coefficients to float simultaneously, but on the other, uses a large set of measurements to constrain all of these possible directions. One can see that the combined effect of the larger dataset leads to tighter constraints in this space than from the individual branching ratio measurements, indicating that the power of Higgs production and decay through $ZH$ and VBF is able to tightly constrain this parameter space by exploiting the correlations between the various Higgs couplings to gauge bosons, and that other couplings that may affect these processes are sufficiently constrained elsewhere to not dilute this power too much. Moreover, the four-fold degeneracy that arises when considering the $\gamma\gamma$ and $Z\gamma$ branching ratio measurements in isolation, discussed in~\cref{sec:BRcomparison}, is not present in the quadratic fit, indicating that the combination of multiple measurements besides $H\gamma$ would also be sufficient to lift it in the SMEFT.
Performing the same exercise but projecting to the one-dimensional $\cza$ case, the bounds outperform the $\mathrm{BR}_{Z\gamma}$ projection in~\cref{tab:1D_bounds}, giving $\cza\supset[-0.0328,0.0334]$ and  $\cza\supset[-0.0325,0.0332]$ in the interference only and interference-plus-square fit cases, respectively. This underlines that there is more information entering the EFT fit to indirectly constrain this direction than just the combined HL-LHC and FCC-ee measurements of this branching ratio.

\section{Conclusion}
\label{sec:conc}

We have presented the first detailed study of the loop-induced \eeha process at the FCC-ee, exploring its sensitivity to the effective $H\gamma\gamma$ and $HZ\gamma$ couplings. The analysis considers the $H\rightarrow b\bar{b}$ and the semi-leptonic $H\rightarrow WW^*$ decays across the planned FCC-ee center-of-mass energies, including a thorough analysis of background processes which ultimately drive the sensitivity to the process of interest. We find that the semi-leptonic $WW^*$ channel is the most sensitive as it suffers to a lesser extent from the radiative-return induced $Z$-boson background, which limits the sensitivity on the $b\bar{b}$ channel. Among the proposed energy stages, we find the $\sqrt{s}=$240 GeV run to have the most promising sensitivity. Over all the projected, combined sensitivities to the $H\gamma$ signal strengths are found to be $80\%$, $26\%$ and $97\%$ at the 160, 240 and \SI{365}{\giga \electronvolt} center-of-mass energies, respectively.

The investigated \eeha production mode is sensitive to both, the magnitude and the sign of the effective couplings. We use our projected sensitivities to determine the ability to constrain these couplings at the FCC-ee. In combination with measurements of the $H\rightarrow Z\gamma$ and $H\rightarrow \gamma\gamma$ branching ratios, the inclusion of \eeha lifts degeneracies that arise from the branching ratio measurements alone. The projections can be interpreted as offering a stand-alone sensitivity to $\kza$ of $\sim15\%$.

Our analysis shows that the projected sensitivities of the \eeha\ process at the FCC-ee in constraining the effective $HZ\gamma$ coupling are about a factor of 4 weaker than the bounds coming from the combined measurements of the $H\to Z\gamma$ branching fractions from HL-LHC and FCC-ee by alternative means and a factor 6 weaker than those coming from a global SMEFT fit. In a more general EFT framework such as the HEFT, where the correlations between Higgs bosons gauge couplings are not present, the former comparison is more appropriate. Ideally, a comparison could also be made with projections from a global analysis in the HEFT framework, which is beyond the scope of this work.

Despite the weaker sensitivity, which can be expected from the comparative rarity of the \eeha\ production mode in the SM, our study presents theoretical inputs and realistic experimental sensitivities for this complementary process that can be used as inputs to future global projection studies in the EFT framework. Moreover, it could serve as a useful cross-check when constraining the $\gamma\gamma$ and $Z\gamma$ effective couplings from global SMEFT analyses that make use of the $ZH$ and VBF processes, since the \eeha\ production mode is independent, at leading order, of modifications of the $WW$ and $ZZ$ couplings of the Higgs boson.

\acknowledgments
We would like to thank Fabio Maltoni and Davide Pagani for initial discussions and significant input that led to this project, Marco Zaro for helpful discussions on the event generation setup, Jorge de Blas for discussions on the HL-LHC and FCC-ee Higgs projections, Luca Mantani and Eugenia Celada for discussions on the SMEFiT results, and Jan Eysermans, Xunwu Zuo, and Andy Mehta for useful discussions in the context of the FCC Higgs physics working group. K.M. is supported by an Ernest Rutherford Fellowship from the STFC, Grant No.~ST/X004155/1 and by the STFC Consolidated Grant  No.\ ST/X000583/1.

\appendix

\section{Key kinematic distributions, alternative energies}
\label{app:recoil_obs}

\begin{figure}[H]
  \centering
  \begin{minipage}[t]{0.49\textwidth}
    \centering
    \includegraphics[width=\linewidth]{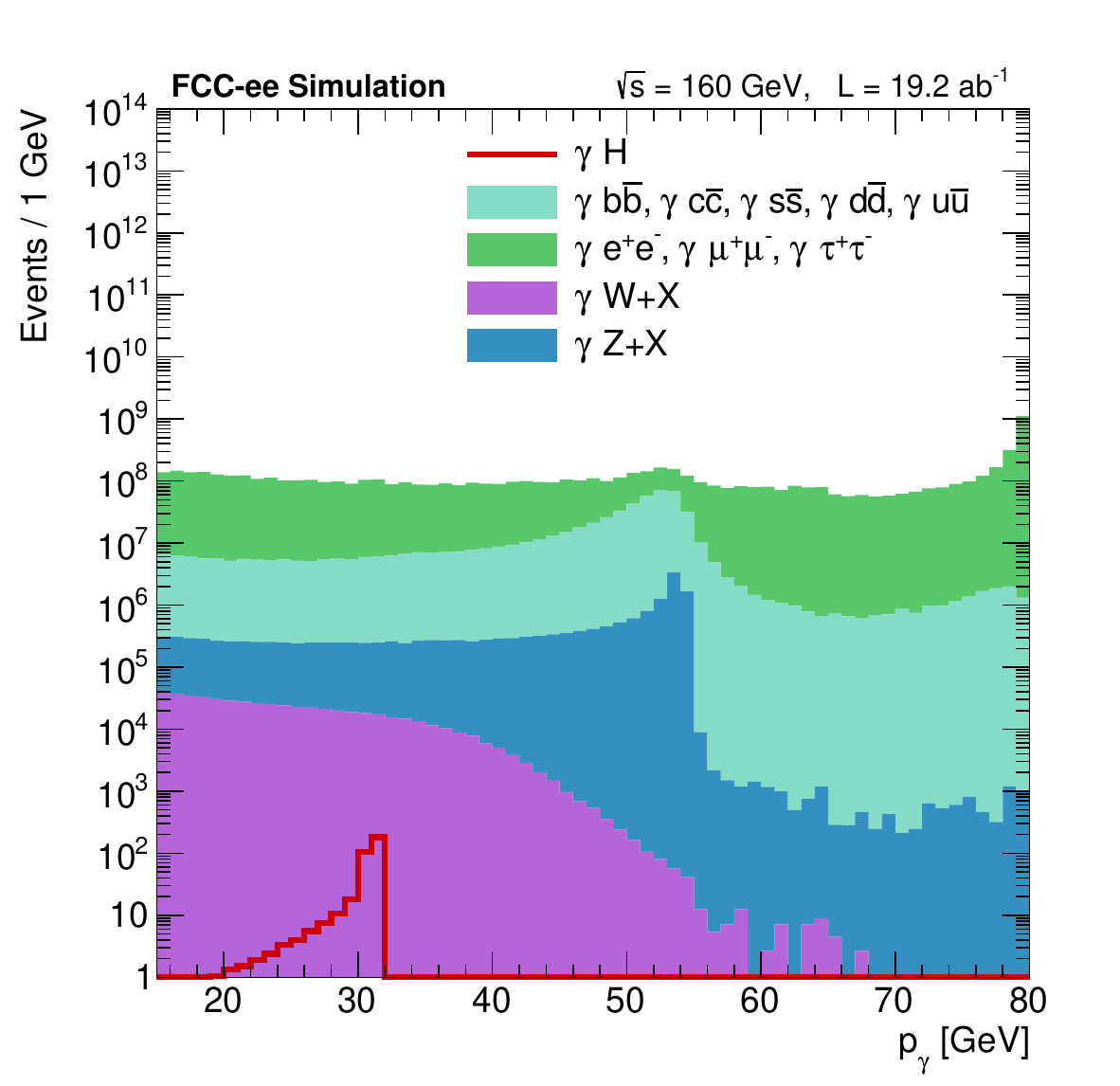}
  \end{minipage}
  \begin{minipage}[t]{0.49\textwidth}
    \centering
    \includegraphics[width=\linewidth]{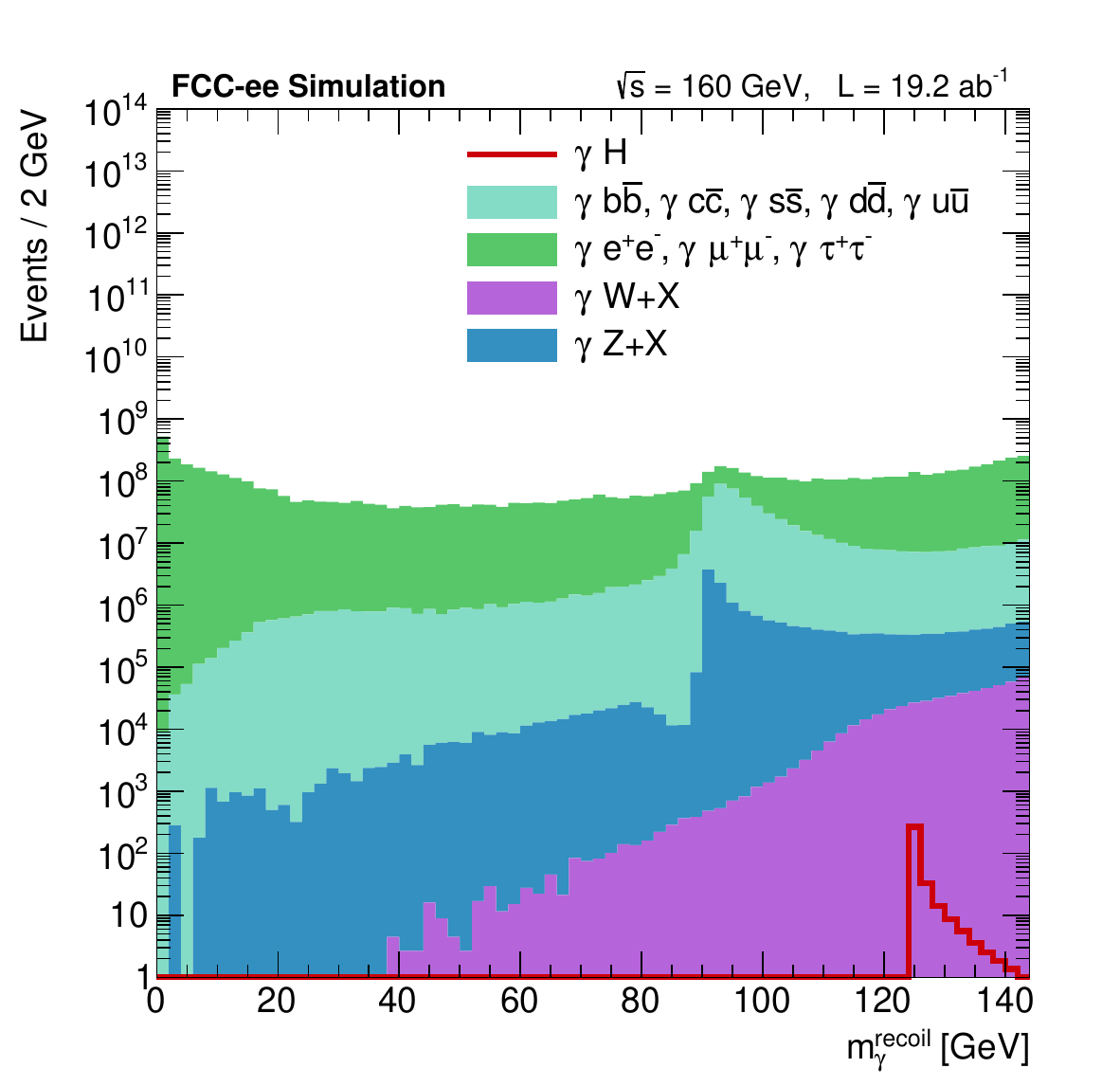}
  \end{minipage}\hfill

\caption{Photon momentum $p_\gamma$ (left) and recoil mass $m_\mathrm{recoil}$ (right), as defined in\cref{eq:p_mono,eq:m_recoil}, after requiring at least one isolated photon with $p_\gamma > \SI{15}{\giga\electronvolt}$ at 
$\sqrt{s}=\SI{160}{\giga\electronvolt}$. The signal (solid line) is shown overlaid on the stacked background contributions.}
  \label{fig:mrec_pa_nosel_160}
\end{figure}

\begin{figure}[H]
  \centering
  \begin{minipage}[t]{0.49\textwidth}
    \centering
    \includegraphics[width=\linewidth]{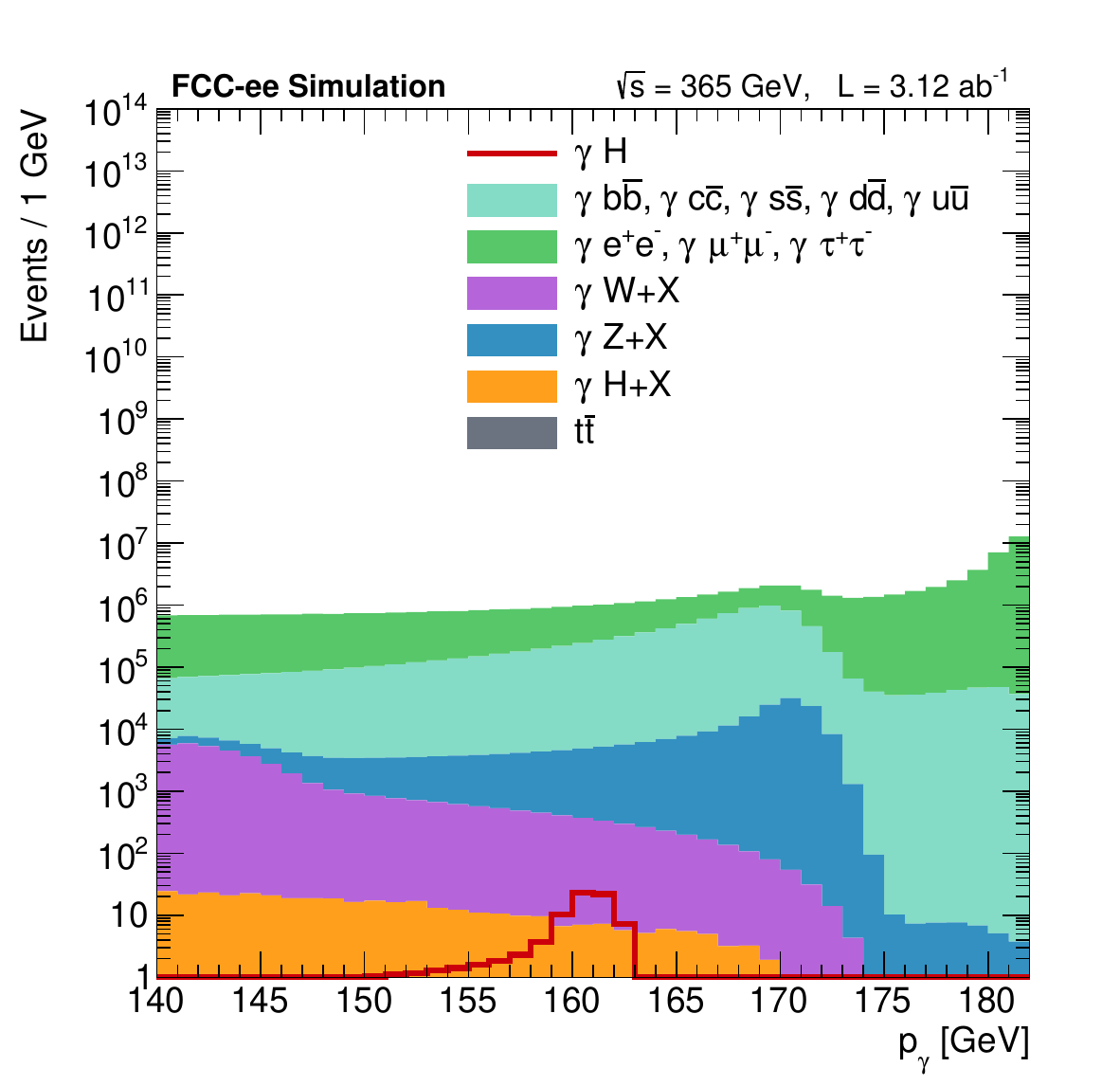}
  \end{minipage}
  \begin{minipage}[t]{0.49\textwidth}
    \centering
    \includegraphics[width=\linewidth]{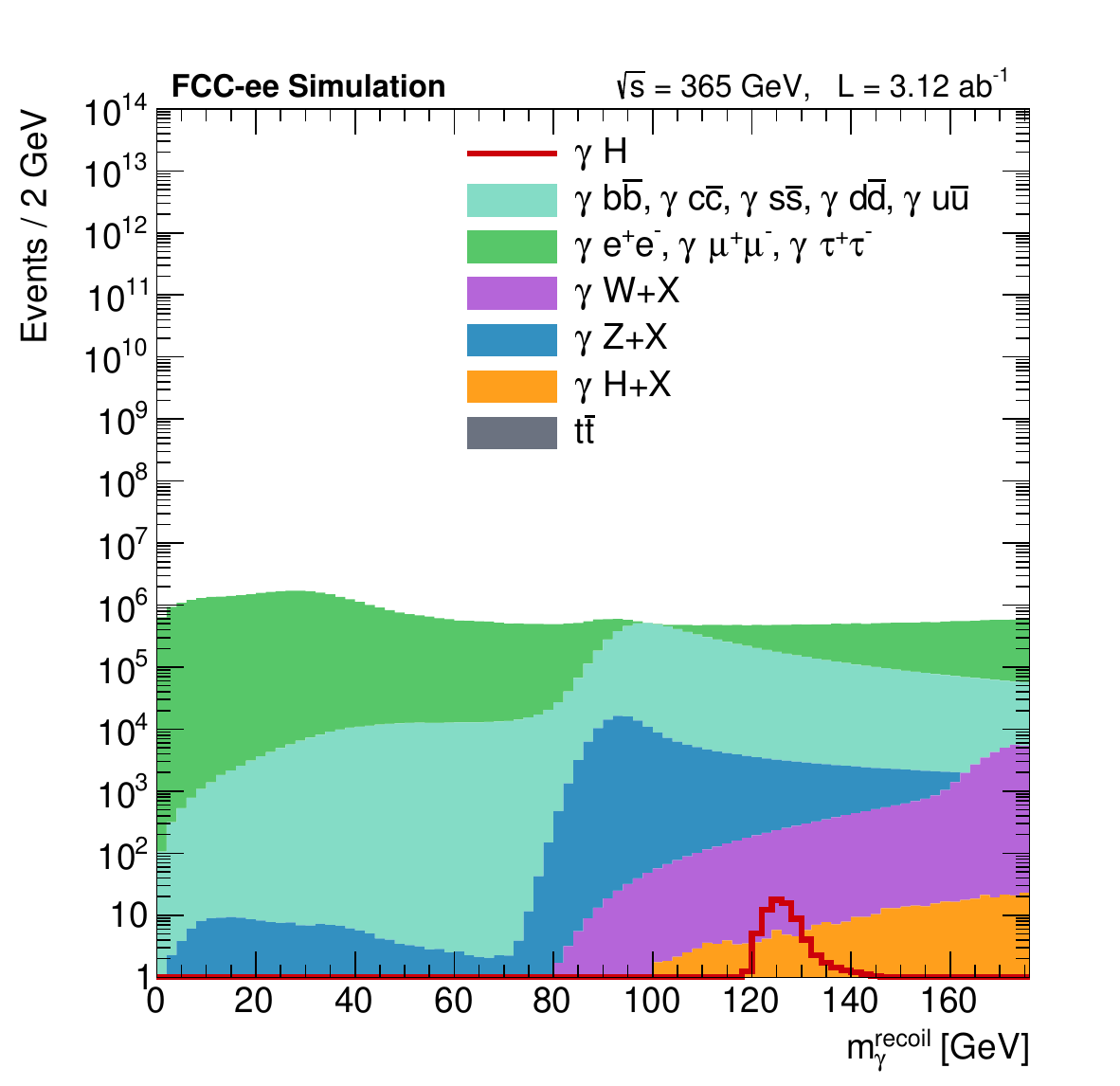}
  \end{minipage}\hfill

\caption{Photon momentum $p_\gamma$ (left) and recoil mass $m_\mathrm{recoil}$ (right), as defined in \cref{eq:p_mono,eq:m_recoil}, after requiring at least one isolated photon with $p_\gamma > \SI{100}{\giga\electronvolt}$ at  $\sqrt{s}=\SI{365}{\giga\electronvolt}$. The signal (solid line) is shown overlaid on the stacked background contributions.}
\label{fig:mrec_pa_nosel_365}
\end{figure}

Photon momentum and recoil mass are key observables to differentiate signal from background processes. The corresponding distributions for $\sqrt{s}=\SI{160}{\giga\electronvolt}$ and $\sqrt{s}=\SI{365}{\giga\electronvolt}$ are shown in \cref{fig:mrec_pa_nosel_160} and 
\cref{fig:mrec_pa_nosel_365} respectively.

\section{Common Selection, alternative energies}
\label{app:preselection}
A base event selection, common to all analysis channels is discussed in \cref{sec:ana_strat}. The resulting event yields and relative efficiencies for the $\sqrt{s}=\SI{160}{\giga\electronvolt}$ and $\sqrt{s}=\SI{365}{\giga\electronvolt}$ scenario are shown in \cref{tab:common_preselection_160} and \cref{tab:common_preselection_365} respectively.

\begin{table}[htb]
  \centering
  \renewcommand{\arraystretch}{1.15}
  \footnotesize
  \resizebox{\textwidth}{!}{
  \begin{tabular}{lcccc|c}
    \toprule
                   selection &              $\gamma\,q\bar q$ &          $\gamma\,\ell^+\ell^-$ &                  $\gamma\,W+X$ &                  $\gamma\,Z+X$ &                     $\gamma H$ \\
\midrule
           $\rm all\,events$ & $7.62 \times 10^{8}$ (100.0)\% & $1.78 \times 10^{10}$ (100.0)\% & $7.13 \times 10^{5}$ (100.0)\% & $2.13 \times 10^{7}$ (100.0)\% & $3.61 \times 10^{2}$ (100.0)\% \\
$\rm p_{\gamma}\,>\,15\,GeV$ &  $6.14 \times 10^{8}$ (80.6)\% &   $7.82 \times 10^{9}$ (44.0)\% &  $5.53 \times 10^{5}$ (77.5)\% &  $1.75 \times 10^{7}$ (82.0)\% &  $3.47 \times 10^{2}$ (96.3)\% \\
         $\rm N_{trk}\,>\,5$ &  $6.12 \times 10^{8}$ (80.3)\% &    $8.65 \times 10^{5}$ (0.0)\% &  $4.68 \times 10^{5}$ (65.6)\% &  $1.36 \times 10^{7}$ (63.8)\% &  $3.16 \times 10^{2}$ (87.6)\% \\

    \bottomrule
  \end{tabular}
  }
  \caption{Event yields and relative efficiencies after the common preselection at  $\sqrt{s}=~\SI{160}{\giga\electronvolt}$
  . The preselection requires an isolated photon with
  \(p_{\gamma}>15\)~GeV and at least six reconstructed tracks, excluding the photon.}
  \label{tab:common_preselection_160}
\end{table}

\begin{table}[htb]
  \centering
  \renewcommand{\arraystretch}{1.15}
  \footnotesize
  \resizebox{\textwidth}{!}{
  \begin{tabular}{lcccccc|c}
    \toprule
                    selection &              $\gamma\,q\bar q$ &         $\gamma\,\ell^+\ell^-$ &                  $\gamma\,W+X$ &                  $\gamma\,Z+X$ &                  $\gamma\,H+X$ &                     $\gamma H$ \\
\midrule
            $\rm all\,events$ & $1.54 \times 10^{7}$ (100.0)\% & $4.11 \times 10^{8}$ (100.0)\% & $6.32 \times 10^{4}$ (100.0)\% & $2.77 \times 10^{5}$ (100.0)\% & $4.81 \times 10^{5}$ (100.0)\% & $1.14 \times 10^{2}$ (100.0)\% \\
$\rm p_{\gamma}\,>\,140\,GeV$ &  $9.25 \times 10^{6}$ (60.3)\% &  $8.37 \times 10^{7}$ (20.3)\% &  $4.10 \times 10^{4}$ (64.8)\% &  $2.15 \times 10^{5}$ (77.5)\% &   $5.29 \times 10^{2}$ (0.1)\% &  $8.68 \times 10^{1}$ (76.5)\% \\
          $\rm N_{trk}\,>\,5$ &  $9.23 \times 10^{6}$ (60.1)\% &   $1.30 \times 10^{4}$ (0.0)\% &  $3.34 \times 10^{4}$ (52.8)\% &  $1.62 \times 10^{5}$ (58.4)\% &   $3.75 \times 10^{2}$ (0.1)\% &  $7.91 \times 10^{1}$ (69.6)\% \\

    \bottomrule
  \end{tabular}
  }
  \caption{Event yields and relative efficiencies after the common preselection at
   $\sqrt{s}=~\SI{365}{\giga\electronvolt}$. The preselection requires an isolated photon with
  \(p_{\gamma}>140\)~GeV and at least six reconstructed tracks, excluding the photon.}
  \label{tab:common_preselection_365}
\end{table}

\section{BDT Input Observables}
\label{app:BDTinput}

All three analysis channels use a BDT to enhance the signal relative to the background. The $H\rightarrow b\bar{b}$ and $H\rightarrow WW^*$ channels employ similar but not identical sets of input observables, while the two semi-leptonic $WW^*$ sub-channels use the same inputs. The full lists are given in~\cref{tab:bdt_inputs_hbb,tab:bdt_inputs_hww}.

\begin{table}[htb]
  \centering
  \renewcommand{\arraystretch}{1.15}
  \footnotesize
  \begin{tabular}{ll}
    \toprule
    Variable & Description \\
    \midrule
    $p_\gamma$, $\cos\theta_\gamma$, $\phi_\gamma$ & Photon momentum and angles \\
    $m_\mathrm{recoil}$, $p_\mathrm{recoil}$, $\cos\theta_\mathrm{recoil}$, $\phi_\mathrm{recoil}$ & Four-momentum of the photon recoil system \\
    $p_\ell$, $p_{\mathrm{T},\ell}$, $\cos\theta_\ell$, $\phi_\ell$ & Lepton momentum and angles \\
    lepton flavor ($e$ or $\mu$) & Exploits different $t$-channel background contributions per flavor \\
    $p_{j_0}$, $\cos\theta_{j_0}$, $\phi_{j_0}$ & Leading jet momentum and angles \\
    $p_{j_1}$, $\cos\theta_{j_1}$, $\phi_{j_1}$ & Subleading jet momentum and angles \\
    $p^\mathrm{miss}$, $p_{\mathrm{T}}^\mathrm{miss}$, $\cos\theta^\mathrm{miss}$, $\phi^\mathrm{miss}$ & Missing momentum four-vector \\
    $m_{W_{qq}}$, $p_{W_{qq}}$, $\cos\theta_{W_{qq}}$, $\phi_{W_{qq}}$ & Hadronic $W$ candidate four-momentum \\
    $m_{W_{\ell\nu}}$, $p_{W_{\ell\nu}}$, $\cos\theta_{W_{\ell\nu}}$, $\phi_{W_{\ell\nu}}$ & Leptonic $W$ candidate four-momentum \\
    $\cos\theta^*_{W_{qq}}$, $\cos\theta^*_{W_{\ell\nu}}$ & Polar angles of $W$ decay products in the Higgs rest frame \\
    $N_\mathrm{trk}$ & Number of reconstructed particles excluding the photon \\
    $y_{23}$, $y_{34}$ & Durham jet clustering scales \\
    jet$_{0}$, jet$_{1}$: $P_b$, $P_c$, $P_s$, $P_u$, $P_d$, $P_g$, $P_\tau$ & Flavour tag probabilities for both jets (7 hypotheses each) \\
    \bottomrule
  \end{tabular}
  \caption{BDT input variables for the $H\rightarrow WW^*$ analysis, common 
  to both the $W(\ell\nu)\,W^*(qq)$ and $W(qq)\,W^*(\ell\nu)$ sub-channels.}
  \label{tab:bdt_inputs_hww}
\end{table}

\begin{table}[htb]
  \centering
  \renewcommand{\arraystretch}{1.15}
  \footnotesize
  \begin{tabular}{ll}
    \toprule
    Variable & Description \\
    \midrule
    $p_\gamma$, $\cos\theta_\gamma$, $\phi_\gamma$ & Photon momentum and angles \\
    $m_\mathrm{recoil}$, $p_\mathrm{recoil}$, $\cos\theta_\mathrm{recoil}$, $\phi_\mathrm{recoil}$ & Four-momentum of the photon recoil system \\
    $m_{jj}$ & Dijet invariant mass \\
    $\cos\alpha_{jj}$ & Three-dimensional opening angle between the two jets \\
    $p_{j_0}$, $\cos\theta_{j_0}$, $\phi_{j_0}$ & Leading jet momentum and angles \\
    $p_{j_1}$, $\cos\theta_{j_1}$, $\phi_{j_1}$ & Subleading jet momentum and angles \\
    $E_{j_1}/E_{j_0}$ & Jet energy ratio \\
    $p^\mathrm{miss}$, $p_{\mathrm{T}}^\mathrm{miss}$, $\cos\theta^\mathrm{miss}$, $\phi^\mathrm{miss}$ & Missing momentum four-vector \\
    $\chi^2_H$ & Signal compatibility variable combining $m_{jj}$ and $p_\gamma$ \\
    $N_\mathrm{trk}$ & Number of reconstructed particles excluding the photon \\
    $y_{23}$, $y_{34}$ & Durham jet clustering scales \\
    \bottomrule
  \end{tabular}
  \caption{BDT input variables for the $H\rightarrow b\bar{b}$ analysis.}
  \label{tab:bdt_inputs_hbb}
\end{table}

\section{$H\rightarrow WW^*$ channel, alternative energies}
\label{app:hwwpreselection}
\begin{figure}[htb]
  \centering
  \begin{minipage}[t]{0.49\textwidth}
    \centering
    \includegraphics[width=\linewidth]{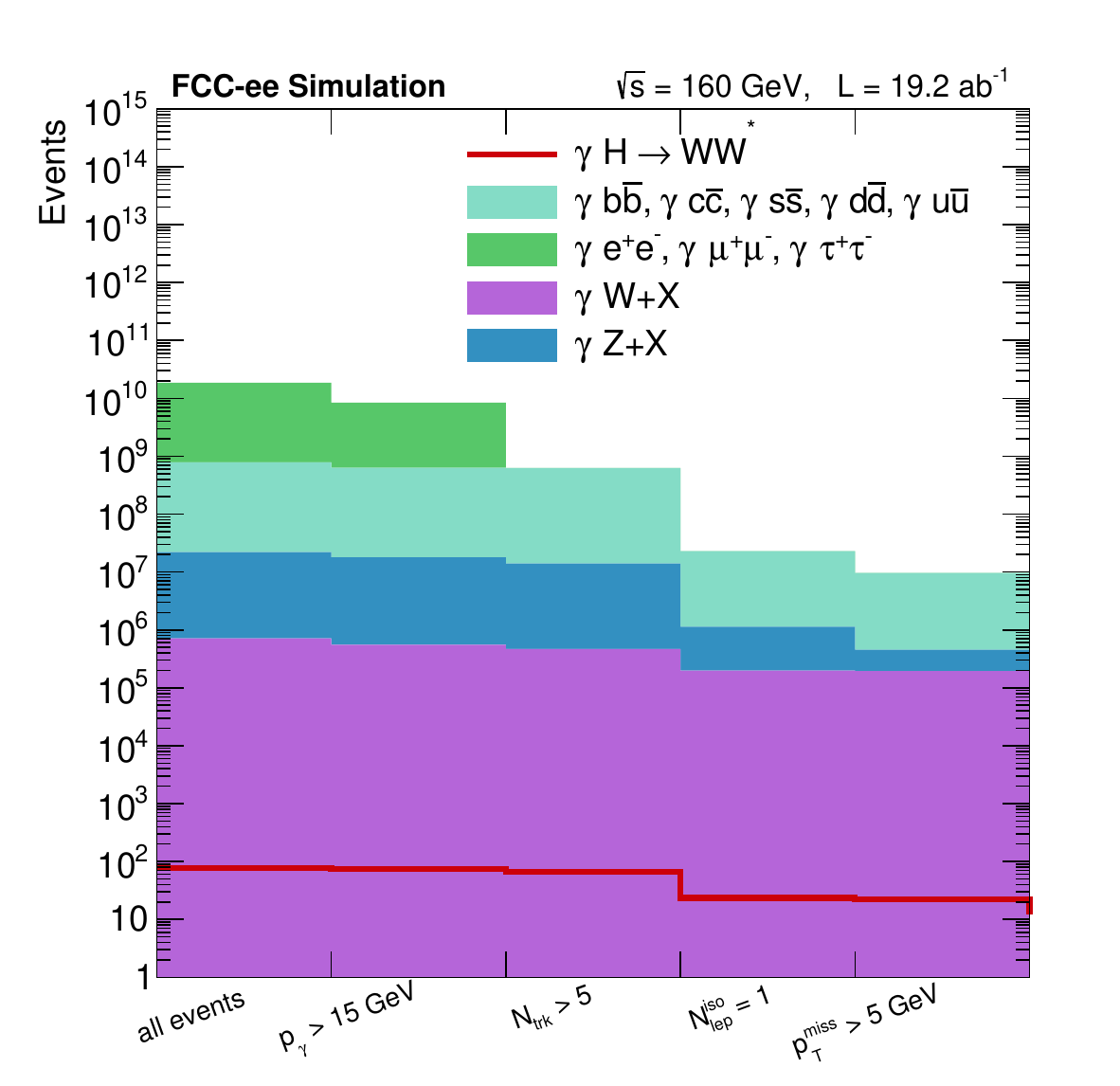}
  \end{minipage}\hfill
  \begin{minipage}[t]{0.49\textwidth}
    \centering
    \includegraphics[width=\linewidth]{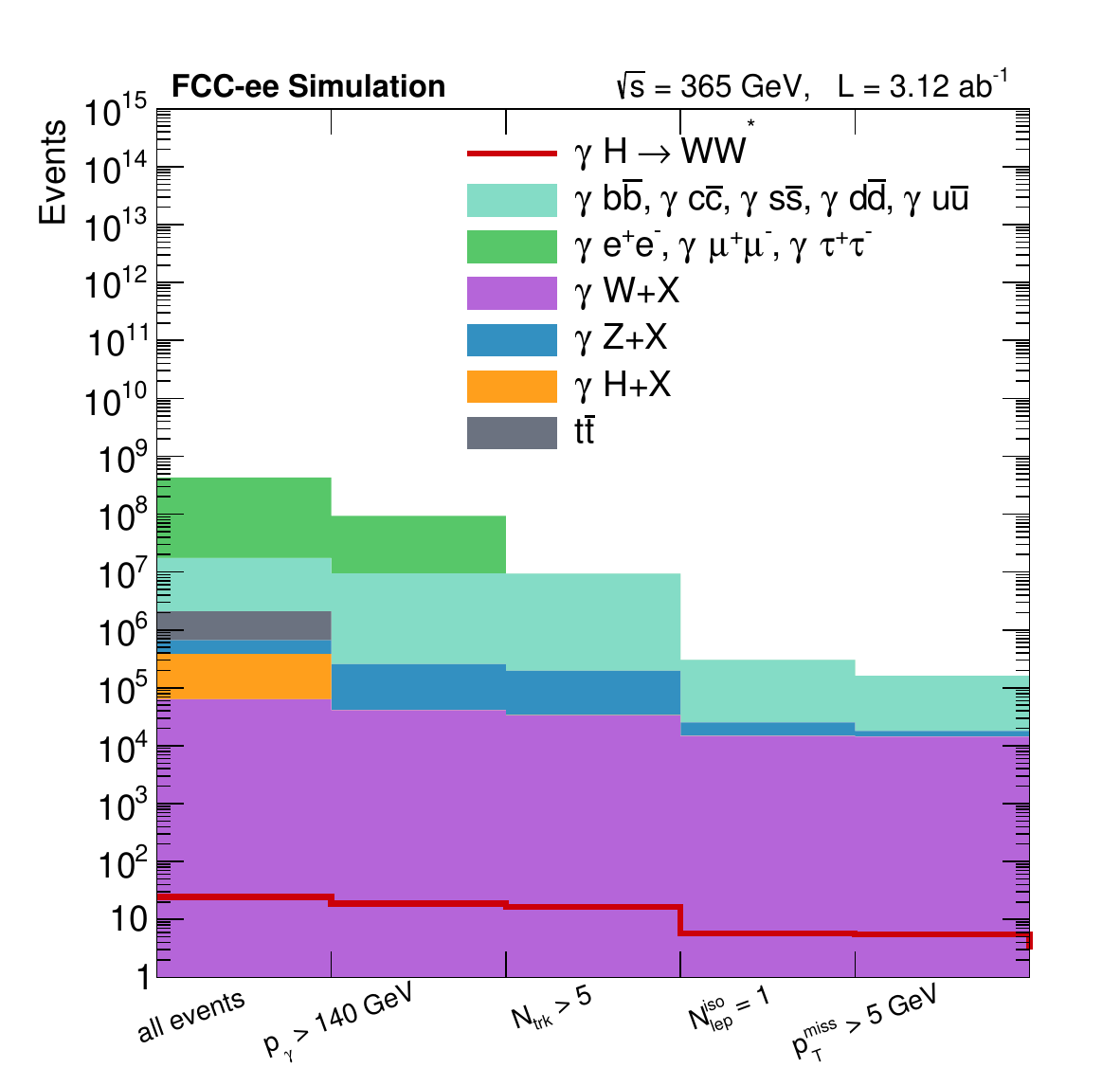}
  \end{minipage}

    \caption{
    Left: Cumulative yields for the signal and dominant background processes in the $WW^*$ final state for \(\sqrt{s}=160\)~GeV and \(\sqrt{s}=365\)~GeV at the preselection stage. 
    }
  \label{fig:hww_presel_160_365}
\end{figure}

\begin{figure}[htb]
  \centering
  \begin{minipage}[t]{0.49\textwidth}
    \centering
    \includegraphics[width=\linewidth]{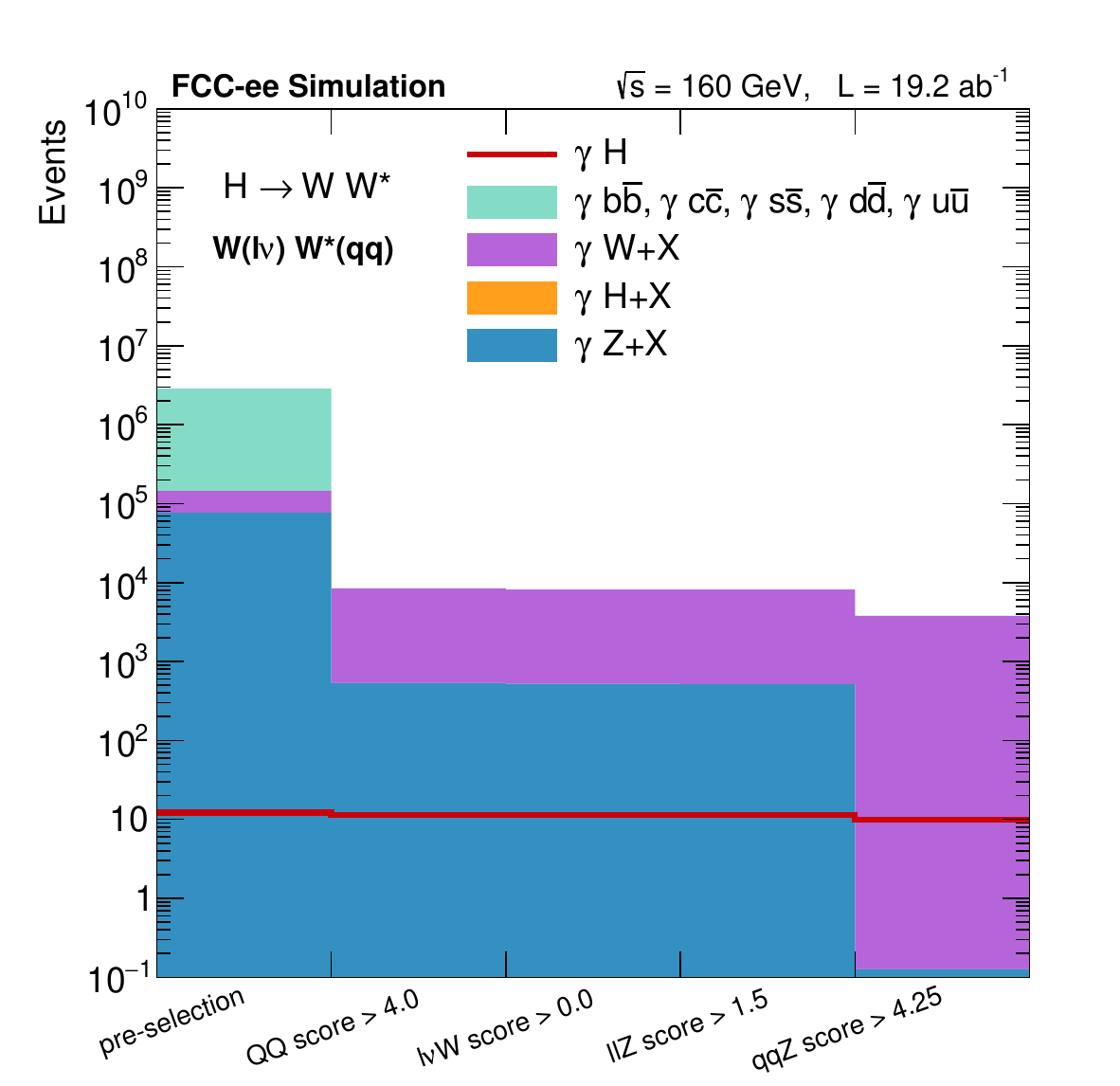}
  \end{minipage}\hfill
  \begin{minipage}[t]{0.49\textwidth}
    \centering
    \includegraphics[width=\linewidth]{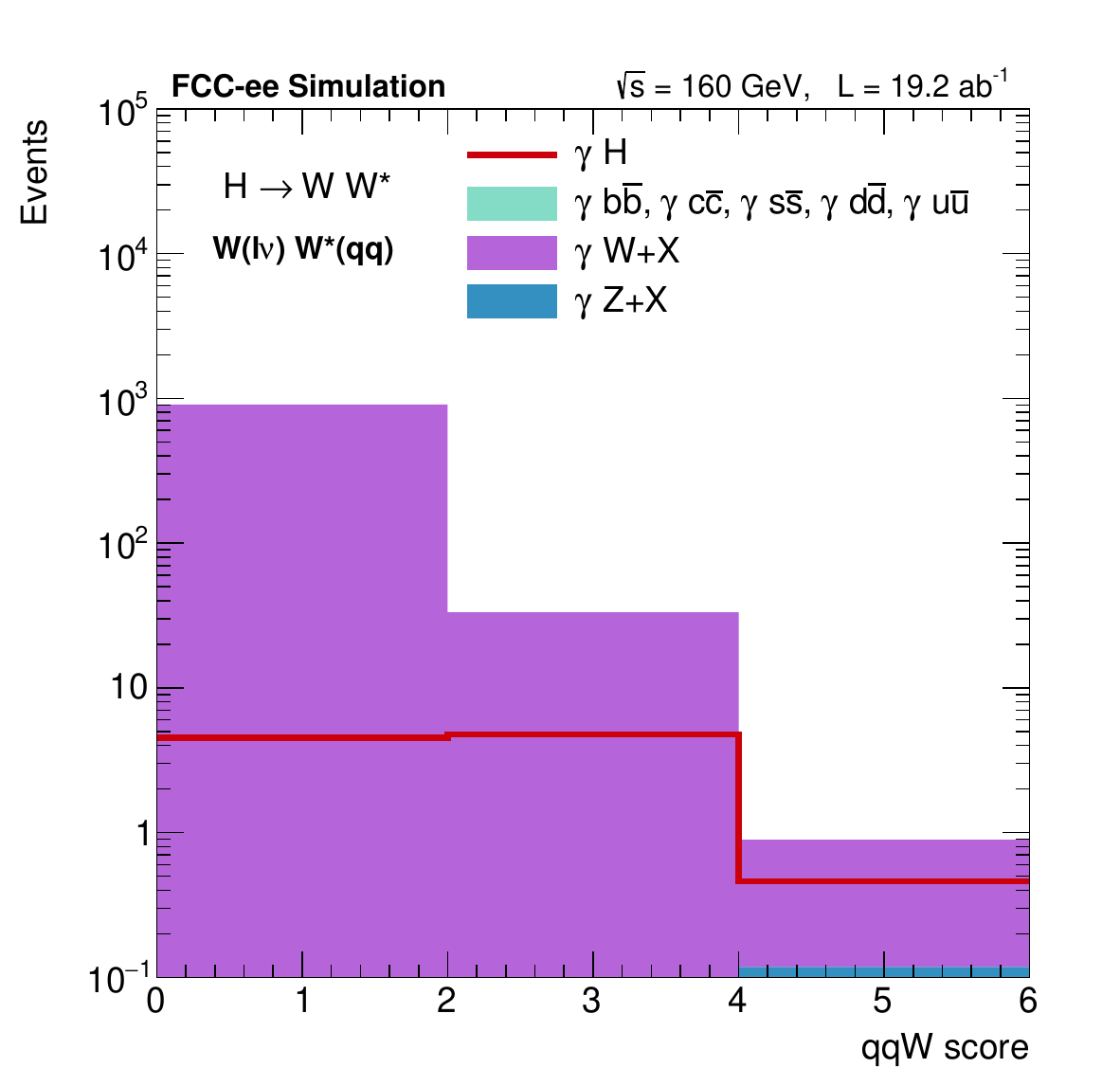}
  \end{minipage}

\caption{
Final selection for the $H\rightarrow W(\ell\nu)\,W^*(qq)$ analysis at
\(\sqrt{s}=160\)~GeV.
Left: cumulative cut flow for the signal and dominant background classes after the multiclass BDT–based selections.
Right: BDT discriminant against the remaining \(\gamma q\bar q W\) background, used as input to the binned likelihood fit.
}
\label{fig:hww_lvqq_final_160}
\end{figure}

\begin{figure}[htb]
  \centering
  \begin{minipage}[t]{0.49\textwidth}
    \centering
    \includegraphics[width=\linewidth]{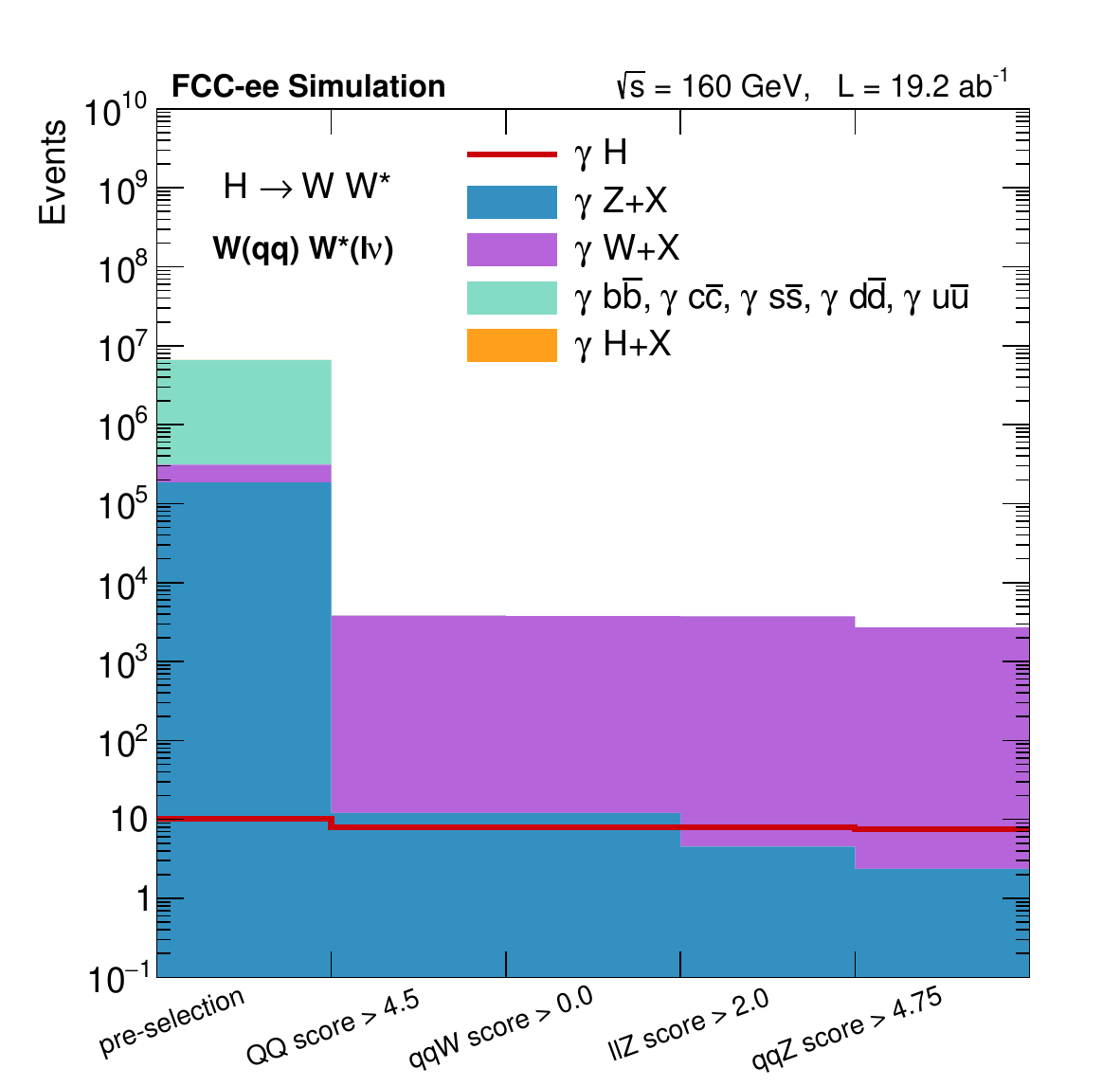}
  \end{minipage}\hfill
  \begin{minipage}[t]{0.49\textwidth}
    \centering
    \includegraphics[width=\linewidth]{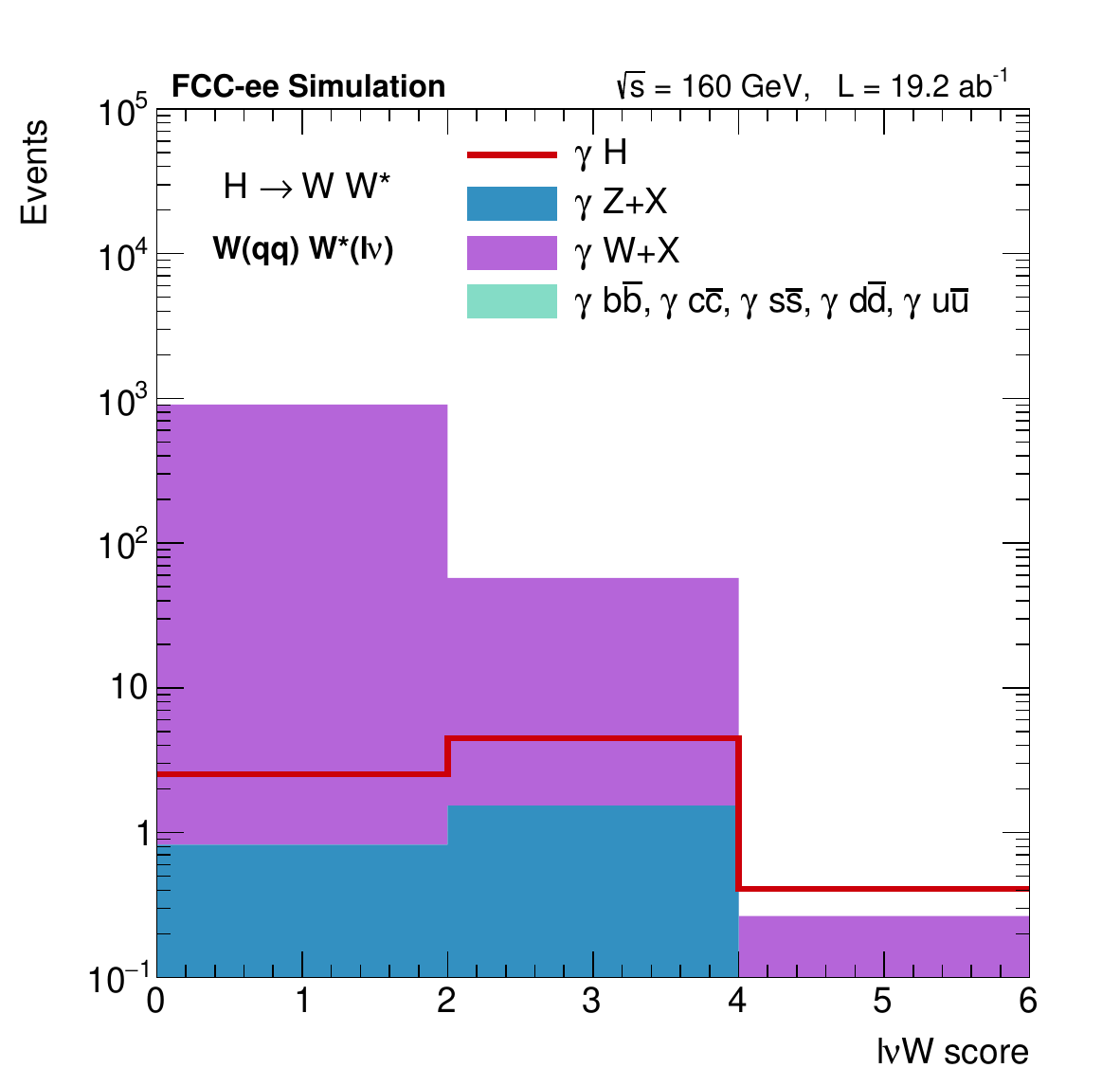}
  \end{minipage}
\caption{
Final selection for the $H\rightarrow W(qq)\,W^*(\ell\nu)$ analysis at
$\sqrt{s}=\SI{160}{\giga\electronvolt}$.
Left: cumulative cut flow for the signal and dominant background classes after the multiclass BDT-based selections.
Right: BDT discriminant against the remaining $\gamma\,\ell\nu W$ background, used as input to the binned likelihood fit.
}
\label{fig:hww_qqlv_final_160}
\end{figure}

\begin{figure}[t]
  \centering
  \begin{minipage}[t]{0.49\textwidth}
    \centering
    \includegraphics[width=\linewidth]{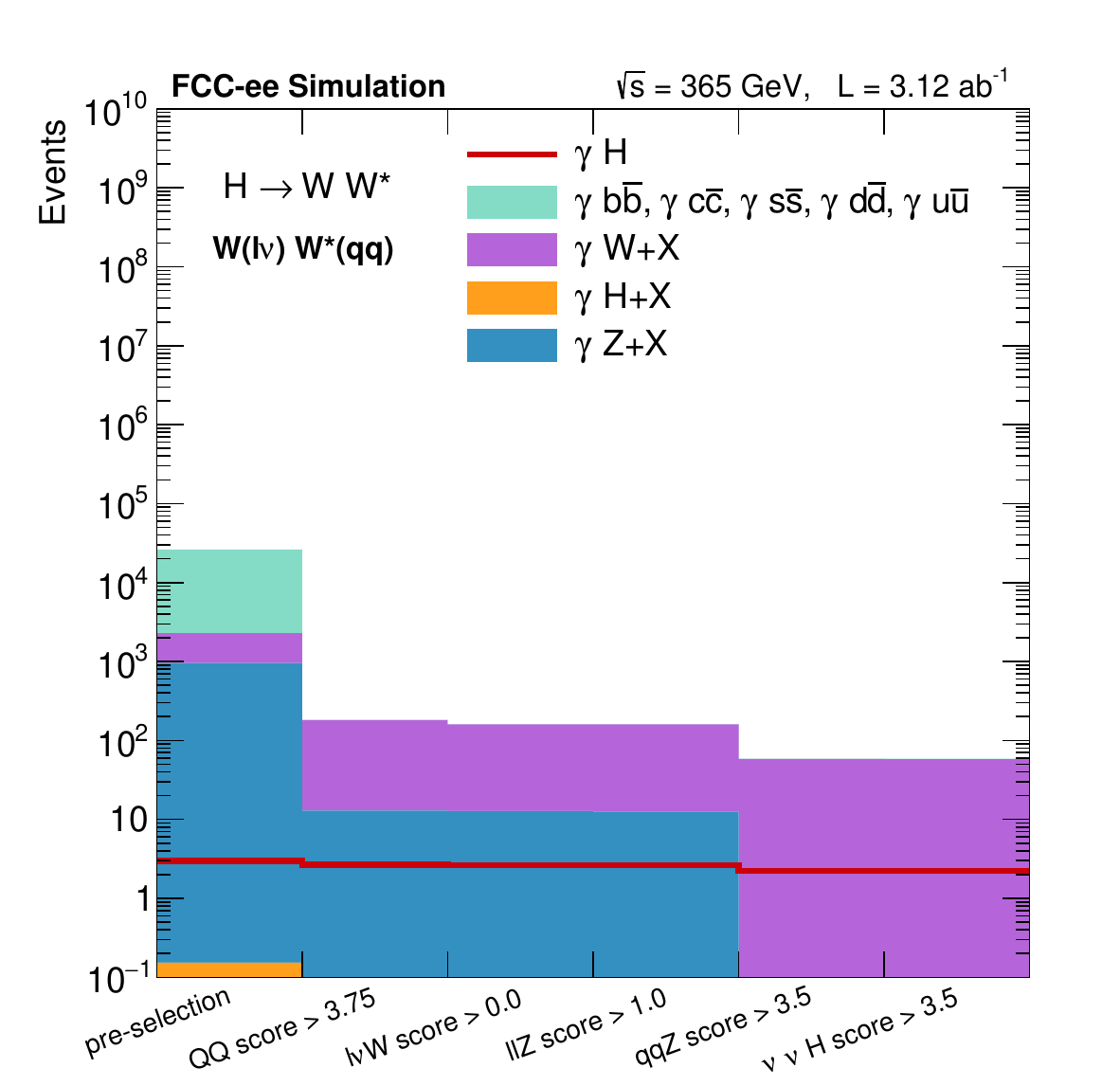}
  \end{minipage}\hfill
  \begin{minipage}[t]{0.49\textwidth}
    \centering
    \includegraphics[width=\linewidth]{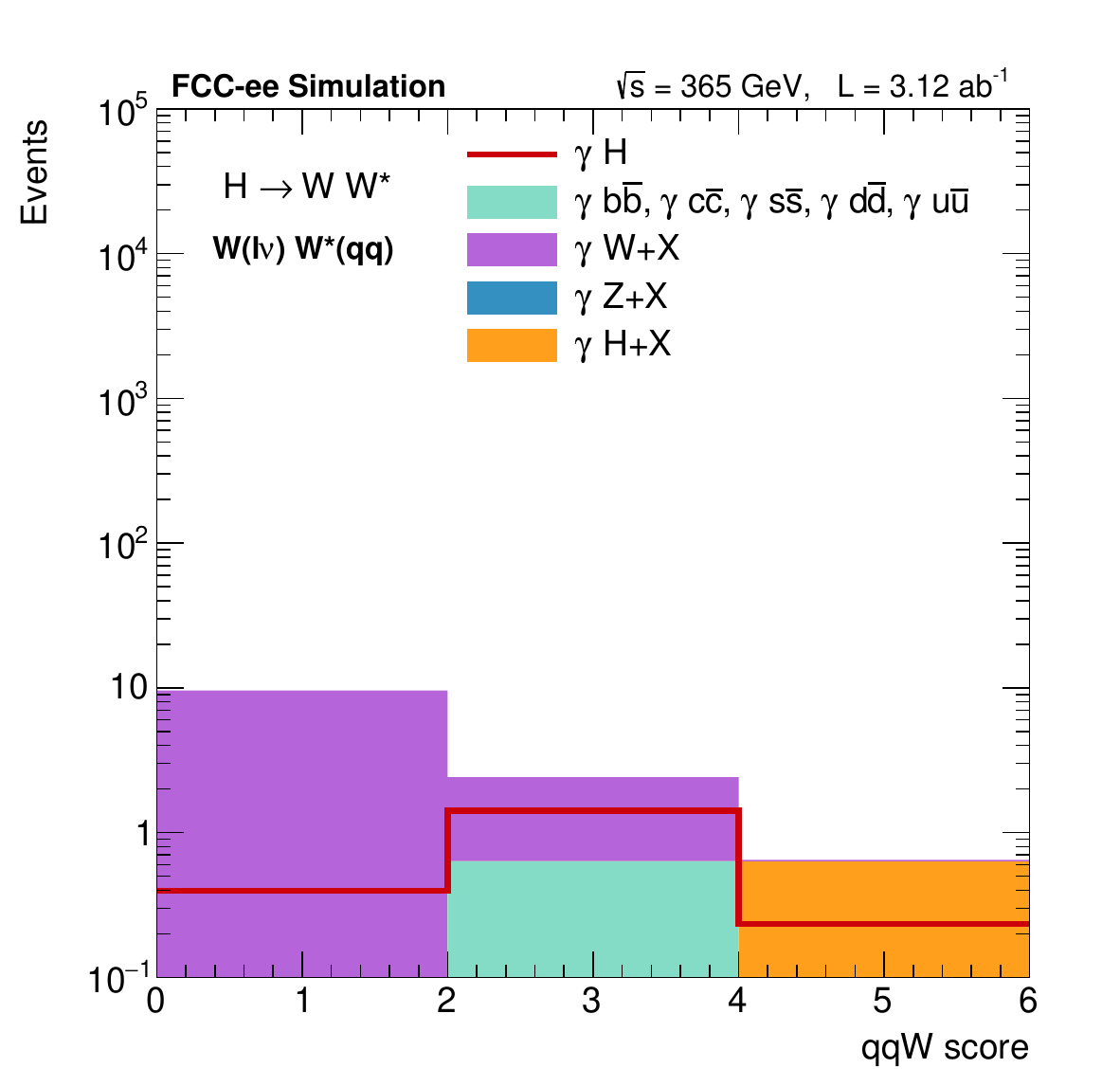}
  \end{minipage}

\caption{
Final selection for the $H\rightarrow W(\ell\nu)\,W^*(qq)$ analysis at
\(\sqrt{s}=365\)~GeV.
Left: cumulative cut flow for the signal and dominant background classes after the multiclass BDT–based selections.
Right: BDT discriminant against the remaining \(\gamma q\bar q W\) background, used as input to the binned likelihood fit.
}
\label{fig:hww_lvqq_final_365}
\end{figure}

\begin{figure}[htb]
  \centering
  \begin{minipage}[t]{0.49\textwidth}
    \centering
    \includegraphics[width=\linewidth]{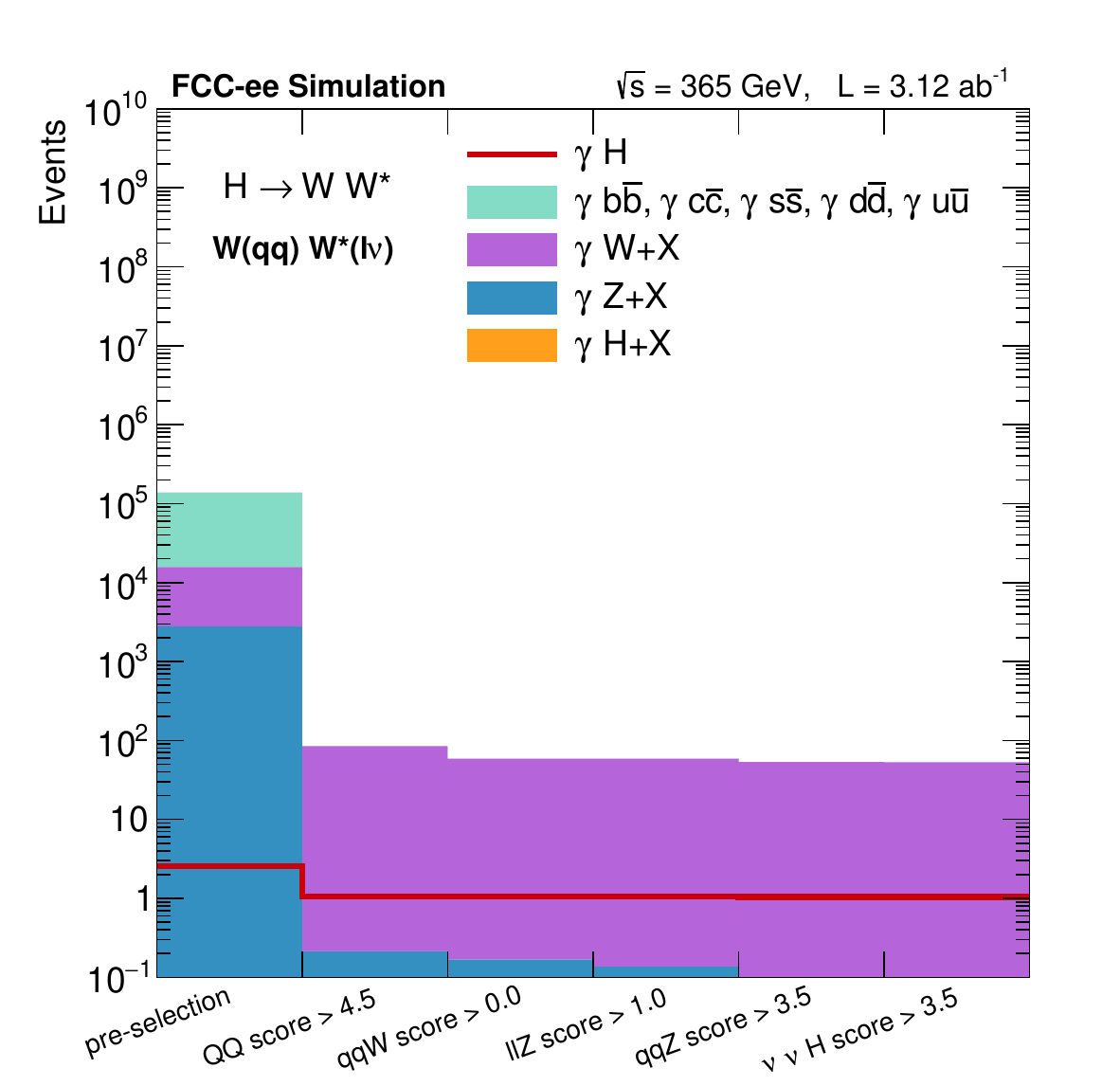}
  \end{minipage}\hfill
  \begin{minipage}[t]{0.49\textwidth}
    \centering
    \includegraphics[width=\linewidth]{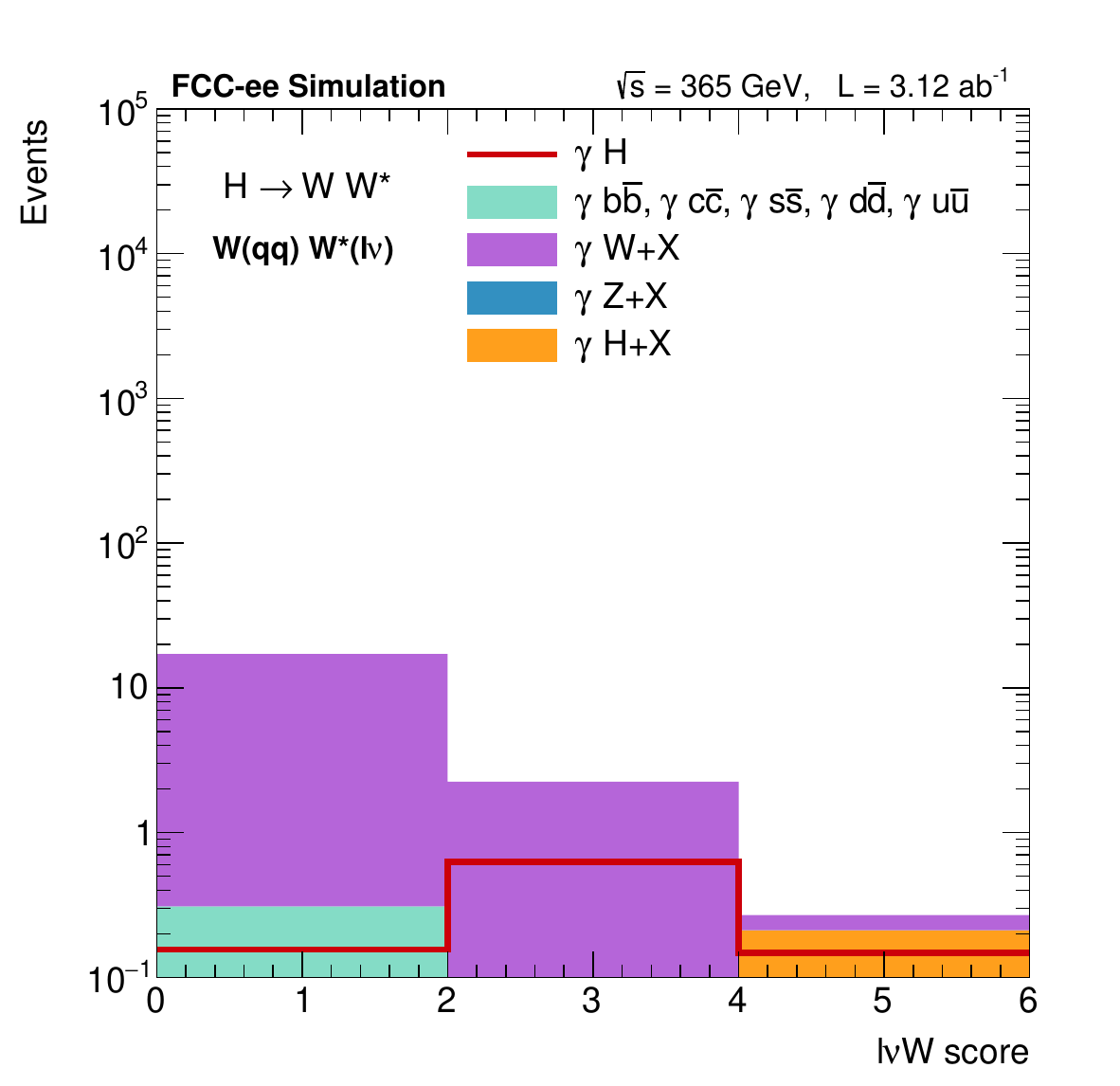}
  \end{minipage}
\caption{
Final selection for the $H\rightarrow W(qq)\,W^*(\ell\nu)$ analysis at
$\sqrt{s}=\SI{365}{\giga\electronvolt}$.
Left: cumulative cut flow for the signal and dominant background classes after the multiclass BDT-based selections.
Right: BDT discriminant against the remaining $\gamma\,\ell\nu W$ background, used as input to the binned likelihood fit.
}
\label{fig:hww_qqlv_final_365}
\end{figure}

The effect of the $WW^*$ channel preselection at $\sqrt{s}=\SI{160}{\giga\electronvolt}$ and $\sqrt{s}=\SI{365}{\giga\electronvolt}$ is shown in~\cref{fig:hww_presel_160_365}.  In a second stage, cuts are applied on the multiclass BDT output scores to suppress all reducible backgrounds, and the dedicated BDT discriminant against the remaining irreducible background is used as the observable in the binned profile likelihood fit. The resulting cutflows and final discriminants are shown in~\cref{fig:hww_lvqq_final_160,fig:hww_lvqq_final_365} for the 
$H\rightarrow W(\ell\nu)\,W^*(qq)$ channel and 
in~\cref{fig:hww_qqlv_final_160,fig:hww_qqlv_final_365} for the 
$H\rightarrow W(qq)\,W^*(\ell\nu)$ channel, at $\sqrt{s}=\SI{160}{\giga\electronvolt}$ 
and \SI{365}{\giga\electronvolt}, respectively.

\FloatBarrier

\section{$H\rightarrow b\bar{b}$ channel, alternative energies}
\label{app:hbbpreselection}

\begin{figure}[htb]
    \centering
    \includegraphics[width=0.45\linewidth]{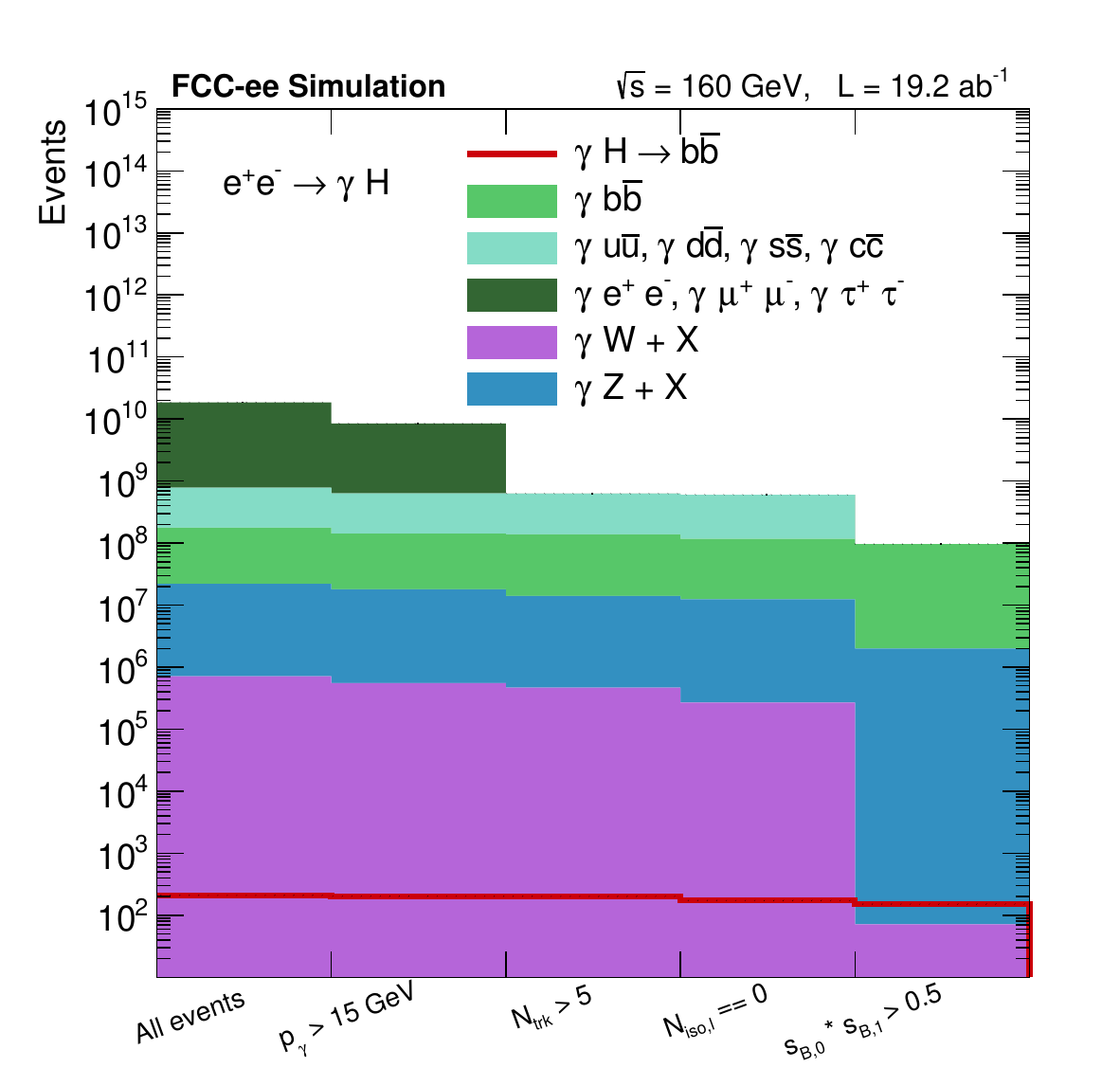}
        \includegraphics[width=0.45\linewidth]{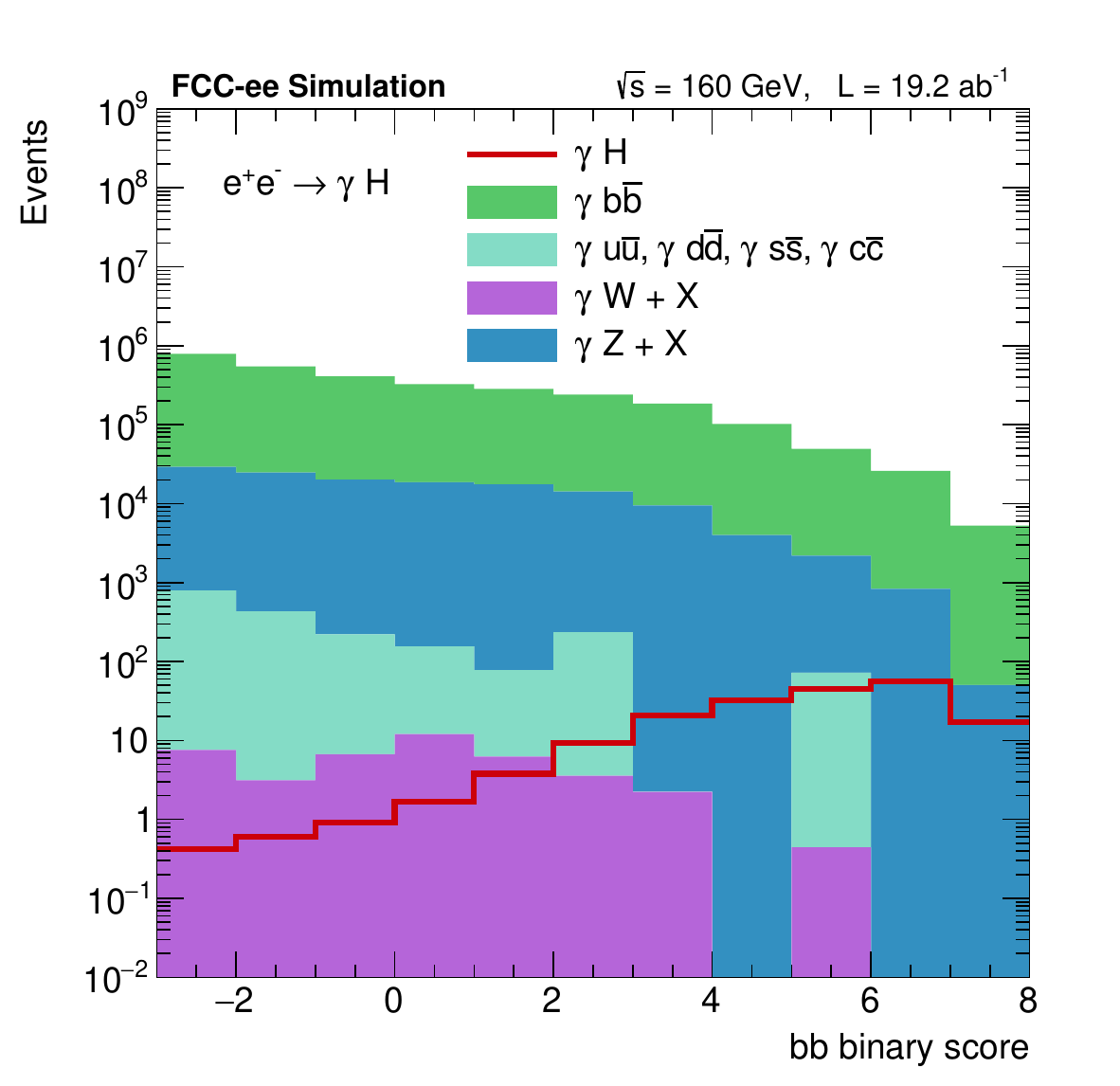}
    \caption{Final selection for the $H\rightarrow b\bar{b}$ analysis at \(\sqrt{s}=160\)~GeV. Left: cumulative cut flow of the applied selection. Right: BDT discriminant, used as input to the binned likelihood fit.}
    \label{fig:fitobs_160}
\end{figure}

\begin{figure}[htb]
    \centering
    \includegraphics[width=0.45\linewidth]{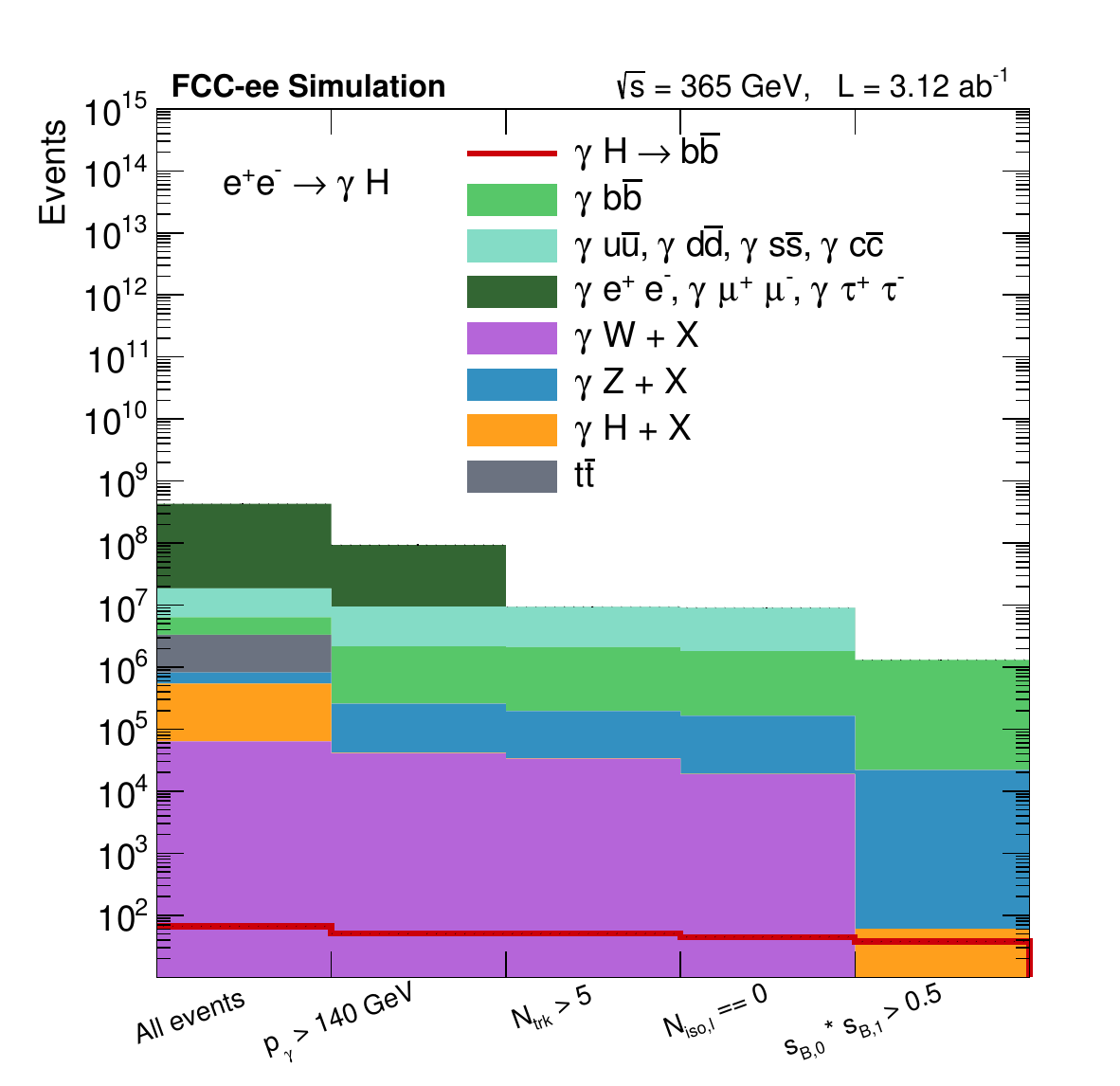}
    \includegraphics[width=0.45\linewidth]{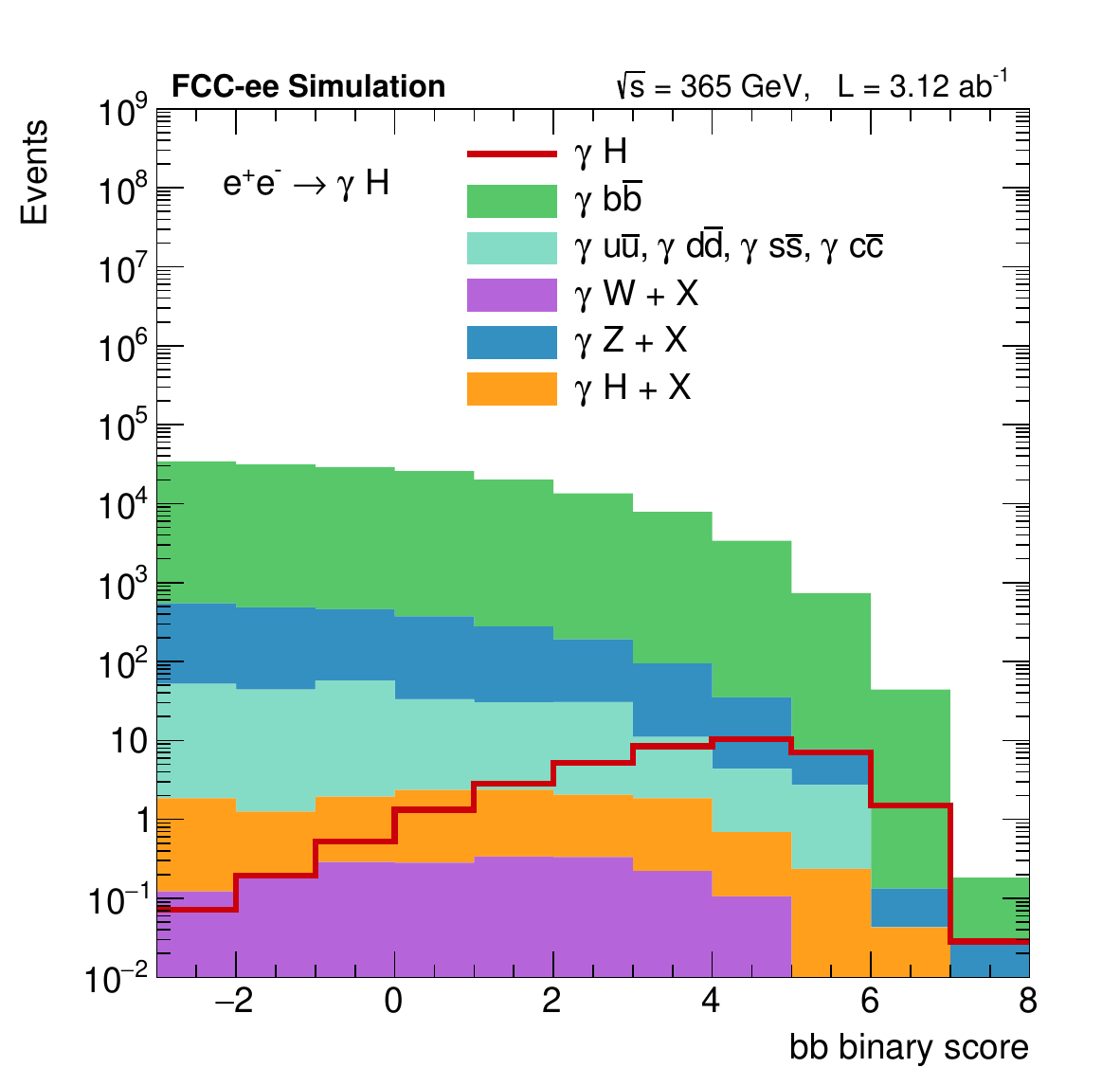}
    \caption{Final selection for the $H\rightarrow b\bar{b}$ analysis at \(\sqrt{s}=365\)~GeV. Left: cumulative cut flow of the applied selection. Right: BDT discriminant, used as input to the binned likelihood fit.}
    \label{fig:fitobs_365}
\end{figure}

In the $H\rightarrow b\bar{b}$ analysis, the common preselection is extended by a channel specific cut on the product of b-tagger scores. A BDT is then trained on the selected events to differentiate signal events from the dominant irreducible $\gamma b\bar{b}$ background. The cumulative cut flow and the logit of the $b\bar{b}$ score are shown in \cref{fig:fitobs_160} and \cref{fig:fitobs_365} for the $\sqrt{s}=~\SI{160}{\giga\electronvolt}$ and $\sqrt{s}=~\SI{365}{\giga\electronvolt}$ scenario, respectively.

\FloatBarrier

\bibliographystyle{JHEP}
\bibliography{sn-bibliography}

\end{document}